\documentclass{aa}  

\pdfoutput=1

\usepackage{amsmath}
\usepackage{bm}

\usepackage{graphicx}
\usepackage[labelfont=bf]{caption}

\usepackage{subfig}
\usepackage{rotating}

\usepackage{txfonts}

\usepackage{makecell}
\usepackage{multirow}

\usepackage{hyperref}
\usepackage{xcolor}
\hypersetup{
    colorlinks,
    linkcolor={blue!80!black},
    citecolor={blue!80!black},
    urlcolor={blue!80!black}
}
\usepackage{natbib}
\bibpunct{(}{)}{;}{a}{}{,}
\usepackage{nicefrac}
\usepackage{dblfloatfix}
\usepackage{boldline}

\newcommand{\Teff}{T_\mathrm{eff}}

\newcommand{\Mbound}{M$_\mathrm{bound}~$}

\newcommand{\dMgw}{$\Delta\dot{M}_{gw}$ }

\newcommand{\Msun}{\textrm{M}$_\mathrm{\odot}$}
\newcommand{\Mearth}{\textrm{M}$_\mathrm{\oplus}$}
\newcommand{\Mj}{\textrm{M}$_\mathrm{J}$}

\begin{document}

   \title{The quest for Magrathea planets I: formation of second generation exoplanets around double white dwarfs.}

   \author{Ledda Sebastiano 
          \inst{\ref{sap}}
          \and
          Danielski Camilla \inst{\ref{iaa}}
          \and
          Turrini Diego \inst{\ref{oato},\ref{inaf}}
          }

   \institute{Sapienza - University of Rome, Physics
            department, Piazzale Aldo Moro 5 - I-00185, Rome, Italy\label{sap}\\
              \email{ledda.1350727@studenti.uniroma1.it}
            \and
            Instituto de Astrof\'isica de Andaluc\'ia, CSIC, Glorieta de la Astronom\'ia s/n, 18008, Granada, Spain.\label{iaa}
            \and
            INAF - Osservatorio Astrofisico di Torino, via Osservatorio 20, 10025, Pino Torinese\label{oato}
            \and
            INAF-IAPS, Via Fosso del Cavaliere 100, 00133, Rome, Italy\label{inaf}
             }

   \date{}
 
  \abstract
   {The formation of planets around binary stars is the subject of ongoing investigations focusing on the early stages of stellar life. The evolution of binaries that become double white dwarf (DWD), however, can cause the ejection of high amounts of dust and gas. Such material can 
   give rise to circumbinary discs and become the cradle of new planets, yet no studies so far have focused on the formation of circumbinary planets around DWDs. These binaries will be the main sources of gravitational waves (GWs) detectable by Laser Interferometer Space Antenna (LISA) mission from the European Space Agency (ESA), opening the possibility to detect circumbinary planets around short-period DWDs everywhere in the Milky Way and in the Large Magellanic Cloud via the modulation of their GW signal.
   }
   {We want to investigate the formation process and its characteristics, e.g., formation times, masses, final locations of \emph{Magrathea} planets within circumbinary discs around detached DWDs, paying particular attention to the formation of gas giant (GG) planets.
   }
   {We simulate multiple planet formation tracks  
   to explore how the planetary formation processes typical of pre-Main Sequence (pre-MS) discs are affected by the disc environments surrounding DWDs. We investigate the mass and orbital evolution of planetary seeds growing first through pebble accretion, then by gas accretion. Our growth tracks account for both the disc accretion rate onto the central binary and the disc photoevaporation rate caused by stellar irradiation.
   }
   {We present both planetary formation tracks taking place in steady-state discs, and formation tracks taking place in discs evolving as a function of time. Our simulations show that planetary formation should be common in circumbinary discs around DWDs, but the formation of GG planets can be hindered by the temperatures of the disc and the rapid disc depletion.
   }
   {Our results show that planetary formation in circumbinary discs around DWDs can be possible. In particular, the extreme planetary formation environment implies three main significant results: (i) the accretion rate and the metallicity of the disc should be high in order to form sub-stellar objects with masses up to $\sim 31$ \Mj, this is achieved only if planet formation starts soon after the onset of the disc and if first generation seeds are present in the disc; (ii) seeds formed within $0.1$ Myr, or within $1$ Myr, from the onset of the disc can only produce sub-Neptune (SN) planets and Neptunian (N) planets, unless the disc accommodates first generation seeds with mass $10$\ \Mearth ; (iii) most of the planets are finally located within $1$ au from the disc centre, while they are still undergoing the gas accretion phase.
   }

    
   \titlerunning{Second generation formation of Magrathea exoplanets}
   \maketitle

\section{Introduction}
\label{introduction}
To date, most of the $5235$ confirmed\footnote{\href{https://exoplanetarchive.ipac.caltech.edu/}{NASA Exoplanet Archive} - Consulted on 28 December 2022.} exoplanets have been found orbiting single stars. 
Yet, a consistent fraction ($\sim$4\% as of May 2022\footnote{Catalogue of exoplanets in binary star systems, maintained by R. Schwarz \citep{schwarz}}) orbits stellar binaries. All planets that form within protoplanetary discs surrounding pre-main sequence (pre-MS) stars (no matter their multiplicity) are usually referred to as `first generation` planets. However, both theory and observations show that discs surrounding binaries can also appear at later times as a consequence of stellar evolution processes, i.e.,  when one of the components of the central binary has already left the Main Sequence (MS) (e.g., \citealt{kashiandsoker, menu, hardy, vanwinckel, oomen2}; and references there-in). 
This disc, formed in the post asymptotic giant branch (post-AGB) phase of one of either the stellar components, is called a ``second generation disc''.  
Any planet forming in such an environment is commonly referred to as ``second generation planet'' \citep{Perets2010}.

In this paper we focus on second generation discs around double white dwarfs (DWDs) systems. The evolutionary stages leading to the formation of DWDs are still uncertain. The overall picture emerging from current studies is that the DWDs progenitors experience at least two mass transfer phases, occurring when one of the stars becomes a giant or a sub-giant, through various processes that depend on the initial stellar masses and binary separation. In the case of an unstable mass transfer, the donor star, i.e.,  the one that is evolving, shares its atmosphere with the companion, fully enveloping it in the ``common envelope'' (CE) phase (e.g., \citealt{paczynski, ivanova}). At the end of this phase, when the shared envelope is ejected, the fraction of its mass that does not reach the escape velocity remains bound to the central binary, eventually forming a second generation circumbinary disc \citep{kashiandsoker}. 

When a post-common envelope (post-CE) disc forms, there is the possibility for it to be the cradle of second generation planets, whose mass could reach up to some Jovian mass \citep{schleicher}. A work by \citet{zorotovic} found that many eclipsing post-common envelope binaries (PCEBs) display eclipse timing variations that could be attributed to circumbinary sub-stellar bodies. Should the latter prove indeed of planetary nature, these authors suggest that it is unlikely that the planets formed when both components of the system were MS stars. Rather, they may have formed following the CE stage and the creation of a second generation disc. The system of NN Serpentis provides an illustrative example of this process. NN Serpentis consists of a white dwarf (WD) of $0.535$ M$_{\odot}$ and a red dwarf of $0.111$ M$_{\odot}$ \citep{mustill}, and several studies agree in suggesting a second generation nature of its two circumbinary gas giant (GG) planets\footnote{It is important to mention that a recent study \citep{pulley} raised the necessity of a refinement of the models used to describe the Eclipse Time Variations (ETVs) observed for NN Ser. Without such refinement, the models fail to predict ETVs more than a year into the future. This do not exclude however the presence of circumbinary planets, rather the authors suggest the importance of deepen other mechanisms affecting ETVs, e.g. magnetic effects associated with the secondary of the binary system.} \citep{zorotovic, mustill, schleicher, volschow, hardy}. 

Recently \citet{kluska} suggested the possible presence of planets around post-AGB binaries. These systems are thought to have experienced a fast CE phase caused by the evolution of an asymptotic giant branch (AGB) star, i.e.,  the progenitor of a WD. Such mass transfer phase produces a circumbinary disc similar to that observed around young stars. However, $\sim 12\%$ of the sample of post-AGB stars considered by \citet{kluska} show the presence of transition discs, i.e.,  discs having large inner dust cavities. Such cavities have been interpreted by the authors as caused (i) by the presence of a first generation GG that survived the mass transfer phase or (ii) by the formation of a second generation GG after the mass transfer. In the first case a variation in the local pressure gradient of the disc, caused by the presence of a surviving planet, could prevent the inward flux of dust to reach the inner disc regions, favouring the formation of a second generation planet in disc regions outside the first generation planet's orbital radius. Both cases (i) and (ii) point that the second generation planet should form within $\sim 1$ Myr, depending on the disc mass, the gas accretion rate onto the binary, and on the initial temperature of the AGB star \citep{oomen}. 
Given that the post-common-envelope age of the NN Serpentis system is $\sim$ 1 Myr, the same concept is applicable to its planets, which should have formed within the age of the system, should they be planets of second generation.

Accounting for planets in evolved stellar binary systems we know that no \emph{Magrathea}\footnote{From ``The Hitchhiker’s Guide to the Galaxy'' book  \citep{Magrathea} Magrathea is a planet orbiting a binary star burning with “white fire”, which we conveniently interpreted to be two white dwarfs, given that they emit white light instead of yellow, like our Sun in the visible band.} planet, i.e., a circumbinary planet orbiting a DWD, has been detected yet \citep{dani2019}. Nor around binaries whose components have both evolved, apart from the planet around PSR B1620-26AB (where the binary is composed by a WD and a pulsar). This planetary system is believed to be the result of a stellar encounter, hence shaped by a very different evolution than a typical isolated circumbinary system \citep{sigurdsson}. Nonetheless, circumbinary planets around evolved binaries are expected to exist, either as first-generation planets surviving one or more CE \citep{kostov,Columba2022}, or by forming directly in circumbinary second-generation discs, as previously discussed. 

To date no dedicated study have been performed on the formation of second generation planets around DWDs. However, due to the similarities that discs around evolved binaries share with their pre-MS counterparts (\citealt{Perets2010} and references there-in), it is reasonable to expect that the processes that regulate the growth of planets in pre-MS discs can also occur in second generation discs. As the evolution of circumbinary discs around evolved binaries is still uncertain, in this study we adopt the well-studied framework developed for the evolution of pre-MS discs \citep{lyndenpringle, hartmann}. Pre-MS discs are expected to viscously evolve such that while the mass of the disc flows inward, the angular momentum flows outward, i.e.,  the disc gets depleted because of accretion onto the central object while increasing its extension.

The circumbinary discs we consider here are formed by the CE material that remains bound to the inner binary, i.e.,  coming from the envelope of the secondary giant evolving star. In the particular case of a DWD, after the second CE we are left with a new-born WD and an older, cooler WD. The temperature of a young WD can be very high (more than $\sim 10^4$ K), and decreases over time down to the order of $10^3$ K \citep{mestel}. This can result in a disc with initially high overall temperatures, such that the condensation temperatures of both dust and ices are reached at larger orbital radii than in pre-MS discs. In turn, this implies that the growing solid core of a planet, which starts its formation in the outer regions of the disc, can have a smaller feeding zone as it migrates toward the inner hotter regions of the disc. Moreover, the irradiation of the central object onto the disc causes photoevaporation of the gas particles of the disc over time, which, together with the accretion onto the binary, contributes to the disc depletion rate, and thus to the disc lifetime. In order for the formation of a giant planet to compensate for such limiting processes, the accretion of solids onto the initial planetary core should be highly efficient.  

In this manuscript, we explore the scenario of planetary formation occurring in second generation discs, i.e.,  the formation processes in circumbinary discs around detached DWDs. 
Short-period (less than $\sim 2$ h) DWDs are expected to be the most common sources of gravitational waves (GWs) in the observational band of the Laser Interferometer Space Antenna mission (LISA, \citealt{LISAcallpaper, AmaroSeoane2022:WP, korol}). Recent studies opened up the possibility for giant exoplanets to be found around these binaries through the detection of the modulation the planet causes in the GW signal produced by the DWD itself \citep{TamaniniDanielski2019,dani2019,DanielskiTamanini2020}. For such, we focus our studies on the formation of giant exoplanets.
This sets our work in the broader context of the studies for the development of the planetary detection science case of the LISA mission. Such a development includes the LISA detection prospects of Sub-Stellar Objects (SSOs) orbiting DWDs through Bayesian analysis by \cite{Katz2022},  the modelling of the long-term evolution of circumbinary systems hosting a giant planet, from MS till the DWD stage by \cite{Columba2022}, and the determination of the LISA detection efficiency by \cite{Danielski-in-prep}.

The paper is organised as follows: 
in Sec. \ref{methods} we describe the planet-forming environment, i.e.,  the specific DWD systems and circumbinary disc models we adopt for our simulations, together with the planetary formation processes. In Sec. \ref{sec:results} we present our results, which consist of growth tracks in steady-state discs (usually referred to as ``analytical formation tracks''), and growth tracks and migration tracks in time-evolving discs.  These tracks allow us to test the kind of planets that can form, and the time needed, depending on the free parameters of the problem. 
Finally, we discuss the results in Sec. \ref{discussion} and draw conclusions in Sec \ref{sec:conclusions}.

\begin{table*}[t]
    \caption{Stellar masses, periods $P$ and eccentricity $e$ of the analysed  DWDs at the end of the second CE, and respective progenitors.}
    \label{tab:DWDparams}
    \centering
    \begin{tabular}{|c|c|c|c|c|c|c|c V{4} c|c|c|c| }
        \hline
         & M$_1$ [\Msun] & M$_2$ [\Msun] & P [h]  & $e$ & $\Delta$t & $\Teff$$_1$ [K] & $\Teff$$_2$ [K] &
         M$_\mathrm{Z1}$ &  M$_\mathrm{Z2}$ & P$_Z$ & e$_Z$ \\
         \hline
        DWD$_1$   & 0.60 & 0.38 & 20.47 & 0 & 135 Gyr & 8600 & 75000 & 2.6 & 1.9 & 172 days & 0\\
        DWD$_2$   & 0.21 & 0.31 &  0.43 & 0 &  9 Myr  & 20400 & 58400 & 1.55 & 1.25 & 1.13 days & 0\\
        DWD$_3^*$ & 0.75 & 0.26 &  1.51 & 0 & 150 Myr & 8800 & 52000 & 3.47 & 1.95 & 52 yr & 0.98 \\
        DWD$_4^*$ & 0.31 & 0.25 &  1.71 & 0 & 430 Myr & 8100 & 50100 & 1.84 & 1.28 & 83.5 days & 0.68 \\
        \hline
    \end{tabular}
    
    \tablefoot{\\While the subscript $Z$ marks the stellar mass and period at the ZAMS, subscripts $1$ and $2$ refers to the primary and the secondary star, respectively. $\Delta$t is the time needed for the merging of the two DWDs computed via Eq. 9 of \citet{dani2019}. $\Teff$$_1$ and $\Teff$$_2$ are the effective temperatures of the WDs immediately after the second CE. The period $P$ corresponds to the period of the DWD binary just after the second CE.
    The systems marked with the $*$ belong to the LISA DWD population presented in \citet{Korol2019}. More details can be found in Sec. \ref{sec:dwdbinary}.
    }
\end{table*}

\section{Methods and models}
\label{methods}
The planet-forming environment we consider in this work, i.e.,  a second generation disc around a DWD, is characterised by a parameter space usually not available in MS systems. Therefore, it can offer an important test bed for the planet formation theories currently used for first generation discs. In this section we describe both the DWD systems and the circumbinary disc model we used.

\subsection{The planet-forming environment: the central DWD}
\label{sec:dwdbinary}
In order to explore differences and similarities among the formation processes in post-AGB and pre-MS discs, we tested four DWD systems. The parameters of each DWD system are reported in Tab. \ref{tab:DWDparams}, together with the values of their progenitors' parameters. The systems span a different range of masses and periods, to account for a variety of circumbinary discs in term of total mass and disc cavity radius (see Sec. \ref{sec:discmodel}).
The evolutionary history of these systems has been simulated using the binary population synthesis code {\sc SeBa}\footnote{The software package simulates the evolution of single and binary stars from the zero-age main-sequence (ZAMS) up to and including remnant phases, and it can be found at \url{https://github.com/amusecode/SeBa}}, originally developed by \citet{por96} and later adapted for DWDs by \citet{nel01} and \citet{too12}. 

The system DWD$_1$ is the main system we used for our simulations, as it is expected to represent a typical DWD system \citep{korol} whose subsequent evolution is such that planet formation has enough time to occur.
After the second CE the DWD$_2$ binary has a period of 0.43 hours  which, accounting for the binary mass and circular orbit, will produce a GW frequency of $\omega_{GW} = 8.11$ mHz falling within the frequency range of the LISA band (0.1 mHz - 1 Hz). This system  will merge in 9 Myr and for such reason we selected two more systems, DWD$_3$ and DWD$_4$, from the LISA DWD population, meaning that, when placed within the Milky Way's history context, their orbits have shrunk enough for them to be 
detectable by LISA \citep{korol,Korol2019,AmaroSeoane2022:WP}. These two systems have been selected with a longer merging time (Tab. \ref{tab:DWDparams}) to make sure we account for enough time for planetary formation before any other binary evolution happen in the system i.e., contact or mass transfer, with final merging in single WD or supernova Ia. In either cases such further evolutionary steps could (i) bring to a non-binary object, or (ii) preclude any detection with LISA, which is the reason why we excluded it from our study.

We use the exploration of the planet formation potential of the system DWD$_1$ as our illustrative case study, while for the remaining systems we exclusively focus on the growth tracks in discs evolving as a function of time, as they provide the more realistic information. 
We note that even for eccentric progenitors, the orbits of all binary circularise throughout evolution: as a consequence all our disc models are based on circular stellar orbits. Moreover, the binaries have very small periods, such that their gravity field can be considered as a central one, and that their perturbations are effective only over a very short distance from the barycentre.
SeBa simulations provided the temperature of the secondary star (i.e.,  the youngest WD), $\Teff$$_2$, immediately after the CE, together with the type of WD formed (i.e., He WDs for all the systems analysed). For our simulations we set the initial temperature, at t$_0 = 0$, to the SeBa $\Teff$$_2$ ones. 
However, we accounted for the variation of the WD temperature after the second CE. In particular, we chose two more temperature values within $1$ Myr from the end of the CE i.e., 0.1 and 1 Myr. Such temperatures have been measured using the synthetic evolutionary sequences for Helium atmospheres\footnote{\url{http://www.astro.umontreal.ca/~bergeron/CoolingModels}}, whose high temperature models are developed by \cite{Bedard2020} and include non-local thermodynamic equilibrium (non-LTE) effects. The reason for which we accounted for a variation of the stellar temperature is to explore different types of disc environments (see Sec. \ref{sec:discmodel}).

\subsection{The second-generation circumbinary disc model}
\label{sec:discmodel}
As mentioned above the disc model is linked to the type of DWDs formation through the CE phase, and the disc mass depends on the binary system characteristics and its evolution. In principle circumbinary discs can form both after the first and the second CEs. However, as the disc fate across the first CE is unclear and its lifetime is highly dependent on the lag of time between the evolution out of MS of both stars, we will focus here on the disc formed after the second CE. 

To date no discs have been detected around DWDs, however simulations have shown the formation of a central cavity in circumbinary discs due to the binary torque arising from the non-axisymmetric part of the binary potential (\citealt{rafikov}, and references therein). The cavity radius is usually found to be about $2a_b$, where $a_b$ is the binary separation (\citealt{artymowicz91, artymowicz94, macfadyen, rafikov} and references there-in), so for our simulations we adopted $r_{in} = 2a_b$ as the inner disc radius. The external radius observed for circumbinary discs around evolved stars is usually found between a few dozen au to 500 au \citep{rafikov, izzard}. Because of this, and due to lack of observational constraints related to the systems studied in this work, we adopted a representative set of values for the characteristic external radius, $r_{\rm c}$, of our disc: $50$, $100$ and $150$ au. Because of the viscous spreading of the gas during the disc evolution, these discs are expected to expand to radii of $200$, $400$ and $600$ au, respectively \citep{lyndenpringle, hartmann}. The first value of $r_{\rm c}$, $50$ au, corresponds to the supposed extension of the disc which gave rise to our Solar System \citep{kretke}; the third value, $150$ au, represents the extension of the wide disc observed around the A-type star HD 163296 \citep{isella}; the second value, $100$ au, is intermediate between the other two. 

Given that no studies exist about planetary formation around DWDs, nor studies about circumbinary discs around DWDs, we assume the disc to behave like passively irradiated circumstellar discs \citep{chiang}.
We adopt the surface density and temperature profiles by \citet{schleicher}. The density profile is described by 
\begin{equation}
\label{eqn:density}
\Sigma(r) = \Sigma_0\bigg(\frac{r_c}{r}\bigg)^n,
\end{equation}
where $n$ is the power law index, which we set to $1$, and where $\Sigma_0$ is the surface density at $r_{\rm c}$.
The temperature profile of the disc is described by two equations: one representative for the inner region where the disc heating is dominated by irradiation, and one for the outer region where irradiation from the stars is negligible.

In the inner disc region where the stellar irradiation is the dominant mechanism, we can find the temperature profile of the disc by equating the fluxes emitted and absorbed by the disc \citep{schleicher}:
\begin{equation}
\label{eqn:temperature_radiation_total}
T = T_b\bigg(\frac{\theta}{4}\bigg)^{1/4}\bigg(\frac{r_{\rm in}}{r}\bigg)^{1/2}.
\end{equation}
In Eq. \ref{eqn:temperature_radiation_total}, $T_b$ is defined by
\begin{equation}
\label{eqn:Tb}
T_b = T_{\rm eq} + T_{\rm eff,2}\bigg(\frac{\tau_c}{t_0}\bigg).
\end{equation}
$T_{eq}$ is the equilibrium temperature of an irradiated object placed at a distance $r_{\rm in}$ from the DWD. The equilibrium temperature is computed considering the total luminosity of the DWD, i.e., the sum of the luminosity of the WDs of the system. Each stellar luminosity was derived by treating the emitted WDs energy as blackbody radiation and by using the WDs parameters, that is,  radius and effective temperature (see Tab. \ref{tab:DWDparams}), obtained by the SeBa simulations (see Section \ref{sec:dwdbinary}). The term $\Teff$$_2$ accounts instead for the initial temperature of the gas once ejected from the star, which we assume to match that of the youngest white dwarf. The variable $t_0$ is the starting time of planetary formation, which we choose to be $10\tau_c$, $0.1$ Myr and $1$ Myr (see Section \ref{sec:accretionmodel}). The parameter $\tau_c$ represents the cooling time of the disc with respect to the initially high post-CE temperatures of the gas. Details about this parameter will be discussed in Section \ref{sec:accretionmodel}.

The parameter $\theta$ is the grazing angle with respect to which the star radiation strikes the disc, and is defined as \citep{chiang}
\begin{equation}
\label{eqn:theta}
\theta = 0.4\frac{r_{\rm in}}{r} + r\frac{d(H/r)}{dr},
\end{equation}
where $H = c_s/\Omega$ is the disc scale height, $c_s$ the speed of sound, $\Omega$ the Keplerian angular velocity, and $H/r$ the disc aspect ratio. The second term in Eq. \ref{eqn:theta} makes Eq. \ref{eqn:temperature_radiation_total} more steep, causing it to reach lower temperatures as $r$ increases. 
In this work we follow \citet{schleicher}, and neglect the second term in Eq. \ref{eqn:theta}. This choice implies a slightly higher overall disc temperature as $r$ increases, but considerably simplifies the disc and planetary evolution calculations. The higher temperatures of the disc reduce the radial range available for the seeds to reach the isolation mass, as dust and ices in the disc will be in solid form at larger radii from the stars \citep{lodders}. Therefore, by neglecting the second term in Eq. \ref{eqn:theta}, we obtain smaller planetary masses than using Eq. \ref{eqn:theta} with the second term, and consequently, a conservative estimate of the planet-forming potential around the DWDs.

Eq. \ref{eqn:temperature_radiation_total} can therefore be written as
\begin{figure}[t]
\centering
\includegraphics[trim=1cm 0cm 2.5cm 2.cm,clip,width=\hsize]{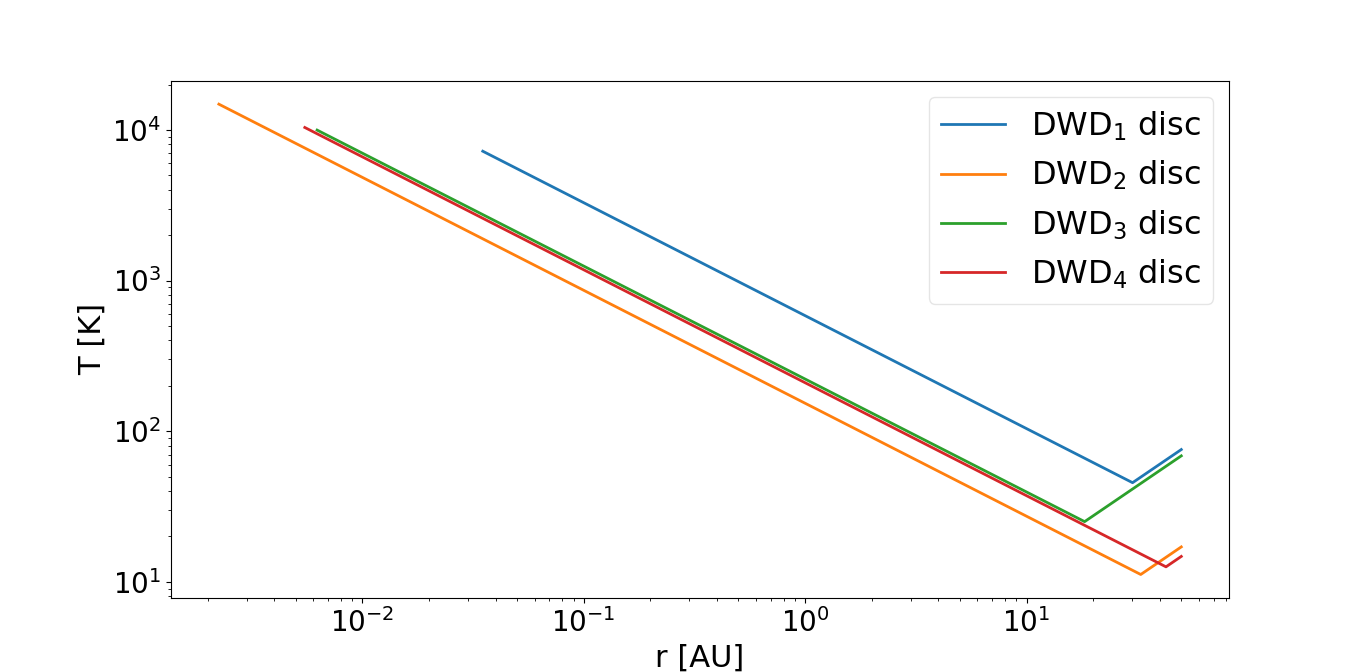}
\caption{Temperature profiles of the circumbinary discs surrounding the DWD systems, colour-coded in the legend. The profiles are estimated at $t_0 = 10\tau_c$ (see Tab. \ref{tab:DWDparams} for details about the systems) for a disc with $r_{\rm c} = 50$ au.}
\label{fig:disc_T_profiles}
\end{figure}
\begin{equation}
\label{eqn:temperature}
\begin{split}
T & = T_b \bigg(0.1\frac{r_{\rm in}}{r}\bigg)^{1/4}\bigg(\frac{r_{\rm in}}{r}\bigg)^{1/2} \\
& = T_b (0.1)^{1/4}\bigg(\frac{r_{\rm in}}{r}\bigg)^{\zeta},
\end{split}
\end{equation}
where $\zeta = 3/4$. By using Eq. \ref{eqn:temperature}, the speed of sound in the irradiated region can be expressed as
\begin{equation}
\label{eqn:soundspeed}
\begin{split}
c_s &= \sqrt{\frac{k_BT}{\mu m_p}} \\
&= c_{s1} \bigg(\frac{r}{\text{au}}\bigg)^{-\zeta/2}
\end{split}
\end{equation}
where $c_{s1}$ is the sound speed at $1$ au, $\mu$ is the mean molecular weight (in units of $m_p$), and $k_B$ and $m_p$ are the Boltzmann constant and the proton mass, respectively.

In the outer disc region where irradiation is negligible, following \citet{schleicher} we consider disc fragmentation as the main heating source. From the Toomre stability criterion, as the so called `Toomre parameter' $Q =  \nicefrac{c_s\Omega}{\pi G \Sigma}$ approaches unity, disc fragmentation occurs, and turbulence and shocks may compensate for the disc cooling in the outer regions of the disc \citep{schleicher}. At the stage of fragmentation ($Q \sim 1$), we can estimate the speed of sound as
\begin{equation}
    \label{eqn:cs_toomre}
    c_s = \frac{\pi G \Sigma}{\Omega} \sim r^{1/2},
\end{equation}
where $\Sigma$ and $\Omega$ are the surface density and the angular velocity of the disc, respectively.
By equating Eq. \ref{eqn:cs_toomre} with $c_s =  \sqrt{\nicefrac{k_BT}{\mu m_p}}$, we obtain the temperature profile in the outer regions of the disc:
\begin{equation}
\label{eqn:temperature_toomre}
T = \frac{2m_p}{k_B} \bigg[\frac{\pi G \Sigma(r)}{\Omega}\bigg]^2 \sim r,
\end{equation}
where, $m_p$ is the proton mass, $k_B$ is the Boltzmann constant, $G$ is the gravitational constant.

In Fig. \ref{fig:disc_T_profiles} we show the temperature profiles for the circumbinary discs surrounding each of the DWD systems under study (see Tab. \ref{tab:DWDparams}). The profiles are estimated when the youngest WD is just born. The negative slope of the curves is due to Eq. \ref{eqn:temperature} while the positive slope, when present, is due to Eq. \ref{eqn:temperature_toomre} with the change in slope identifying the orbital distance where the dominant heating process changes. Since the disc structure depends on the evolution of the system during the last CE (i.e., $r_{in} = 2a_b$), the values of $r_{in}$ are different for each system. Since planet formation in the outer turbulent disc regions is governed by different physical processes than the pebble accretion scenario we consider in this study, we focus only on the inner irradiated region. The turning points between the two temperature regimes in Fig. \ref{fig:disc_T_profiles} therefore mark the outer boundaries of the planet-forming regions we investigate.

\subsection{Accretion and photoevaporation rates of the disc}
\label{sec:accretionmodel} 
The lifetime of protoplanetary discs around pre-MS stars is set by their gas accretion rate on the central star and by their photoevaporation by stellar irradiation. Both processes act to remove mass from circumstellar discs, limiting the duration of the planetary formation process. Since the temporal evolution of discs around evolved stars, and DWD in particular, is poorly known, we fill the gaps in our understanding working by analogy with pre-MS discs.

Around 95$\%$ of pre-MS stars within the spectral type interval K0-M5 stops accreting material from their circumstellar discs before $5$ Myr, reaching an accretion rate of $\approx 10^{-11}$ M$_{\odot}$/yr \citep{fedele}. We consequently set the upper limit of the disc lifetime and the planetary formation process to be 5 Myr. Also by analogy, the planetary formation process in pre-MS discs is known to start within the first $1$ Myr of the disc life \citep{scott, manara, mulders, lichtenberg, bernabo}, so we consider three representative starting times in our analysis: $10\tau_c$, $0.1$ Myr, and $1$ Myr.  
These initial times $t_0$ account for the time needed to assemble the Moon-sized planetary seeds at the beginning of the simulations (see Section \ref{sec:results}) and span the range revealed by the meteorites formation times \citep{scott, lichtenberg}.

Each of these values of $t_0$ implies a different scenario for the formation of the planetary seeds and is associated with a different temperature and luminosity of the youngest WD as it cools over time (see Section \ref{sec:dwdbinary}). The scenario with $t_0=10\, \tau_c$ assumes that the planet formation process starts as soon as the gas in the second-generation discs cools down to allow the condensation of dust. The scenario with $t_0=0.1$ Myr assumes that the planetary seed requires about 10$^5$ years to assemble, compatible with the fastest formation timescales from meteorites \citep{scott, lichtenberg}. Finally, the scenario with $t_0=1$ Myr is consistent with the formation timescale of massive planets suggested by the collisional dust regeneration process in circumstellar discs \citep{bernabo}.

\begin{figure}[t]
\centering
\subfloat[][]{\includegraphics[trim=0.cm 0.cm 1cm 2.cm, clip,width=0.52\textwidth]{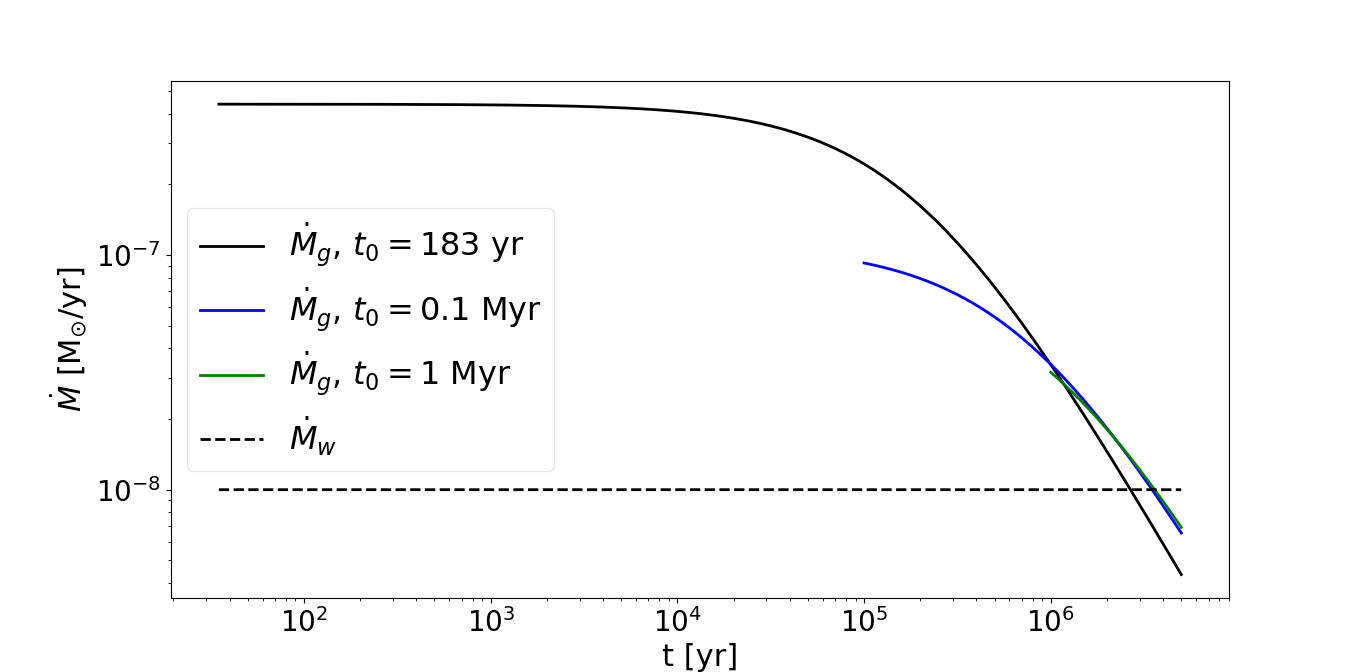}}\\
\subfloat[][]{\includegraphics[trim=0.cm 0.cm 1cm 2.cm, clip,width=0.52\textwidth]{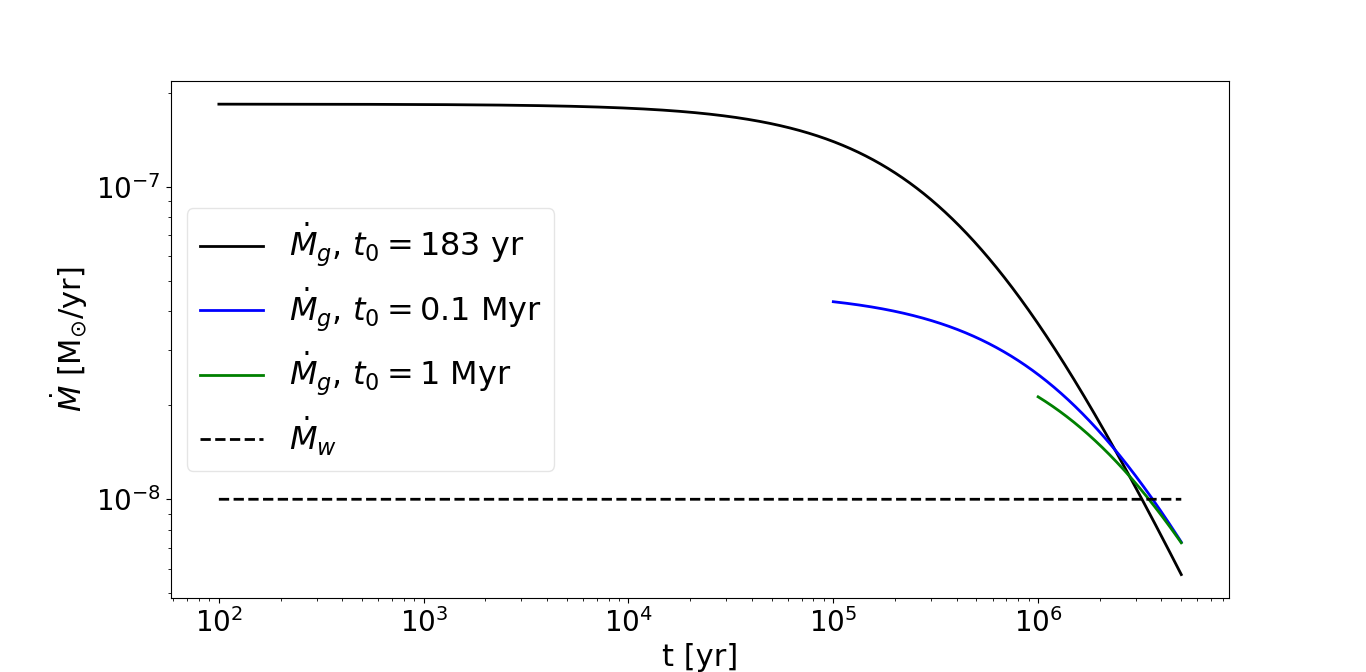}}\\
\subfloat[][]{\includegraphics[trim=0.cm 0.cm 1cm 2cm, clip,width=0.52\textwidth]{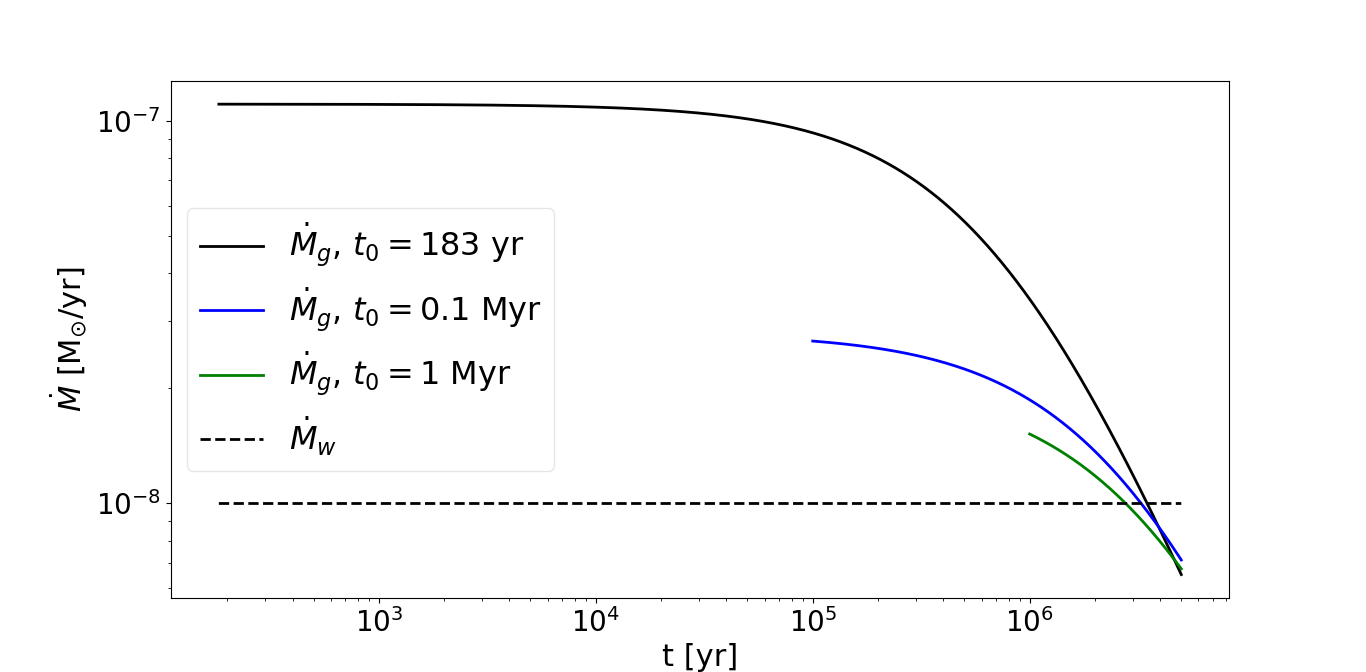}}
\caption{Gas accretion rates ($\dot{M}_g$) and photoevaporation rate ($\dot{M}_w$) of the disc surrounding the system DWD$_1$. The accretion rates start at $10\tau_c$, $0.1$ Myr and $1$ Myr after the birth of the youngest WD (colour-coded in the legend), which has a temperature of $75000, 45650, 33950$ K, respectively. From top panel to bottom panel, $r_{\rm c} = 50, 100, 150$ au, respectively.}
\label{fig:accr_rates_dwd1}
\end{figure}

The parameter $\tau_c$ used to time the first scenario is the characteristic cooling time of the disc and is defined as $\tau_c = C \cdot P_d$, where $C = 0.01$ and $P_d$ is the orbital period of the disc at $r_{\rm c}$ \citep{nelson, woitke, rabago}. The adoption of $t_0 = 10\,\tau_c$ (see Section \ref{sec:discmodel}) means that we consider a disc whose gas has cooled down significantly with respect to the initial post-CE phase, but still retains some of its initial thermal budget while having undergone negligible physical evolution, as $\tau_c \sim$ 1\% of $P_d$ (i.e. the dynamical time of the disc). Moreover, $10\, \tau_c < 246$ yr throughout all our simulations, meaning that this time is negligible both with respect to the cooling time of the youngest WD and with respect to the pebble accretion rates. Consequently, we associate to $\Teff$$_2$ the temperatures indicated in Tab. \ref{tab:DWDparams} when 
$t_0 = 10\, \tau_c$. 

Following \citet{hartmann} we describe the accretion rate of our disc by
\begin{equation}
\label{eqn:accr_rate}
\dot{M}_g = \dot{M}_g(t_0)\bigg(\frac{t}{t_s} + 1\bigg)^{-(5/2 - \delta)/(2 - \delta)},
\end{equation}
where $\dot{M}_g(t_0) = -2\pi r_{\rm c}u_g(r_{\rm c})\Sigma_0$ is the accretion rate at $t = t_0$, with $u_g(r_{\rm c})$ the inward radial velocity of the gas at $r_{\rm c}$ and $\Sigma_0$ the density at $r_{\rm c}$ (see Section \ref{sec:pebbleaccretion}). The parameter $\delta$ represents the power law index of the viscosity, $\nu$. 
Specifically, $\nu = \alpha c_s H \propto r^{3/2}r^{-\zeta} = r^{\delta}$, where $\alpha$ characterises the efficiency of angular momentum transfer in the disc \citep{shakura}. For our disc, $\delta = 3/2 - \zeta = 0.75$. The parameter $t_s$ is the viscous timescale of the disc
\begin{equation}
\label{eqn:viscous_timescale}
t_s = \frac{1}{3(2 - \delta)^2}\frac{r_{\rm c}}{\nu_c},
\end{equation}
with $r_{\rm c}$ the characteristic disc size and $\nu_c = \nu(r_{\rm c})$ \citep{johansen}. 

In their recent study of giant planets formation, \citet{tanaka} consider photoevaporation rates between $10^{-9}-10^{-8}$ Myr. Given that our central object is a binary composed by two WDs, we adopted one of their highest values, i.e., $\dot{M}_w = 10^{-8}$ \Mearth/yr.
\\
In Fig. \ref{fig:accr_rates_dwd1} we show as an example the accretion rates, $\dot{M}_g$, and the photoevaporation rate, $\dot{M}_w$, of the disc surrounding the system DWD$_1$. On each plot, we show one track for each value of adopted starting time ($t_0 = 10\tau_c$ and $t_0 = 0.1, 1$ Myr). From top panel to bottom panel, the value of $r_{\rm c} = 50, 100, 150$ au, respectively (i.e. from top panel to bottom panel the density of the disc decreases). Therefore, in each individual plot, every accretion rate track is associated to the same value of $r_{\rm c}$. As the time increases, the accretion rate decreases (see Eq. \ref{eqn:accr_rate}). The later planetary formation starts, the closer $\dot{M}_g(t_0)$ is to $\dot{M}_w$. By decreasing $t_0$ the disc becomes colder, as the youngest WD cools significantly over time lowering the disc equilibrium temperature (i.e. the second term of Eq. \ref{eqn:Tb} decreases). Consequently, both the radial speed of the gas and  $\dot{M}_g(t_0)$ decrease. In our exploratory study, however, once we set $t_0$ (i.e., $\Teff$$_2$), the temperature profile of the disc does not change throughout the simulation.
As we will see in Sec. \ref{sec:gas_accretion}, the difference between $\dot{M}_g$ and $\dot{M}_w$ determines whether a planet can undergo runaway gas accretion or not. Specifically, circumbinary discs characterised by $\dot{M}_g - \dot{M}_w = 0$ cannot form GGs.

\subsection{Planetary formation processes}
\label{planetformation}
It is reasonable to expect that the same processes that regulates the growth of planets in circumstellar protoplanetary discs can occur also in second-generation circumbinary discs. 

The core formation is modelled starting from a seed mass of $0.01$ \Mearth, based on theoretical \citep{chambers, johansenandlambrechts} and observational constraints (e.g., \citealt{scott, brasser, lammer}). We account for the formation time of the starting seed with the different starting time of our simulations, that are also linked to the temperature evolution of the younger WD (see Sec. \ref{sec:dwdbinary}). 

The core accretion phase is described through the pebble accretion scenario \citep{johansen}. The gas accretion phase is described following \cite{tanaka}. We consider both a gas contraction phase, starting at the end of pebble accretion, and the subsequent runaway gas accretion.

\subsubsection{Pebble accretion}
\label{sec:pebbleaccretion}
In the following, we consider accretion of solids onto our seeds as described by the faster pebble accretion regime i.e.,  the 2D Hill regime \citep{lambrechts1}. The accretion of pebbles by a protoplanetary core can follow different regimes, namely drift accretion (or Bondi accretion) and Hill accretion regimes. Both regimes are characterised by the protoplanet having a specific radius of influence (Bondi radius and Hill radius, respectively), which in turn determines how effective the protoplanet is in deviating and accreting a pebble embedded in the gas flux of the disc. The transition between drift regime and Hill regime occurs at $0.001-0.01$ M$_{\oplus}$ \citep{lambrechts1, lambrechts2, johansenandlambrechts}. When the mass of the protoplanet is small and its Hill radius is smaller than the scale height of the pebbles, accretion through Hill regime occurs in a 3D way. If the Hill radius is larger than the scale height of the pebbles, accretion through Hill regime occurs in a 2D way (\citealt{bitsch1}, and references therein). As the protoplanet cannot accrete pebbles from outside its Hill radius, 2D Hill accretion is faster than 3D Hill accretion \citep{lambrechts1, lambrechts2}. Starting from a seed mass of $0.01$ \Mearth, we account for the initial 3D Hill accretion phase through the starting time, $t_0$.

The pebble accretion rate is
\begin{equation}
\label{eqn:pebble_accr}
\frac{dM}{dt} = 2(S_t/0.1)^{2/3} \Omega r_H^2 \Sigma_p,
\end{equation}
where $M$ is the mass of the growing planet, $\Sigma_p$ is the pebble surface density of the disc, $r_H = [M/(3M_b)]^{1/3}r$ is the Hill radius of the protoplanet, with $M_b$ the total mass of the binary, and $S_t$ is the Stokes number of the pebbles, which is proportional to the pebbles size. 

Following \citet{johansen}, we consider accretion of mm-sized pebbles ($S_t \ll 1$). The inward gas and pebbles radial velocities can be defined respectively as
\begin{subequations}
\label{eqn:velocities}
\begin{align}
u_g &= -\frac{3}{2} \frac{\nu}{r} \label{eqn:velocities1}\\
u_p &= -2S_t\Delta v + u_g,	\label{eqn:velocities2}
\end{align}
\end{subequations}
where $\Delta v \approx  v_K$ is the relative velocity between the pebble and the core, with $v_K$ the Keplerian velocity at orbital radius $r$ \citep{lambrechts1}. Coupling Eqs. \ref{eqn:velocities} with the formula of inward gas and pebble fluxes
\begin{subequations}
\label{eqn:fluxes}
\begin{align}
\dot{M}_g &= -2\pi r u_g \Sigma_g \label{eqn:fluxes1}\\
\dot{M}_p &= -2 \pi r u_p\Sigma_p, \label{eqn:fluxes2}
\end{align}
\end{subequations}
we can express the ratio between the pebble and the gas surface densities, $\Sigma_p/\Sigma_g$, as
\begin{equation}
\label{eqn:sigmaratios}
\frac{\Sigma_p}{\Sigma_g} = \frac{\xi}{(2/3)(S_t/\alpha)\chi + 1}.
\end{equation}
In the last equation, 
$\chi = \nicefrac{-\partial \ln{P}}{\partial \ln{r}} = n + \nicefrac{\zeta}{2} + \nicefrac{3}{2}$ 
and 
$\xi = \dot{M}_p/\dot{M}_g$.
Following \citet{johansen}, we set $S_t \sim \alpha$, which implies: (i) approximately the same radial speed for gas and pebbles, i.e., same depletion time for both components, which in turn results in $\xi$ approximately equal to the metallicity of the disc\footnote{For this reason, despite in general it would be more correct referring to $\xi$ as dust-to-gas ratio, we will refer to $\xi$ as the metallicity of the disc from now on.}; (ii) pebbles whose dimensions are consistent with observations of mm-cm pebbles in protoplanetary discs over a wide range of ages \citep{perez, huang}. In particular, we set $S_t \sim 0.01$ ($\sim$mm-sized pebbles).

As the temperature of the disc increases with decreasing orbital distances from the central stars, inward drifting pebbles will lose growing fractions of their mass due to the sublimation of their composing materials \citep{lodders}. As a result, as the growing planet migrates toward the disc inner regions, a lower mass of pebbles becomes available to support its growth and pebble accretion eventually stops. The planet migration toward the inner regions of the disc is caused by the torque between the surrounding disc material and the planet itself. Such migration mechanism, which is called `Type I migration', is defined by
\begin{equation}
\label{eqn:typeI}
\frac{dr}{dt} = -k_{mig}\frac{M}{M_b}\frac{\Sigma_g \Omega r^3}{M_b}\bigg(\frac{H}{r}\bigg)^{-2},
\end{equation}
where $M_b$ is the total mass of the binary and $k_{mig} = 2(1.36 + 0.62\beta + 0.43\zeta)$ \citep{dangelo, johansen}. 

Dividing Eq. \ref{eqn:pebble_accr} by Eq. \ref{eqn:typeI}, we obtain the equation for pebble accretion in a steady-state disc:
\begin{equation}
\label{eqn:analytical_pebble}
\frac{dM}{dr} = -2\frac{M_b^2}{k_{mig}}\bigg(\frac{S_t}{0.1}\bigg)^{2/3}\bigg(\frac{Hr_H}{r^2}\bigg)^2\frac{1}{Mr}\frac{\xi}{(2/3)(S_t/\alpha)\chi + 1}.
\end{equation}
The solution of the previous equation is
\begin{equation}
\label{eqn:analytical_pebble_sol}
\begin{split}
M^{4/3} = M_0^{4/3} -&\frac{2(S_t/0.1)^{2/3} M_b (3M_b)^{-2/3}}{k_{mig} Gc_{s1}^{-2} \text{AU}^{-\zeta}}\frac{(4/3)\xi}{(2/3)(S_t/\alpha)\chi + 1} \\
&\times \frac{(r^{1-\zeta} - r_0^{1 - \zeta})}{1 - \zeta},
\end{split}
\end{equation}
where $M_0$ and $r_0$ are the initial mass and position of the seed. Eq. \ref{eqn:analytical_pebble} provides the growth tracks in steady-state disc for pebble accretion in our results. The time-dependent pebble accretion phase is instead described by Eq. \ref{eqn:pebble_accr}, which can be expressed also as
\begin{equation}
\label{eqn:time_pebble}
\begin{split}
\frac{dM}{dt} = &2\bigg(\frac{S_t}{0.1}\bigg)^{2/3}\frac{GM_b}{(3M_b)^{2/3}} \\
&\times \frac{\xi \dot{M}_g(t)}{2\pi c_{s1}^2[\chi S_t + (3/2)\alpha]\text{AU}^{\zeta}}M^{2/3}r^{\zeta - 1}.
\end{split}
\end{equation}
The last equation clearly shows that $\xi$ and $\dot{M}_g$ are important parameters for an efficient pebble accretion phase to occur. In particular, as $\dot{M}_g$ decreases over time, the sooner pebble accretion starts, the higher its rate will be, and the quicker the planetary mass will grow. On the other hand, discs surrounding young DWDs will be characterised by temperatures up to $10^4$ K. This results in more inflated discs, i.e., into larger $H/r$ (see Eqs. \ref{eqn:temperature}-\ref{eqn:soundspeed}), with condensation lines shifted toward larger orbital radii. This  aspect is important because pebble accretion stops once the mass of the planet reaches the so-called `isolation mass'. As the protoplanet mass grows, its gravitational influence on the surrounding material grows as well. The local gas density is perturbed 
such that the pressure gradient out of the planet-forming region tends to zero, stopping the drift of pebbles toward the protoplanet. In parallel, the planet perturbs the pressure gradient inside its orbital location and creates a dust trap that commoves with the planet while it migrates inward. As a result, the planet is prevented from accreting the pebbles that reside inside its orbit. Considering typical orders of magnitude of $H/r \sim 10^{-2}$, $\alpha \sim10^{-3}-10^{-4}$ and $c_s \sim 10^2$ m/s, we have $u_p \approx 0$ if the pressure gradient, and thus $\Delta v$, tends to zero (see Eqs. \ref{eqn:velocities}) \citep{lambrechts2}. We estimate the isolation mass as \citep{lambrechts3, bitsch2, johansen}
\begin{equation}
\label{eqn:isomass}
M_{iso} = 20\ \text{M}_{\oplus} \bigg[\frac{H/r}{(H/r)_{\rm 5 au}}\bigg]^6 \bigg[0.34 \bigg(\frac{\log \alpha_3}{\log \alpha_v}\bigg)^4 + 0.66 \bigg] \bigg(1 - \frac{2.5 - \chi}{6}\bigg).
\end{equation}
We replaced the exponent and the normalisation of the term in $H/r$ from \citep{johansen} consistently with our temperature profile power low index, i.e. $\zeta = 3/4$. That is, in our case $H/r \sim r^{1/8}$.
The higher $H/r$, the higher the isolation mass, and the harder can be for the planet to reach the gas accretion phase. The parameter $\alpha_3 = 10^{-3}$ is a constant and $\alpha_v = 10^{-4}$  is the turbulent viscosity parameter \citep{johansen}. Once the pebble isolation mass is reached, the gas accretion phase of planetary formation can start.

\subsubsection{Gas accretion}
\label{sec:gas_accretion}
During the pebble accretion phase, gas accumulates in the Hill sphere of the planet. However, the energy deposited in the gas by the flux of pebbles prevents it from collapsing because of the planet's gravity: as a result, the planet accumulates an inflated gaseous envelope. Once pebble accretion is halted, the thermal energy provided by pebbles ceases. The gas pressure in the envelope prevents it from quickly collapsing onto the planet, but the envelope starts a slow contraction phase while more gas is accreted. When the planet reaches a mass $M_{crit} \sim 2M_{iso}$, i.e.,  when the mass of the envelope becomes roughly the same as the mass of the rocky core, the pressure of the gas is not able to counteract gravity, and a runaway gas accretion phase starts (\citealt{dangelo}, and references there-in; \citealt{johansen, dangelo2}).

During the gas contraction phase, the gas component of the planet-forming region is slowly depleted due to the accreting planet. This causes a variation in the torque between the gas in the planet-forming region and the planet itself, which in turn affects the migration rate of the planet. Following \citet{johansen} and \citet{tanaka}, we adopt Type I migration for the planet during gas contraction, and Type II migration for the planet during runaway gas accretion. In particular, following \citet{tanaka} we adopt a new accurate formulation for Type II migration, where the torque exerted on the planet is proportional to the gas surface density inside the gap, because the planet mainly interacts with the gas at its bottom \citep{kanagawa, tanaka}. In this case the Type II migration rate is described by
\begin{equation}
\label{eqn:typeII}
\frac{dr}{dt} = -6\frac{M}{M_b}\bigg(\frac{H}{r}\bigg)^{-2} \frac{\Sigma_{gap} \Omega r^3}{M_b},
\end{equation}
where $\Sigma_{gap}$ is the surface density of the gap carved by the growing planet. The gap surface density can be expressed as
\begin{subequations}
\label{eqn:Sigmagap}
\begin{align}
\Sigma_{gap} &= \frac{\Sigma_{out}}{1 + 0.04K} \label{eqn:Sigmagap1}\\
K &= \bigg(\frac{M}{M_b}\bigg)^2 \bigg(\frac{H}{r}\bigg)^{-5} \alpha^{-1} \label{eqn:Sigmagap2},
\end{align}
\end{subequations}
where $\Sigma_{out}$ is the unperturbed surface density of the planet-forming region \citep{kanagawa}.

The formalism used to describe the gas contraction phase and the runaway gas accretion phase is reported as follows
\begin{itemize}

\item {\it Gas contraction:} the envelope contraction occurs on a Kelvin-Helmholtz timescale, whose duration depends on the ratio between the masses of the gaseous envelope and of the planetary core, and on the envelope opacity (\citet{tanaka}, and references therein). Such timescale can be expressed as
\begin{equation}
\label{eqn:KHtimescale}
\tau_{KH} \simeq 1\times10^3\bigg(\frac{M}{30\ \text{M}_{\oplus}}\bigg)^{-5/2}\bigg(\frac{k}{0.05\ \text{cm}^2/\text{g}}\bigg)\ \text{yr},
\end{equation}
where $k$ is the envelope opacity, which we set to $0.05\ \text{cm}^2/\text{g}$ \citep{tanaka}. The gas contraction rate can be determined by the ratio between the mass of the planet and Eq. \ref{eqn:KHtimescale}, such that
\begin{equation}
\label{eqn:time_gascontraction}
\frac{dM}{dt} = 3\times10^{-2}\bigg(\frac{M}{30\ \text{M}_{\oplus}}\bigg)^{7/2}\bigg(\frac{k}{0.05\ \text{cm}^2/\text{g}}\bigg)^{-1} \frac{\text{M}_{\oplus}}{\text{yr}}.
\end{equation}
The ratio between Eq. \ref{eqn:time_gascontraction} and Eq. \ref{eqn:typeI}, brings to the formula for the gas contraction phase in a steady-state disc
\begin{equation}
\label{eqn:analytical_gascontraction}
\begin{split}
\frac{dM}{dr} &= -\frac{3\times10^{-2}c_{s1}^2\ \text{AU}^{5/2}}{k_{mig}(GM_b)^{3/2}\ \text{yr}}\bigg(\frac{k}{0.05\ \text{cm}^2/\text{g}}\bigg)^{-1}\bigg(\frac{M_b}{30\ \text{M}_{\oplus}}\bigg)^2
\\
&  \times \bigg(\frac{\Sigma_0}{30\ \text{M}_{\oplus}\ \text{AU}^{-2}}\bigg)^{-1}\bigg(\frac{r}{r_{\rm c}}\bigg)^{-n}\bigg(\frac{r}{\text{AU}}\bigg)^{-\zeta-1/2}\ \frac{M_{\oplus}}{\text{AU}}.
\end{split}
\end{equation}

\item {\it Runaway gas accretion:} once the planet reaches $M_{crit}$, the runaway gas accretion phase starts. For more details of this process, we refer to the work of \citet{tanaka}. The runaway gas accretion rate is
\begin{equation}
\label{eqn:time_runawaygas}
\begin{split}
\frac{dM}{dt} &= \bigg[0.29 \bigg(\frac{M}{M_b}\bigg)^{4/3} \bigg(\frac{H}{r}\bigg)^{-2}\Omega r^2\bigg]\Sigma_{gap} \\
&= D\Sigma_{gap},
\end{split}
\end{equation}
where now $\Sigma_{gap} \propto \dot{M}_g - \dot{M}_w =$ \dMgw. When the latter term is $0$, the gas accretion rate of the disc $\dot{M}_g$ cannot compensate the gas loss due to the photoevaporation rate, $\dot{M}_w$, and the planet-forming region is not replenished by the gas flux. This halts runway gas accretion and planet migration \citep{tanaka}. 

Dividing Eq. \ref{eqn:time_runawaygas} by Eq. \ref{eqn:typeII} we have the equation for runaway gas accretion in a steady-state disc
\begin{equation}
\label{eqn:analytical_runawaygas}
\frac{dM}{dr} \simeq -0.48\frac{M_b}{r}\bigg(\frac{M}{M_b}\bigg)^{1/3}.
\end{equation}

\end{itemize}

\section{Results}
\label{sec:results}
In this section we present the growth tracks in steady-state disc (or time-independent tracks) and growth tracks in time-evolving disc (or time-dependent tracks) for the model of circumbinary disc described in Section \ref{sec:discmodel}, following the formalism introduced in Section \ref{methods}. 

The time-independent tracks show the growth of the planets neglecting the time evolution of the disc, i.e.,  the planets produced are not affected by, e.g., the decrease over time of the accretion rate of the disc onto the binary. Therefore, such tracks represent the maximum planet-forming potential of the disc. On the other hand, the time-dependent tracks are affected both by the accretion rate and the photoevaporation rate of the disc. The simulations follow the evolution of seeds with initial masses of 0.01 M$_{\odot}$ placed at different starting radial distances from the central binary. We make different assumptions for the origin of such seeds depending on the starting time of planet formation, i.e. $t_0 = 10\tau_c$, $0.1$ Myr and $1$ Myr. When $t_0 = 10\tau_c$, we assume the seeds have a first generation nature, i.e. they formed before the CE which originate the discs we study in this work. When $t_0 \geq 0.1$ Myr we assume the seeds formed within $t_0$ in the second-generation disc (see Section \ref{sec:accretionmodel}).

The number of seeds in the simulations depends on the extension of the radial region between $r_{in}$ and the orbital radius outside of which planet formation occurs through disc instability (i.e.,  due to the collapse of self gravitating clumps of gas originated by disc gravitational instabilities). Specifically, we do not consider the region of the disc where Eq. \ref{eqn:temperature_toomre} holds, as it represents a planet formation channel governed by different physical processes (e.g., \citealt{helled}) than those we consider in this work. As a result, the spatial region sampled by both kind of formation tracks always stops before the transition radius between Eq. \ref{eqn:temperature_toomre} and Eq. \ref{eqn:temperature}. The disc temperature profile is always described by Eq. \ref{eqn:temperature} in the sampled region. The spatial region for planetary formation is further reduced during the pebble accretion phase, as solids experiences almost complete sublimation in disc regions where $T > 1200$ K \citep{lodders}. If a planet has a mass $M < M_{\rm iso}$ at $r \leq r_t$, with $r_t$ the radial coordinate where $T = 1200$ K, its growth will stop and it will never reach $M_{\rm iso}$.

\subsection{Exploring formation within a post-CE disc around the system DWD$_1$}
\label{sec:dwd1_planetformation}
In our first analysis we study planet formation around the system DWD$_1$ in two steps, first by analysing the steady-state disc and then the time-evolving disc. The set of parameters we consider in our analysis is summarised in Table \ref{tab:DWDparams}.

Following \citet{schleicher}, the mass that remain bound to the binary after the last CE is $\geq 10\%$ the envelope of the last-evolving star (\citealt{kashiandsoker, passy, schleicher} and references there-in). For the DWD$_1$ system we compute \Mbound = 0.26 M$_{\odot}$. We assume this to be also the total mass of the circumbinary disc, i.e. M$_d = $ \Mbound $= 0.26$ \Msun. 

We explore planet formation as a function of the disc metallicity $\xi$ (i.e.,  the pebbles-to-gas ratio, see Section \ref{sec:pebbleaccretion}), the disc characteristic radius $r_{\rm c}$, and the effective temperature $\Teff$$_2$ of the youngest WD. The gas surface density values at $r_{\rm c}$, i.e. $\Sigma_0$, are $150$, $38$, and $17$ g/cm$^2$ for $r_{\rm c} = 50, 100, 150$ au, respectively.

\begin{table}[h]
\caption{Free parameters used in the simulations for the DWD$_1$ system.} 
\label{table:freeparams}   
\centering                          
\begin{tabular}{c c}       
\hline\hline                 
Parameter & Values  \\    
\hline        
   $t_0$ [Myr]	        & $10\tau_c$, $0.1$, $1$ \\      
   $\Teff$$_2$ [K]		& $75000, 45700, 33950$ \\      
   $\xi$               & $0.01, 0.015, 0.02$ \\
   $r_{\rm c}$ [au]	& $50, 100, 150$ \\
\hline                                   
\end{tabular}
\tablefoot{\\The parameter $t_0$ represents the starting time of the simulation, $\Teff$$_2$ is temperature of the youngest WD, $\xi$ is the disc metallicity and $r_{\rm c}$ is the characteristic external radius of the disc. The choice of $r_{\rm c}$ is explained in Section \ref{sec:discmodel}. Note that the higher value of $\Teff$$_2$ corresponds to the smaller value of $t_0$ and viceversa. Moreover, $10\tau_c = 35, 100, 183$ yr for $r_{\rm c} = 50, 100, 150$ au, respectively (see Section \ref{sec:accretionmodel}).}
\end{table}

\begin{figure*}[t]
\centering
\subfloat[][]{\includegraphics[trim=0cm 0.cm 1cm 1cm, clip,width=0.52\textwidth]{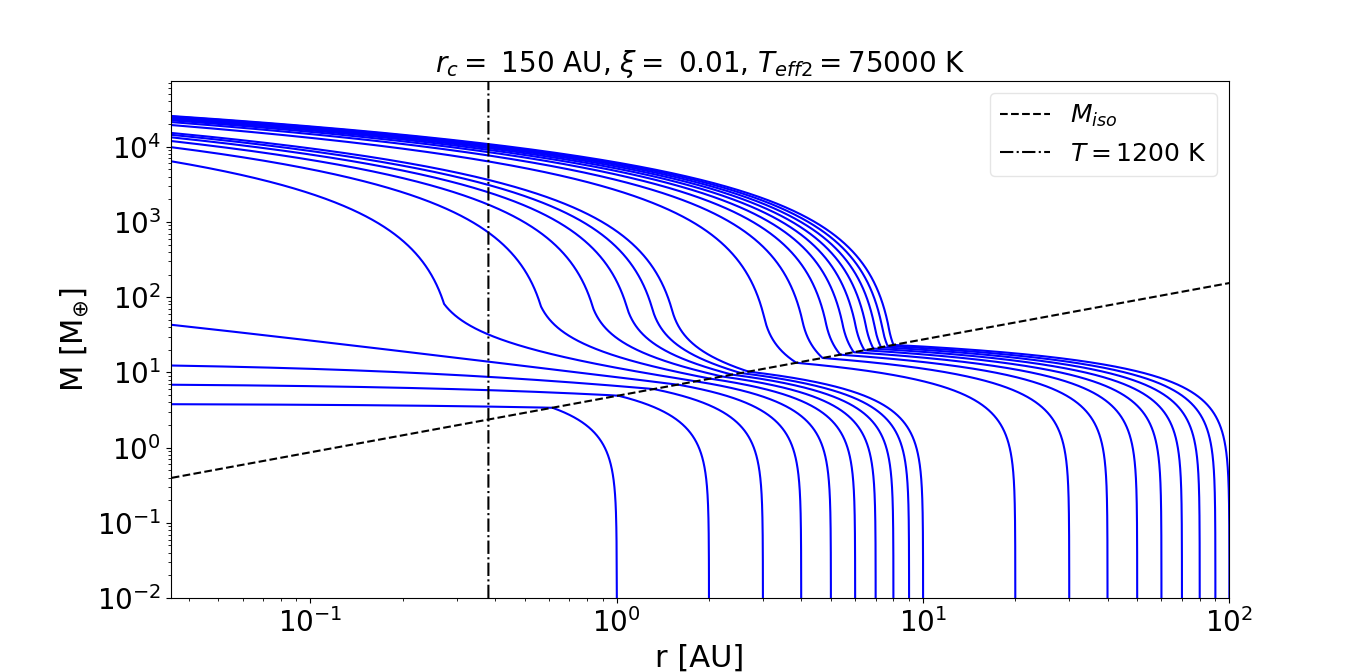}}
\subfloat[][]{\includegraphics[trim=0cm 0.cm 1cm 1cm, clip,width=0.52\textwidth]{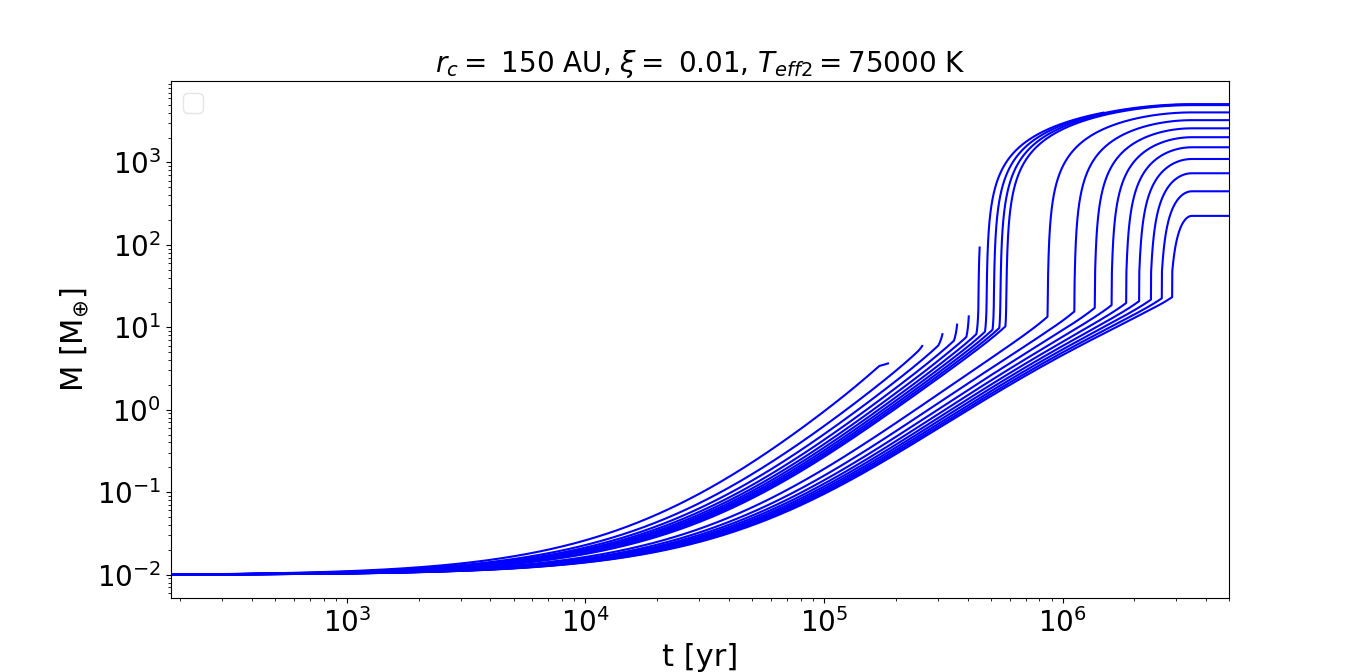}}\\
\subfloat[][]{\includegraphics[trim=0.cm 0.cm 1.cm 1cm, clip,width=0.52\textwidth]{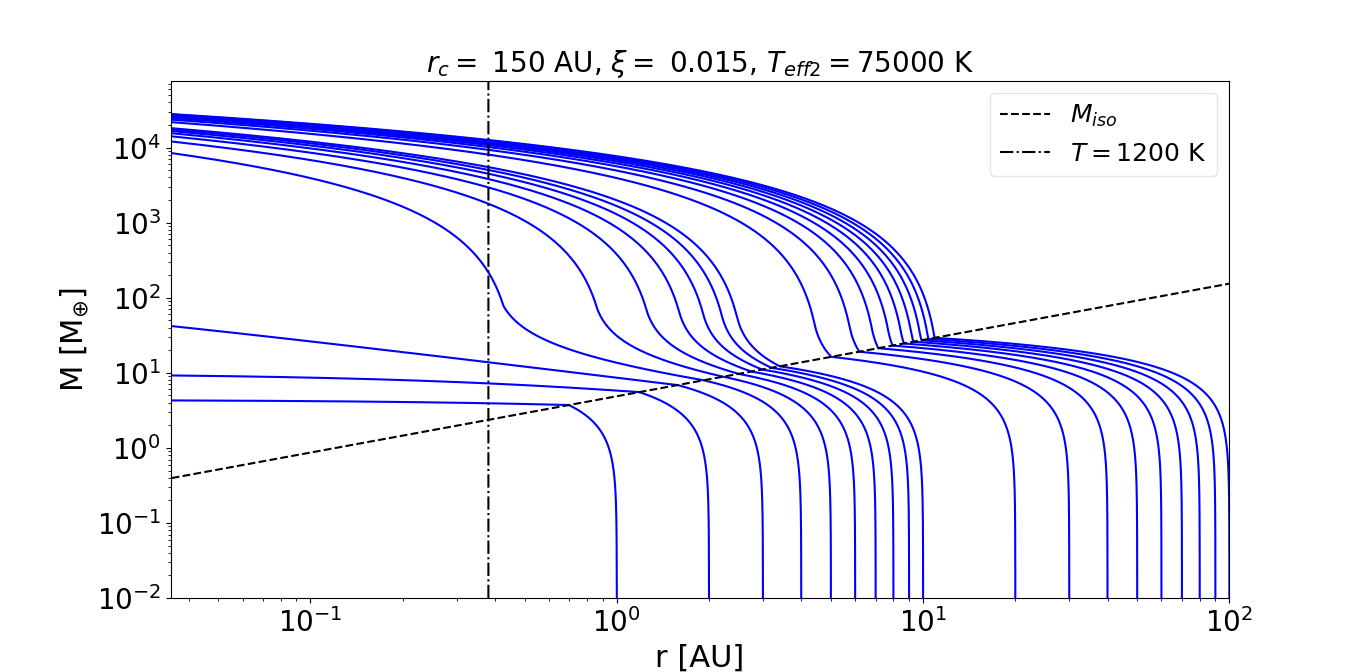}}
\subfloat[][]{\includegraphics[trim=0.cm 0.cm 1cm 1cm, clip,width=0.52\textwidth]{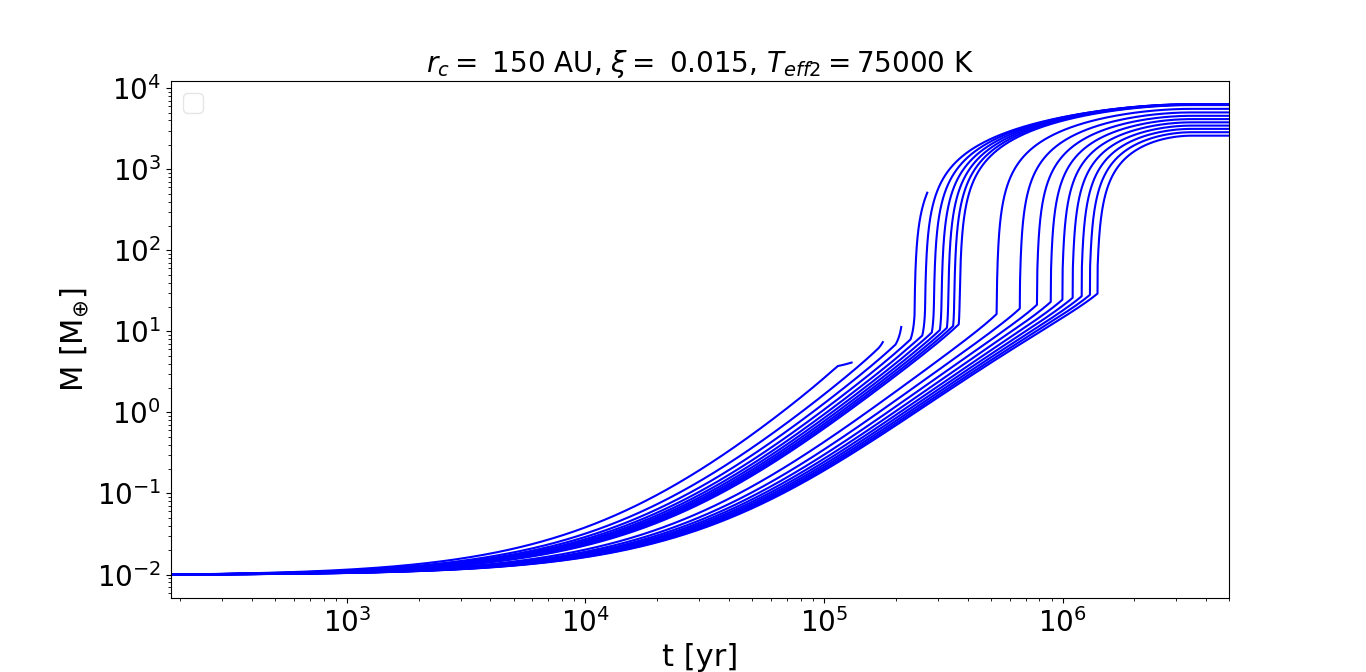}}\\
\subfloat[][]{\includegraphics[trim=0.cm 0.cm 1cm 1cm, clip,width=0.52\textwidth]{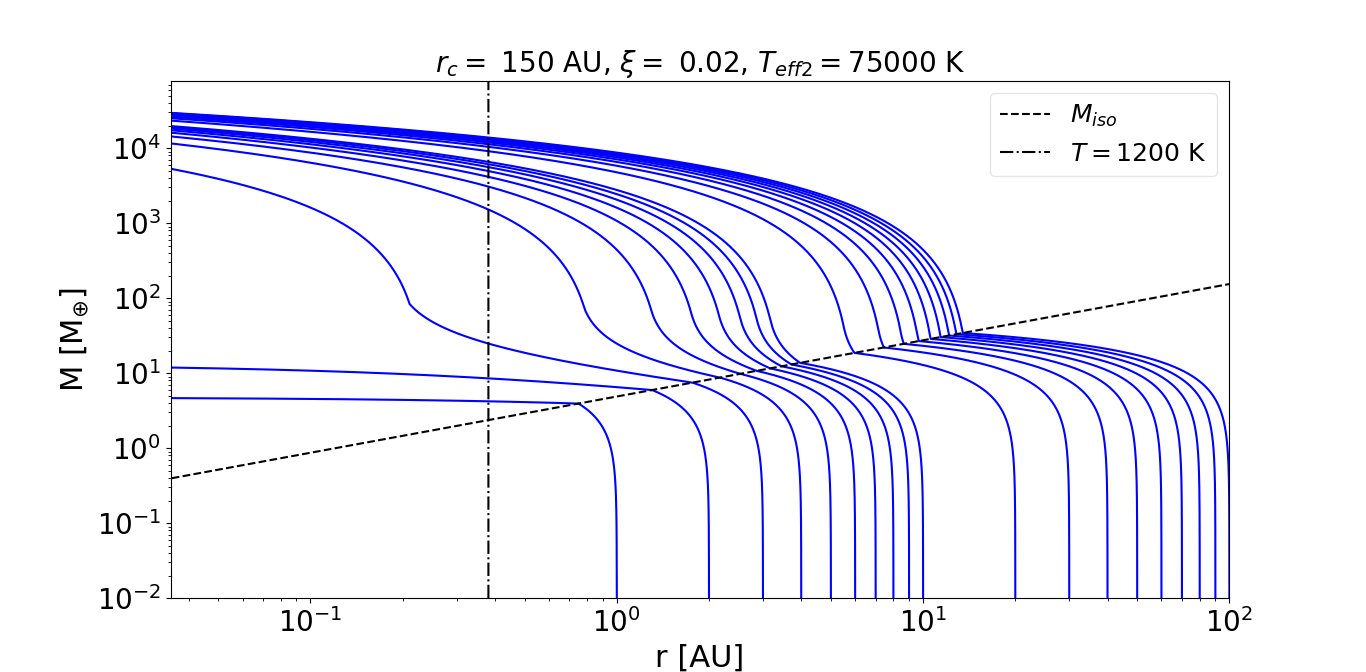}}
\subfloat[][]{\includegraphics[trim=0.cm 0.cm 1cm 1cm, clip,width=0.52\textwidth]{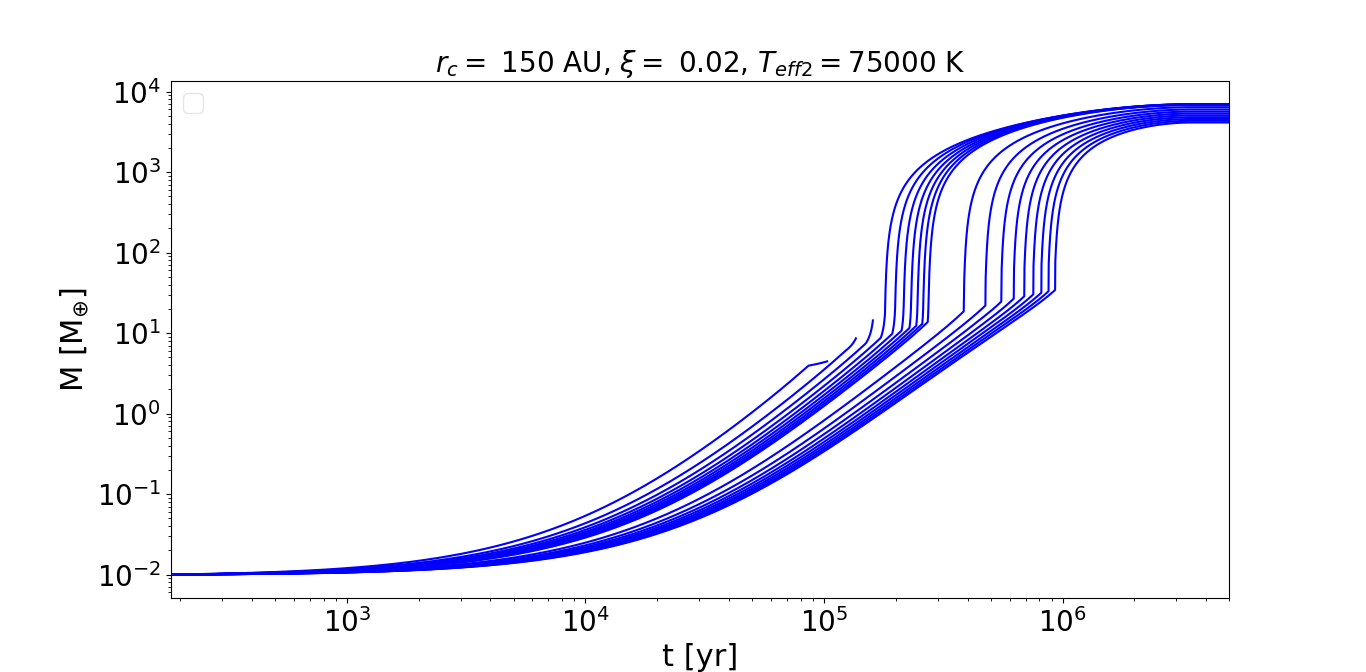}}
\caption{Planetary growth tracks for DWD$_1$ disc model with $r_{\rm c} = 150$ au and $\Teff$$_2$ = 75000 K. From top to bottom, $\xi = 0.01, 0.015, 0.02$, respectively. \emph{Left panels}: time-independent growth tracks. The vertical, dot-dashed line indicates the radial coordinate, $r_t$, where $T = 1200$ K. The dashed line represents the isolation mass profile for the given disc model (Eq. \ref{eqn:isomass}). \emph{Right panels}: time-dependent growth tracks. Starting from $t_0$, the tracks reach $M_{\rm iso}$ when they change slope (e.g. leftmost track at $\sim 170000$ yr).}
\label{fig:example_dwd1}
\end{figure*}

\subsubsection{Parameters adopted and definition of planetary classes}
\label{sec:classesdefinition}
We derive the disc metallicity $\xi$ from the stellar one. We choose three typical stellar values: a solar type one (Z = $0.014$) and two higher values which fall into the range of planet-hosting stars metallicity, i.e.,  Z = $0.021$ and Z = $0.028$. This means we consider cases with metallicity values $1\times$ solar, $1.5\times$ solar and $2\times$ solar.
To account for the condensation gradients created in the disc by its temperature profile, and the resulting  heavy elements (dust and pebbles) incomplete condensation \citep{turriniA, turriniB}, we set the disc metallicities $\xi$  to be $70\%$ of the adopted stellar values, i.e.,  $\xi = 0.7\ Z$ (Tab. \ref{table:freeparams}). The values of $\Teff$$_2$ when $t_0 > 0$ have been estimated by interpolation of the cooling tracks described in Section \ref{sec:dwdbinary}. 
 
We classify the simulated planets and their representative mass ranges as following:
\begin{itemize}
\item {\it Sub-Neptunes } (SNs, $1 < M$/\Mearth $< 9$): planets that complete the gas contraction phase but do not undergo runaway gas accretion. The mass of these planets is equally composed by gas and solids but, depending on their final orbit and atmospheric irradiation, they could lose most if not all their primary atmospheres. In the first scenario they could evolve into Mini-Neptunes, while in the second into Super-Earths.
\item {\it Neptunians} (Ns, $9 < M$/\Mearth $< 40$): their mass is equally composed by gas and solids as well as sub-neptunes, but they are usually originated by seeds starting their evolution in the outer regions of the disc. Therefore they can reach higher isolation masses, i.e. higher critical masses. However, they do not undergo runaway gas accretion.
\item {\it Gas Giants} (GGs, $40 <  M$/\Mearth $ < 10000$): planets that undergo runaway gas accretion and whose mass is dominated by gas.
\end{itemize}

\subsubsection{Planetary formation in the circumbinary disc aroud DWD$_1$: the steady-state disc case}
\label{sec:tracks_analytical}
The tracks described in this section implicitly assume a stationary disc structure lasting for as long as needed for the planets to form. As such, they provide an indication of the maximum planet-forming potential of the disc, but not of the real planet population it can produce during its lifetime.

The shape of these time-independent tracks is mainly affected by the radial position where the planet reaches $M_{\rm iso}$ and $M_{\rm crit}$, and by the transition radius $r_t$ where the disc temperature is $1200$ K. However, despite the temperatures of the disc surrounding DWD$_1$ can reach up to $\sim 16000$ K (see Fig. \ref{fig:disc_T_profiles}), $r_t < 0.5$ au (vertical dot-dashed line) for every simulation. Moreover, such high temperatures imply pebble accretion rates (see Eq. \ref{eqn:analytical_pebble} and Fig. \ref{fig:accr_rates_dwd1}) which are high enough to compensate for the high isolation masses (see Eq. \ref{eqn:isomass}). Therefore every seed reaches the isolation mass, whichever the value of $\xi$ and $r_{\rm c}$. 

The left panels of Fig. \ref{fig:example_dwd1} show the time-independent tracks for the combination of parameters $r_{\rm c} = 150$ au, $\Teff$$_2 = 75000$ K. From top to bottom, $\xi = 0.01, 0.015, 0.02$, respectively. The dashed line represents the isolation mass profile of the disc. Therefore, the part of each track under the dashed line represents the pebble accretion phase; the part of each track above the dashed line represents the gas accretion phase. Every seed reaches the isolation mass before the sublimation line at $\sim 0.38$ au. However, the lowest isolation masses are not able to reach the critical mass and do not undergo runaway gas accretion (e.g., Fig. \ref{fig:example_dwd1} top-left panel). 

If instead the seeds are initially located farther away from $r_{in}$, they can accrete pebbles across a larger distance, reaching higher isolation masses than seeds closer to $r_{in}$. Consequently, their gas contraction phase is fast (see Eq. \ref{eqn:KHtimescale}) and they trigger runaway gas accretion.
An increase in $\xi$ speed up the pebble accretion phase (see Eq. \ref{eqn:analytical_pebble}) for every seed. From top-left to bottom-left panel of Fig. \ref{fig:example_dwd1}, we can see some of the seeds located closer to $r_{in}$ reaching the isolation mass and eventually runaway gas accretion. The seeds located farther out reach higher isolation masses and consequently higher final masses at the end of the runaway gas accretion.

It is important to stress that the results shown in Fig. \ref{fig:example_dwd1} do not account for the temporal dimension, therefore they represents the maximum planet-forming potential of that particular disc model. If we account for the temporal dimension, the results are the right panels of Fig. \ref{fig:example_dwd1}, which will be described in the next section.

\begin{figure*}[t]
\centering
\subfloat[][]{\includegraphics[trim=1cm 0.cm 1cm 1cm, clip,width=0.52\textwidth]{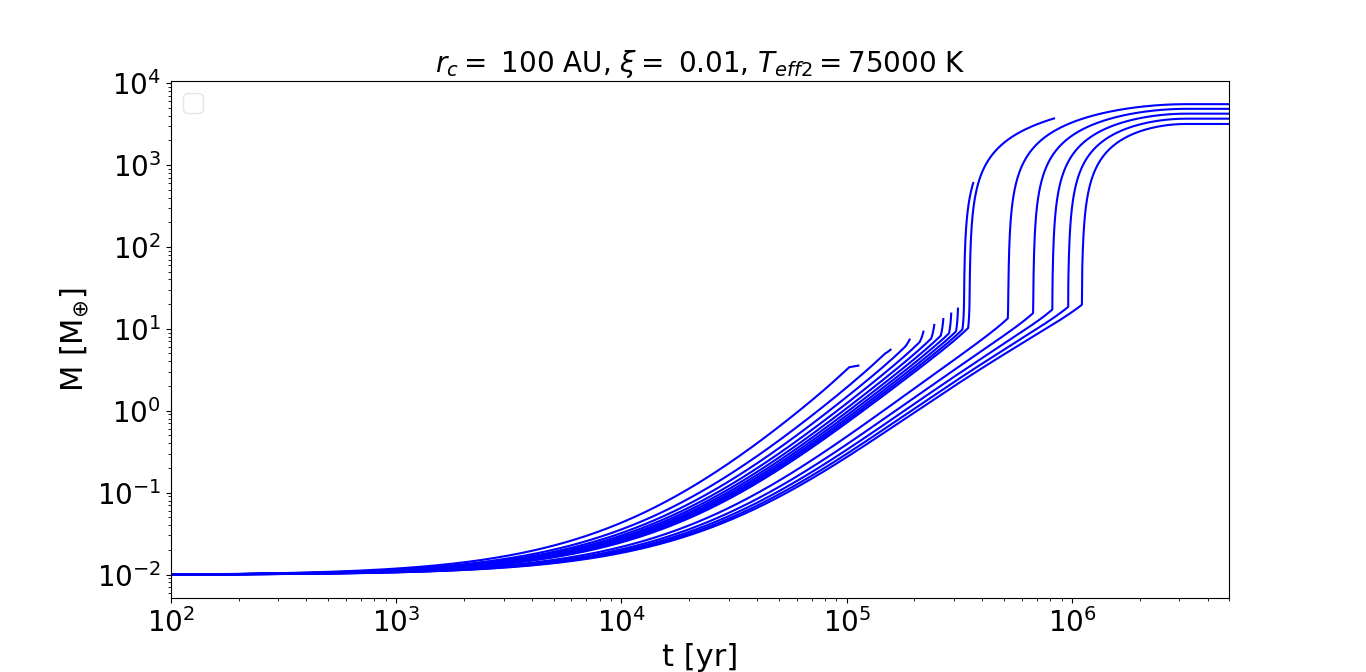}}
\subfloat[][]{\includegraphics[trim=1cm 0.cm 1cm 1cm, clip,width=0.52\textwidth]{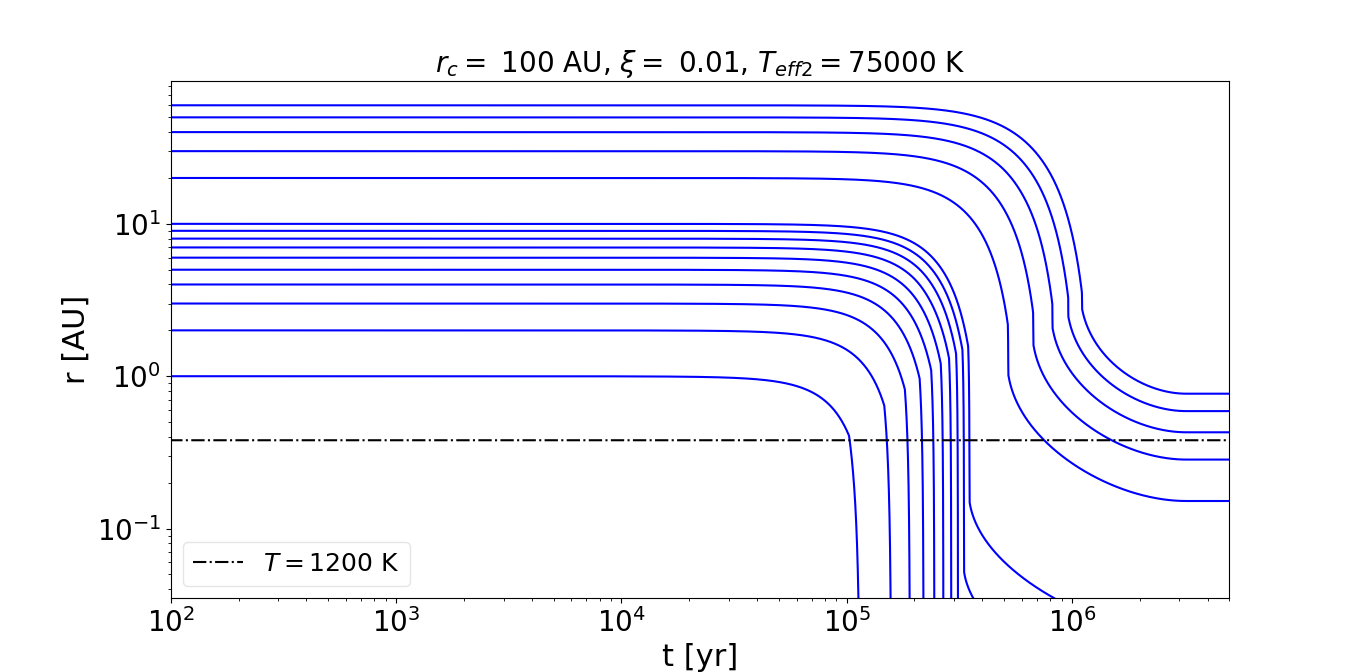}}\\
\subfloat[][]{\includegraphics[trim=1cm 0.cm 1cm 1cm, clip,width=0.52\textwidth]{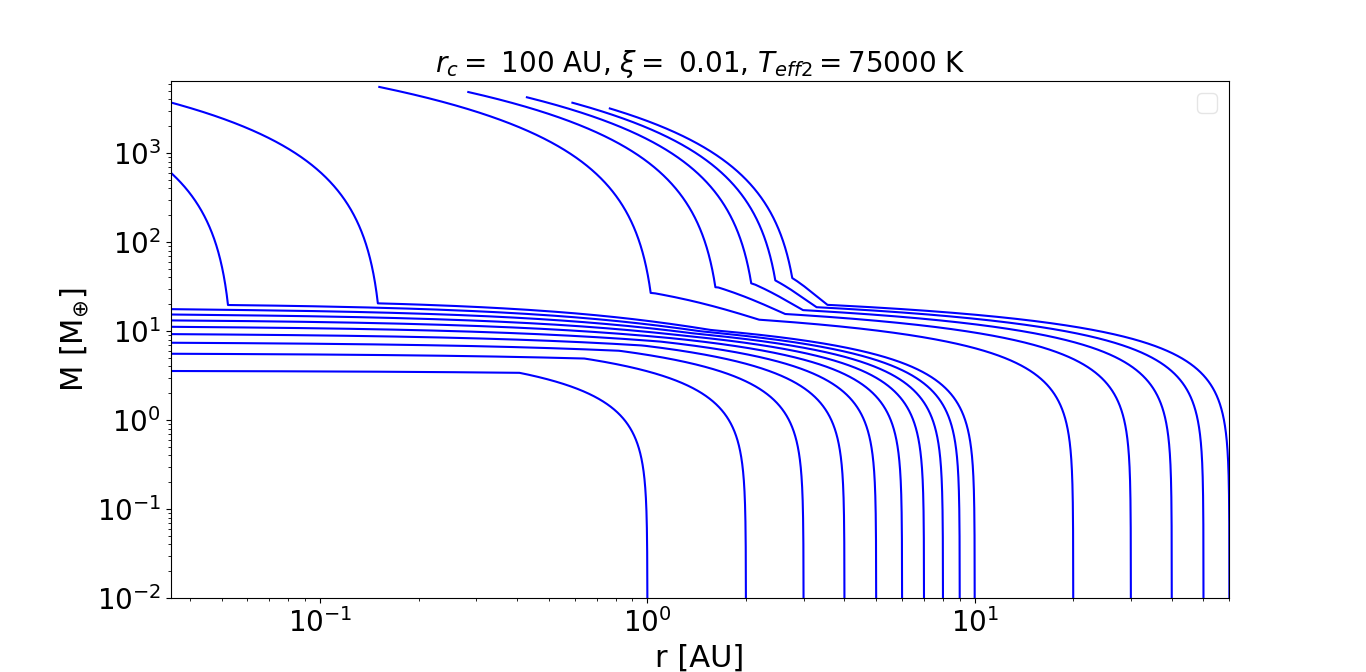}}
\subfloat[][]{\includegraphics[trim=1cm 0.cm 1cm 1cm, clip,width=0.52\textwidth]{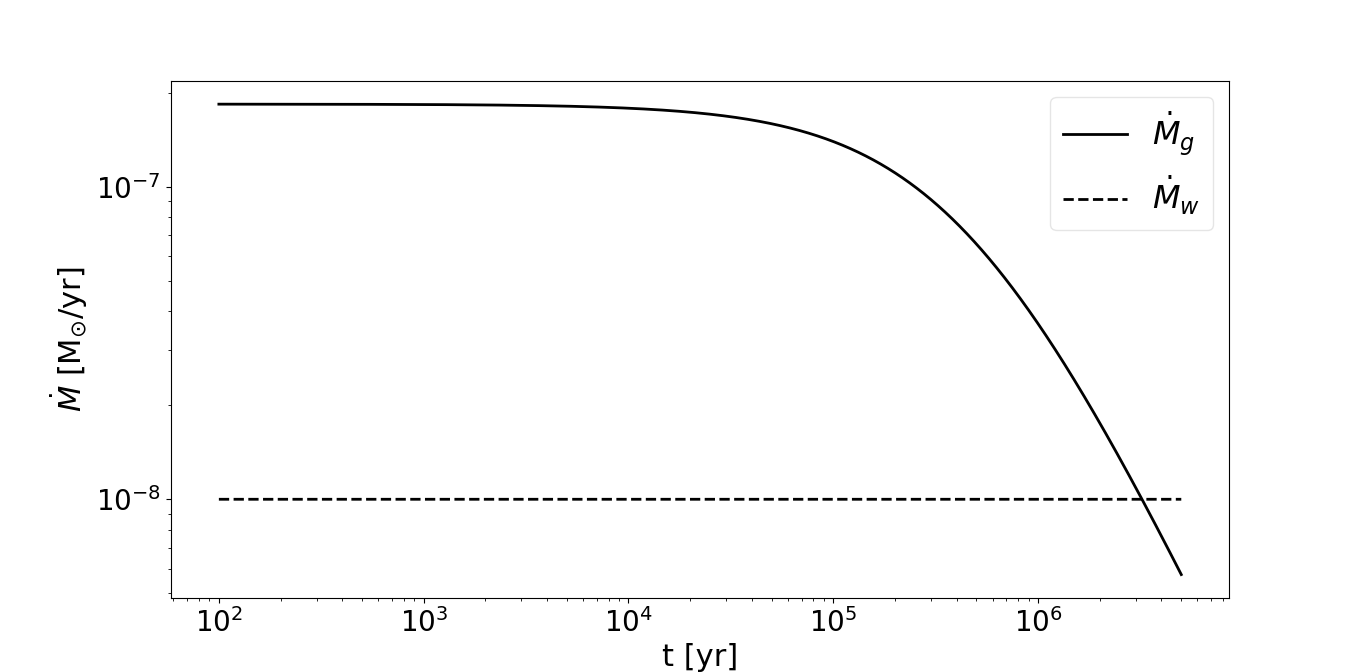}}
\caption{Time-dependent growth tracks (\emph{top-left panel}) and orbital tracks (\emph{top-right panel}) for the disc model with $r_{\rm c}, \xi$, $\Teff$$_2$ indicated at the top of the panels (DWD$_1$ system). The \emph{bottom-left panel} shows the same tracks in top-left and top-right panel but with the mass on the y-axis and the orbital radius on the x-axis. The \emph{bottom-right panel} shows the accretion rate of the disc described by the adopted parameters ($\Sigma_0 = 37.53$ g/cm$^2$). In the top-right panel, the dot-dashed line corresponds to the orbital radius where $T = 1200$ K. The apparently vertical lines corresponds to the gas contraction phase (see Section \ref{sec:tracks_time}). The fractions of track before and after the ``vertical'' section correspond to the pebble accretion and the runaway gas accretion phase, respectively.}
\label{fig:T75000_rc100}
\end{figure*}

\subsubsection{The time-evolving disc case}
\label{sec:tracks_time}
The time-dependent tracks include both mass growth tracks and migration tracks as a function of time in discs where $\dot{M}_g$ decays over time. Therefore they provide realistic information about the different masses, final positions and formation times of the planets. Once $t_0$ is fixed, the corresponding temperature of the stars and the temperature profile of the circumbinary disc are held constant throughout the entire simulation. In the time-dependent tracks the difference between $\dot{M}_g$ and $\dot{M}_w$ (i.e. the photoevaporation rate) is crucial: the closer to $0$ is \dMgw (see Section \ref{sec:gas_accretion}), the smaller the amount of gas supplied to the planet-forming region during the runaway gas accretion phase (see Eq. \ref{eqn:time_runawaygas}). 

The planet-forming potential of the DWD$_1$ disc emerging from the time-dependent tracks (Fig. \ref{fig:example_dwd1}, right panels) depicts a different picture from the time-independent tracks we previously discussed (Fig. \ref{fig:example_dwd1} (left panels). The growth tracks in each of the right panels can be read left-to-right: the leftmost track is the one related to the seed starting at $r_0 = 1$ au; the rightmost track is related to the seed initially located closer to $r_{\rm c}$ (the disc region we are sampling is the one where Eq. \ref{eqn:temperature} holds - see Section \ref{sec:discmodel} and Fig. \ref{fig:disc_T_profiles}). The fast-growing part of the tracks represents the gas accretion phase, while the part of track below gas accretion is the pebble accretion phase. Between the two, the gas contraction is represented by a path which becomes steeper for higher isolation masses. As an illustrative example, the leftmost track in the right panel with $\xi = 0.015$ almost do not increase in mass after reaching the isolation mass, which is $\sim 4$\ \Mearth. On the other hand, in the same panel, isolation masses higher than $\sim 10$\ \Mearth are characterised by a fast gas contraction followed by the runaway gas accretion once $M_{\rm crit} \sim 2M_{\rm iso}$ is reached.

The tracks represented in Fig. \ref{fig:example_dwd1} start at $t_0 = 10\tau_c$ (from top to bottom, $t_0 = 35, 100, 150$ yr, respectively - see Section \ref{sec:accretionmodel}), therefore the accretion rate of the disc is the highest possible (see Fig. \ref{fig:accr_rates_dwd1}). However, even when $\xi = 0.02$, the leftmost tracks do not undergo runaway gas accretion. This is due to the migration of the seeds, whose initial positions are the closest to $r_{in}$, and to their isolation masses, which are the lowest available for the seeds (see dashed lines in the left panels of Fig. \ref{fig:example_dwd1}). The combination of these two factors imply a slow gas contraction phase combined to a migration rate higher than during pebble accretion (see Eq. \ref{eqn:typeI} and \ref{eqn:time_gascontraction}), which eventually lead the planets to $r_{in}$. For the same reason, there are two tracks which interrupt runaway gas accretion in the right panels with $\xi = 0.01, 0.015$. When runaway gas accretion starts, the closer to $0$ is \dMgw, the less is the gas supplied to the planet-forming region, and the lower is the mass the planet can reach. Despite the high accretion rate of the disc, when $\xi$ = 0.01 or 0.015 the planets are already close to $r_{in}$ when the gas accretion phase starts.

Given that the isolation mass, and therefore the critical mass, increase with the radius of the disc, it takes more time for the seeds starting farther away from the binary stars to end pebble accretion and trigger gas accretion. When $t_0 = 10\tau_c$ and $r_{\rm c} = 150$ au, the rightmost tracks in Fig. \ref{fig:example_dwd1} start runaway gas accretion always after $\sim 0.7$ Myr, when the gas accretion rate has already decreased by $\sim 35\%$ with respect to its initial value (see Fig. \ref{fig:accr_rates_dwd1}, bottom panel).  They therefore experience a shorter runaway gas accretion rate with respect to planets which started their evolution closer to the stars, e.g. at $50$ au. This effect becomes more marked with decreasing $\xi$, i.e. with decreasing pebble accretion rate. Consequently, the mass range of the GGs in the right panels of Fig. \ref{fig:example_dwd1} narrows with increasing $\xi$.

When $r_{\rm c}= 50$ or $100$ au, the number of SN and N planets increases with respect to that of GGs, as the highest isolation masses cannot be reached. However, all planets reach higher masses with decreasing $r_{\rm c}$, i.e. with increasing disc density (as the total disc mass is fixed, see Section \ref{sec:dwd1_planetformation}). The reduced number of GGs is due to the smaller extensions of the discs, which limit the size of the regions where seeds could eventually form GGs as we saw previously.

In Fig. \ref{fig:T75000_rc100} we show the case of the disc model with parameters $r_{\rm c} = 100$ au, $\xi = 0.01$, $\Teff$$_2 = 75000$ K. The top-right panel shows the evolution of the orbital radii of the planets, where the topmost tracks are related to planets initially located farther from $r_{\in}$. The pebble accretion phase is mostly represented by the initially horizontal part of the tracks. Due to the high temperatures of the disc, the isolation masses are high. When a planet is close to its isolation mass, its orbital track quickly deviates from the constant path, and the following gas contraction phase eventually leads the planet to $r_{in}$. The farther the initial position of the seeds, the higher the isolation masses they can reach, and the time scale of their gas contraction phase decreases. Therefore the gas contraction phase in such orbital tracks is apparently vertical. The runaway gas accretion phase, which occurs with Type II migration, can be identified by a significantly slow down in the planet migration, which halts if the GG reaches its final mass before the disc lifetime. 

By comparing the top-right panel with the top-left panel of the same figure, we can see that the migration rate increases proportionally to the mass of the planet. Eventually this can lead the planets to $r_{in}$ before completing the gas accretion phase. Two tracks in the top-left panel of Fig. \ref{fig:T75000_rc100} reach $r_{in}$ at $\sim 0.38$ and $\sim 0.82$ Myr, during runaway gas accretion, as we can see from the top-right panel. The latter also shows that the later the runaway gas accretion starts, the closer the accretion rate of the disc is to the photoevaporation rate (bottom-right panel of the same figure). Therefore, the planet-forming region is depleted faster and the planet stops migrating and increasing its mass.

The bottom-left panel of Fig. \ref{fig:T75000_rc100} shows masses and radii of the top-left and top-right panel on the y axis and on the x axis, respectively. That is, the bottom-left panel is analogous to the time-independent tracks (e.g. left panels of Fig. \ref{fig:example_dwd1}), but it is made of time-dependent masses and radii. Therefore, it shows the actual planet-forming potential of the disc model adopted for that tracks.

\begin{figure}
\centering
\subfloat[][]{\includegraphics[trim=1.cm 0.cm 1.5cm 1.cm,clip,width=0.52\textwidth]{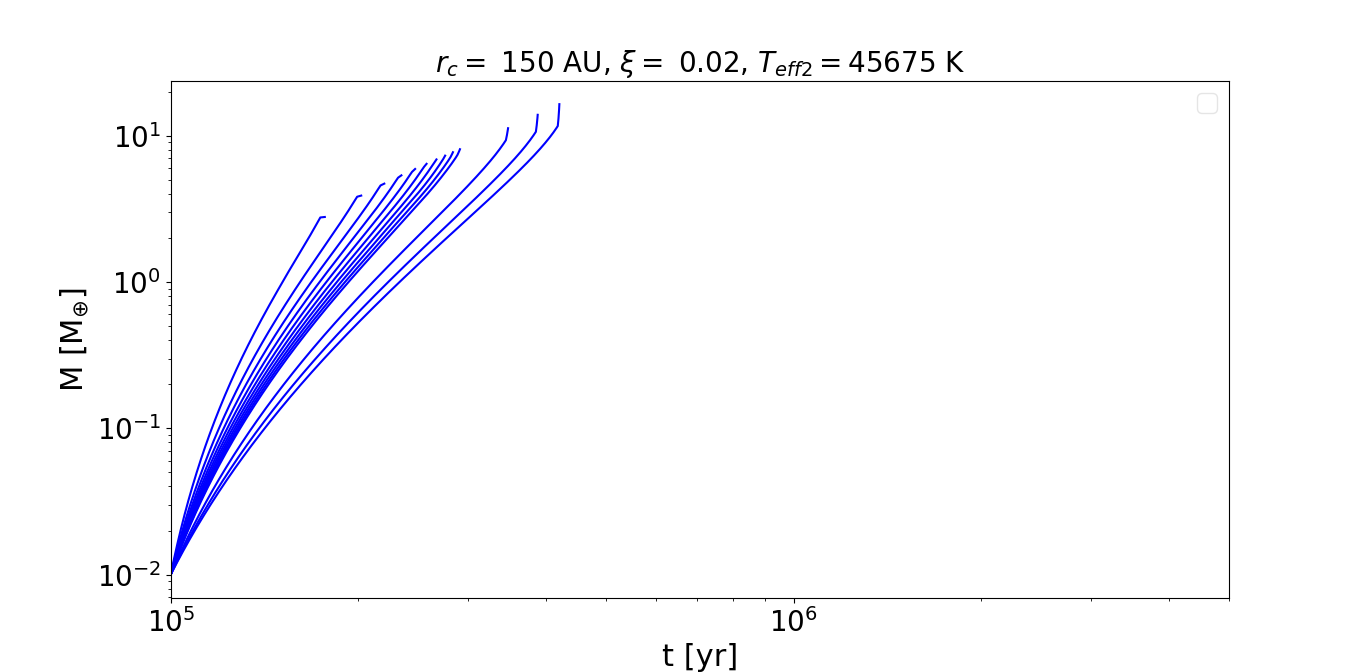}} \\
\subfloat[][]{\includegraphics[trim=1.cm 0.cm 1.5cm 1.2cm,clip,width=0.52\textwidth]{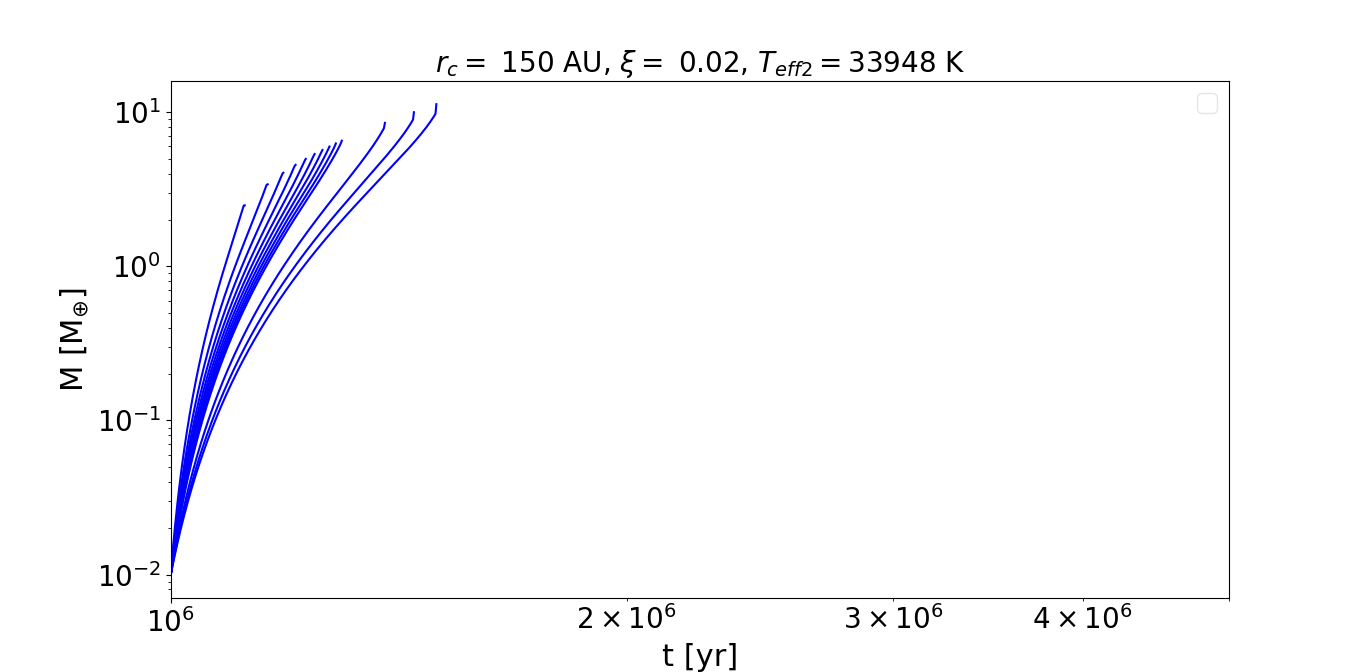}}
\caption{Time-dependent growth tracks of the planets formed by the circumbinary disc surrounding DWD$_1$ when planetary formation starts at $t_0 = 0.1$ Myr (top panel) and $t_0 = 1$ Myr (bottom panel).}
\label{fig:mostly_SNs}
\end{figure}

When $t_0 = 10\tau_c$ the planetary formation is boosted by the residual heat of the post-CE phase (see Eq. \ref{eqn:Tb}). Therefore, \dMgw and the isolation masses are the largest possible, which in turn facilitates the formation of GGs or Ns of large mass (see Section \ref{sec:formation_efficiency}). On the contrary, when $t_0 = 0.1$ or $1$ Myr, the \dMgw is close to $0$ already from the beginning of the simulations (see Fig. \ref{fig:accr_rates_dwd1}). Even if a planet reaches an isolation mass larger than $\sim 10$\ \Mearth, runaway gas accretion cannot be triggered. Therefore, no GGs form in these cases and the disc can only produce SNs and Ns. In particular, when $t_0 = 1$ Myr, the accretion rate is so low that the disc mostly forms SNs, and only a few Ns when $r_c = 150$ au (see Fig. \ref{fig:accr_rates_dwd1}, bottom panel, and Fig. \ref{fig:mostly_SNs}).

\begin{table}[t]
\caption{Formation times, fraction and positions of the planets produced by the circumbinary disc of DWD$_1$ when $t_0 = 10\tau_c$.} 
\label{table:taudwd1}
\centering
\begin{tabular}{|c|c|c|r|c|} 
\hline
\multicolumn{5}
{|c|}{\rule{0pt}{3.2ex}\large{$t_0$}= 10$\tau_c$,\ $\Teff$$_2$ = 75000 K} \\ [1.5ex]
\cline{1-5}
\hline
\hline
$\xi$ & $\Delta t_1$ & $\Delta t_g$ & \makecell[c]{$n_F$} & $n_p$\\ 
& [Myr] & [Myr] & \makecell[c]{[\%]} & [\%] \\
\hline\hline
\multicolumn{5}{|c|}{$\boldsymbol{r_{\rm c}} = 50$ au} \\ 
\hline

\cline{1-3} 
&   &   & \makecell[l]{\textbf{SN}} 33 & \makecell[l]{$\bf{\Delta_1}$}  100 \\

0.01 & 0.05 & 2 & \makecell[l]{\textbf{N}} 50 & \makecell[l]{$\bf{\Delta_{14}}$}  $-$ \\

 &  &  & \makecell[l]{\textbf{GG}} 17 & \makecell[l]{$\bf{\Delta_4}$}  $-$ \\
\hline

\cline{1-3} 
 &  &  & \makecell[l]{\textbf{SN}} 25 & \makecell[l]{$\bf{\Delta_1}$} 100 \\

0.015 & 0.03 & 2 & \makecell[l]{\textbf{N}} 58 & \makecell[l]{$\bf{\Delta_{14}}$} $-$ \\

 &  &  & \makecell[l]{\textbf{GG}} 17 & \makecell[l]{$\bf{\Delta_4}$} $-$ \\
\hline

\cline{1-3}
&  &  & \makecell[l]{\textbf{SN}} 17 & \makecell[l]{$\bf{\Delta_1}$} 100 \\

0.02 & 0.03 & 0.15 & \makecell[l]{\textbf{N}} 42 & \makecell[l]{$\bf{\Delta_{14}}$} $-$ \\

 &  &  & \makecell[l]{\textbf{GG}} 41 & \makecell[l]{$\bf{\Delta_4}$} $-$ \\
\hline\hline
\multicolumn{5}{|c|}{$\boldsymbol{r_{\rm c}} = 100$ au} \\ 
\hline\hline

\cline{1-3} 
&  &  & \makecell[l]{\textbf{SN}} 20 & \makecell[l]{$\bf{\Delta_1}$} 100 \\

0.01 & 0.12 & 0.36 & \makecell[l]{\textbf{N}} 33 & \makecell[l]{$\bf{\Delta_{14}}$} $-$ \\

 &  &  & \makecell[l]{\textbf{GG}} 47 & \makecell[l]{$\bf{\Delta_4}$} $-$ \\
\hline

\cline{1-3} 
&  &  & \makecell[l]{\textbf{SN}} 13 & \makecell[l]{$\bf{\Delta_1}$} 100 \\

0.015 & 0.08 & 0.4 & \makecell[l]{\textbf{N}} 20 & \makecell[l]{$\bf{\Delta_{14}}$} $-$ \\

 &  &  & \makecell[l]{\textbf{GG}} 64 & \makecell[l]{$\bf{\Delta_4}$} $-$ \\
\hline

\cline{1-3}
&  &  & \makecell[l]{\textbf{SN}} 13 & \makecell[l]{$\bf{\Delta_1}$} 100 \\

0.02 & 0.06 & 0.8 & \makecell[l]{\textbf{N}} 13 & \makecell[l]{$\bf{\Delta_{14}}$} $-$ \\

 &  &  & \makecell[l]{\textbf{GG}} 74 & \makecell[l]{$\bf{\Delta_4}$} $-$ \\ 
\hline\hline
\multicolumn{5}{|c|}{$\boldsymbol{r_{\rm c}} = 150$ au} \\ 
\hline\hline

\cline{1-3} 
&  &  & \makecell[l]{\textbf{SN}} 10 & \makecell[l]{$\bf{\Delta_1}$} 68 \\

0.01 & 0.19 & 0.45 & \makecell[l]{\textbf{N}} 16 & \makecell[l]{$\bf{\Delta_{14}}$} 37 \\

 &  &  & \makecell[l]{\textbf{GG}} 74 & \makecell[l]{$\bf{\Delta_4}$} $-$ \\
\hline

\cline{1-3}
&  &  & \makecell[l]{\textbf{SN}} 10 & \makecell[l]{$\bf{\Delta_1}$} 74 \\

0.015 & 0.14 & 0.27 & \makecell[l]{\textbf{N}} 5 & \makecell[l]{$\bf{\Delta_{14}}$} 26 \\

 &  &  & \makecell[l]{\textbf{GG}} 85 & \makecell[l]{$\bf{\Delta_4}$} $-$ \\
\hline

\cline{1-3}
&  &  & \makecell[l]{\textbf{SN}} 5 & \makecell[l]{$\bf{\Delta_1}$} 74 \\

0.02 & 0.1 & 2.5 & \makecell[l]{\textbf{N}} 10 & \makecell[l]{$\bf{\Delta_{14}}$} 26 \\

 &  &  & \makecell[l]{\textbf{GG}} 85 & \makecell[l]{$\bf{\Delta_4}$} $-$ \\ 
\hline
\end{tabular}
\tablefoot{\\The values related to the vertically arranged items of columns $n_F$ and $n_p$ are presented on their right, respectively. See Section \ref{sec:classesdefinition} for the definition of the three classes of planets. Following, a brief description of the parameters reported in the table (see Section \ref{sec:discmodel}, \ref{sec:accretionmodel} and \ref{sec:timing} for details).\\
\tablefoottext{a}{$\Delta t_1$: formation time of the first planet reaching the end of its formation process.} \\
\tablefoottext{b}{$\Delta t_g$: formation time of the first gas giant planet.}\\
\tablefoottext{c}{$n_F$: percentage of (i) Sub-Neptunes (SN), (ii) Neptunes (N), (iii) Gas Giants (GG) formed with respect to the total number of initial seeds.} \\
\tablefoottext{e}{$n_p$: percentage of planets located within the ranges $\Delta_1 = r_{in}-1$ au, $\Delta_{14} = 1-4$ au and $\Delta_4 = 4\ \text{AU}-r_c$.}
} 
\end{table}

\subsubsection{Timing of planetary formation and orbital evolution}
\label{sec:timing}
We now draw the global picture and discuss the implications of the characteristic formation times of the planets born from the circumbinary disc surrounding DWD$_1$.
For each adopted value of $t_0$, we report results in three sets of tables: Tab. \ref{table:taudwd1}, Tab. \ref{table:0.1dwd1} and Tab. \ref{table:1dwd1}. Every set is composed by three sub-tables, one for each value of $r_{\rm c}$. Each sub-table has three rows corresponding to different values of $\xi$, and seven columns. From left to right, such columns show the formation time of the first planet ($\Delta t_1$); if a GG exists, the formation time of the first GG ($\Delta t_g$); the percentage of SNs, Ns and GGs formed with respect to the total number of planets ($n_F$); the percentage of planets ($n_p$) whose final positions are within $r_{\rm in}$ and $1$ au (i.e. within $\Delta_1$), within $1$ and $4$ au (i.e. within $\Delta_{14}$) and between $4$ au and $r_{\rm c}$ (i.e. within $\Delta_4$). We focus on the orbital interval within 4 au as it is associated to the range of orbital periods where the LISA mission can detect giant planets within its nominal mission lifetime \citep{dani2019}. We consider as planets all the seeds that reached at least $0.1$ M$_{\oplus}$ \citep{sinurkoff}, i.e. the mass of Mars.

When $t_0 = 10\tau_c$, the effective temperature of the youngest, hotter WD is $\Teff$$_2$ = 75000 K, and the disc is the hottest possible. Also the accretion rate is the highest possible, and in this case we obtained GGs for each value of $r_{\rm c}$. However, a few GGs do not complete runaway gas accretion as they reach the inner radius of the disc (e.g., Fig. \ref{fig:T75000_rc100}). The formation time of such GGs coincides with their crossing times of the disc, which span $0.15$ to $0.8$ Myr (see Tab. \ref{table:taudwd1}). The GGs left complete runaway gas accretion within at least $2$ Myr, whatever the extension of the disc and the value of $\xi$. 

Due to the high accretion rate of the disc every seed can reach the isolation mass and starts the gas contraction. Consequently, the first planets formed are always SNs, and they formed very fast (between $0.03$ and $0.1$ Myr in $7$ cases over $9$). However, given that such SNs grew up from seeds initially located close to $1$ au, they spend most of their evolution within $1$ au. Moreover, it take at least $2$ Myr for the farther seeds to complete runaway gas accretion. Therefore, in every case most, if not all, the evolution of the planets end within $1$ au from the centre of the disc (see Tab. \ref{table:taudwd1}). Only the case with $r_{\rm c} = 150$ au shows between $26\%$ and $37\%$ of planets within $1$ and $4$ au, and they are all GGs.

When $t_0 = 0.1$ Myr, \dMgw decreases significantly with respect to the previous case (e.g., Fig. \ref{fig:accr_rates_dwd1}), therefore no GGs are formed. However, the accretion rate is high enough to allow for a quick formation of the first planets (SNs), which is still within $0.035$ and $0.1$ Myr in almost all cases (see Tab. \ref{table:0.1dwd1}). However, as no planet starts the runaway gas accretion and transitions to the slower Type II migration, every planet formed in this case end its evolution within $1$ au from the disc centre.

Finally, after $1$ Myr, when $\Teff$$_2 = 33950$ K and the disc is the coolest one, the disc can only form SNs, which complete their growth process between 0.15 and 0.35 Myr. Since these planets form early and rapidly migrate due Type I migration, all planets end their evolution within $1$ au from the centre of the disc.

\subsubsection{Planetary formation efficiency of the adopted disc models}
\label{sec:formation_efficiency}
\begin{figure}
\centering
\subfloat[][]{\includegraphics[trim=1.cm 0.cm 1.5cm 1.cm,clip,width=0.52\textwidth]{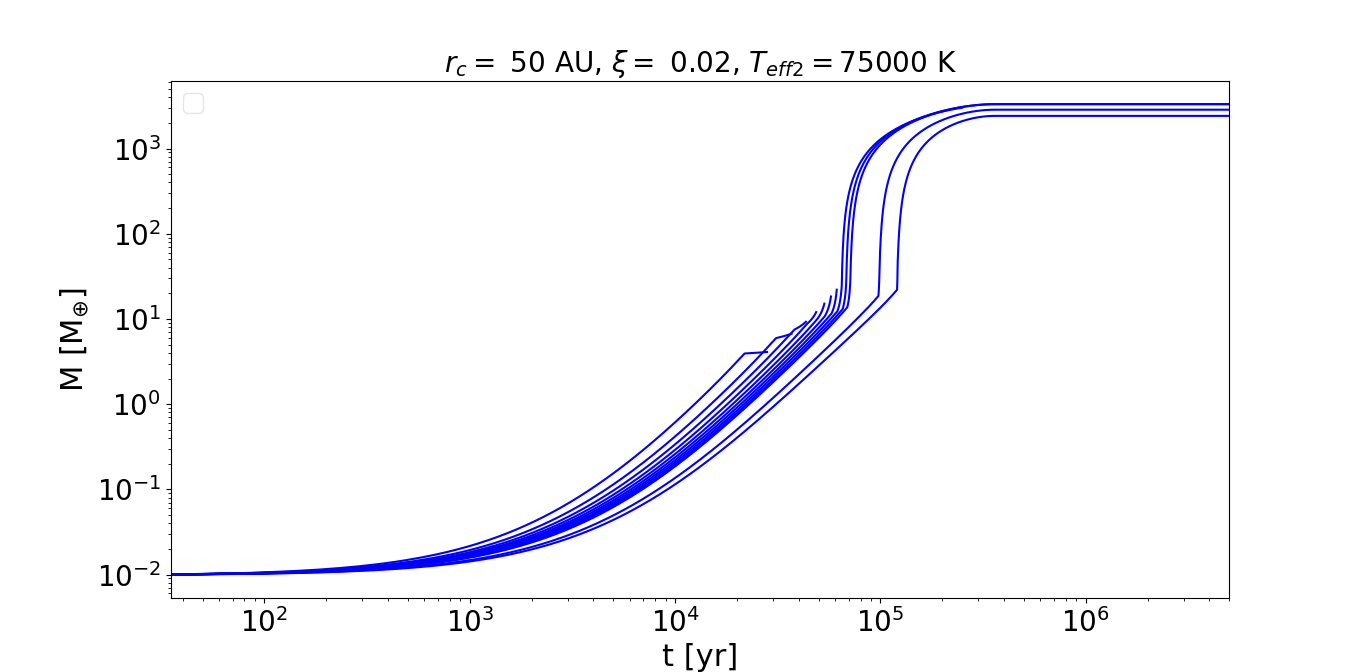}} \\
\subfloat[][]{\includegraphics[trim=1.cm 0.cm 1.5cm 1.5cm,clip,width=0.52\textwidth]{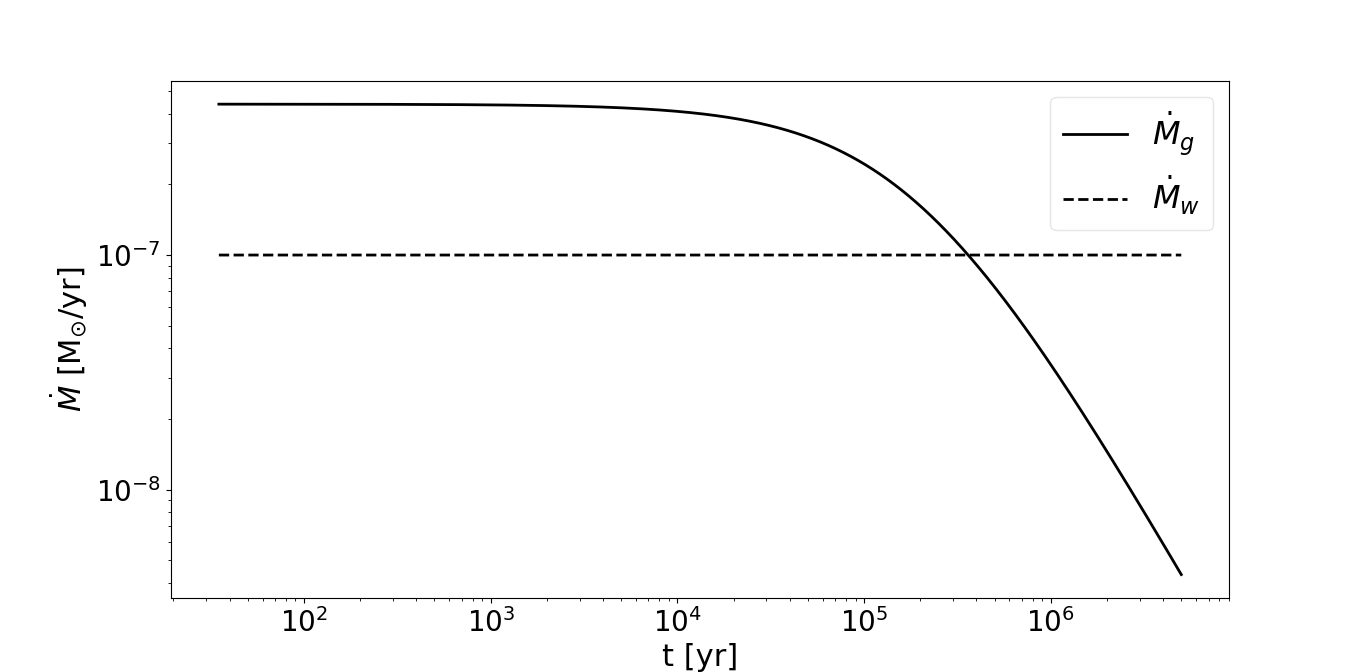}}
\caption{Time dependent growth tracks starting at $t_0 = 10\tau_c$ for the circumbinary disc of DWD$_1$ with parameters $r_{\rm c} = 50$ au, $\xi = 0.02$ (top panel), considering a constant photoevaporation rate of $\dot{M}_w = 10^{-7}$ \Msun/yr. Accretion rate and photoevaporation rate are shown in the bottom panel. In this case \dMgw (see Section \ref{sec:gas_accretion}) is smaller than in the top panel of Fig. \ref{fig:accr_rates_dwd1}, as well as the mass of the GGs in the top panel, which are smaller than those indicated in Section \ref{sec:formation_efficiency}.}
\label{fig:Mw_higher}
\end{figure}

The mass range and percentage of planets produced by the adopted disc models for system DWD$_1$ can be summarised as the following:
\begin{enumerate}
\item \textbf{\boldmath $t_0$ = 10\,$\tau_c$} ($\Teff$$_2$ = 75000 K): as \dMgw is the larger possible (e.g., Fig. \ref{fig:accr_rates_dwd1}), we obtain GGs in almost every case apart from the two models with $r_{\rm c} = 50$ au, $\xi = 0.01$ and $0.015$, where most of the planets are Ns (see Tab. \ref{table:taudwd1}). However, the latter class of planets forms in every case, and their mass range is $8.5-26$\ \Mearth ($0.027-0.082$ \Mj). The GGs which grew up from seeds closer to $r_{in}$ only complete their evolution in the case $r_{\rm c} = 50$ au, and their mass is within $4500-10000$\ \Mearth ($14.15-31.45$ \Mj). This is due to the highest disc density when $r_{\rm c} = 50$ au. In the 
other cases ($r_{\rm c} = 100, 150$ au), where the disc extension is larger, new seeds are able to begin runaway gas accretion, even if a fraction of them do not complete it. The mass range is therefore larger than the case $r_{\rm c} = 50$ au: $100-8800$\ \Mearth ($0.31-27.68$ \Mj). Given the higher percentage of Ns and GGs with increasing $r_{\rm c}$ (see Tab. \ref{table:taudwd1}), the percentage of SNs decreases from $33\%$ ($r_{\rm c} = 50$ au, $\xi = 0.01$) to $5\%$ ($r_{\rm c} = 150$ au, $\xi = 0.02$), and their total mass range is between $3.5$ and $8$\ \Mearth.

\item \textbf{{\boldmath $t_0 = 0.1$} Myr} ($\Teff$$_2$ = 45700 K). In this case most of the planets are SNs for every choice of the adopted parameters (see Tab. \ref{table:0.1dwd1}). The discs form no GGs and only a limited number of Ns. Their mass range is within $9-17$\ \Mearth\ ($0.03-0.05$ \Mj). This outcome, and the consequent high percentage of SNs, is due to the decreased disc accretion rate, which is not able to support a fast pebble accretion phase in most of the cases. Moreover, the isolation mass profile of the disc includes smaller masses with respect to $t_0 < 0.1$ Myr, where the disc temperatures are the highest (see Eq. \ref{eqn:isomass}). Therefore, the seeds cannot reach high isolation masses and the gas contraction phase is slow (see Eq. \ref{eqn:KHtimescale}). The mass range of such SNs is $2-2.9$\ \Mearth.

\item \textbf{\boldmath $t_0 = 1$ Myr} ($\Teff$$_2$ = 33950 K). As the temperatures of the disc are the lowest possible, this case is a more extreme version of the one at $t_0=0.1$ Myr. The planets formed in this case are all SNs, apart from the models $r_{\rm c} = 150$ au, $\xi = 0.015, 0.02$ (see Tab. A.2). In these cases the farther seeds become Ns and their mass range is $9-13$\ \Mearth\ ($0.030.04$ \Mj). The masses of SNs are globally between $2-2.6$\ \Mearth.
\end{enumerate}
Scenarios including high $\dot{M}_w$ produce a smaller number of GGs with smaller masses and a larger population of Ns than with a smaller $\dot{M}_w$. The new Ns are planets which reached the critical mass but that could not undergo runaway gas accretion due to \dMgw $\approx 0$. On the other hand, a small value of $\dot{M}_w$ can facilitate the formation of GGs at any time, as their planet-forming region would be supplied mass for a longer time through $\dot{M}_g$. 
Therefore by using among the highest value of $\dot{M}_w$ from the study of \citet{tanaka} we obtained conservative results. If we consider $\dot{M}_w = 10^{-7}$ \Msun/yr, which is the highest value considered by \citet{tanaka}, the model which produced the GGs with the highest masses (i.e., the disc surrounding the DWD$_1$ system at $t_0 = 10\tau_c$ with parameters $\xi = 0.02$, $r_{\rm c} = 50$ au) produces less massive GGs (see top panel and bottom panel of Fig. \ref{fig:Mw_higher}, respectively).

\subsection{Planetary formation around LISA-observable DWDs}
\label{sec:LISAsystems}
As previously discussed, the GW signal of the compact systems DWD$_2$, DWD$_3$, DWD$_4$ can be detected by the LISA space mission. Therefore, the formation of planets within their post-CE circumbinary discs will be the subject of this section. In particular, we explore the similarities and discrepancies among the class of planets formed, and among the timescale needed for their formation.
In the following we only analyse the time-dependent planetary formation tracks. The values used for the free parameters $r_{\rm c}$ and $\xi$ are listed in Tab. \ref{table:freeparams}, and the stellar parameters in Tab. \ref{tab:DWDparams}.

\subsubsection{The DWD$_2$ case} 
\label{sec:dwd2system}
The DWD$_2$ system has a circumbinary disc with mass $M_d = 0.09$\ \Msun\ and $\Sigma_0 = 51, 13, 6$ g/cm$^2$ for $r_{\rm c} = 50, 100, 150$ au, respectively. The three main temperatures along the cooling track of the youngest WD are $\Teff$$_2$ = 58400, 32000, 25000 K at $t_0 = 10\tau_c$, $0.1$ Myr and $1$ Myr, respectively. The results of the simulations are shown in Tab. \ref{table:taudwd2}, \ref{table:0.1dwd2} and \ref{table:1dwd2}.

Fig. \ref{fig:disc_T_profiles} shows the circumbinary disc of the DWD$_2$ system is the coldest we consider already at $t_0 = 10\tau_c$. Consequently, even in the most favourable case, i.e. $t_0 = 10\tau_c$ and $\Teff$$_2 = 58400$ K, the planets formed in the disc have low masses and are mostly SNs. 

The case $t_0 = 10\tau_c$ is the only one producing all kind of planets (see Section \ref{sec:classesdefinition}), and in particular it is the only case producing Ns and GGs. The latter are produced only when $r_{\rm c} = 100$ au and $\xi = 0.02$, with a mass within $130-160$\ \Mearth\ ($0.41-0.5$ \Mj). The disc with $r_{\rm c} = 150$ au does not form GGs (see Tab. \ref{table:taudwd2}). In this disc seeds initially located farther away from the stars could reach higher isolation mass, and thus experience a quick gas contraction phase and trigger runaway gas accretion. However, \dMgw $= 0$ (see Section \ref{sec:gas_accretion}) before runaway gas accretion is triggered, therefore the Ns reach the critical mass but the depleted planet-forming region is not replenished by $\dot{M}_g$ and the Ns cannot accrete more gas. When $r_{\rm c} = 150$ au the masses of Ns are among the highest of every simulation (between $16$ and $28$\ \Mearth\, i.e. $0.05-0.09$ \Mj). Globally, Ns reach masses between $9$ and $28$\ \Mearth\ ($0.03-0.09$ \Mj). The rest of the planets are SNs, with masses in the range $2-2.8$\ \Mearth.

Given that the accretion rate is low since the beginning of every simulation, when $t_0 = 0.1$ or $1$ Myr the discs form only SNs, i.e. planet formation has limited efficiency. Therefore, regardless of $t_0$, all the planets produced end their evolution within $1$ au and the formation times are shorter than $0.1$ Myr only when $r_{\rm c} = 50$ au (see Tab. \ref{table:taudwd2}-\ref{table:1dwd2}). Otherwise it can take up to $0.45$ Myr to form a SN, when the disc density, and $\xi$ are the lowest ones and $t_0 = 1$ Myr (see Tab. \ref{table:1dwd2}). When $t_0$ = 0.1 or 1 Myr the mass range of the SNs is $1.2-8.5$\ \Mearth.

\subsubsection{The DWD$_3$ case} 
\label{sec:dwd3system}
The circumbinary disc surrounding the DWD$_3$ system has mass $M_d$ = 0.25 \Msun \ and $\Sigma_0 = 143, 34, 16$ g/cm$^2$ for $r_{\rm c} = 50, 100, 150$ au, respectively.
We simulated the planetary formation tracks considering the three different temperatures $\Teff$$_2$ = 52000, 32200, 24900 K corresponding to $t_0 = 0, 0.1, 1$ Myr, respectively. The results of the simulations are shown in Tab. \ref{table:taudwd3}, \ref{table:0.1dwd3} and \ref{table:1dwd3}.

This disc is more massive and dense than the DWD$_2$ case, and it is similar but colder than the one surrounding the DWD$_1$ system. At $t_0 = 10\,\tau_c$ we have a great diversity of planets. When $r_{\rm c} = 50$ au, despite its high density the disc can form Ns only when $\xi = 0.02$. This is due (i) to the accretion rate, which is three times smaller than the one of the disc surrounding the DWD$_1$ system with the same parameters; (ii) to the temperature of the disc, which is globally lower than the DWD$_1$ case (see Fig. \ref{fig:disc_T_profiles}). Consequently, the isolation masses the seeds can reach are smaller and they cannot complete the gas contraction phase, i.e. the disc forms only SNs and Ns (see Tab. \ref{table:taudwd3}). When $r_{\rm c}$ = 100 or 150 au, more space is available for the seeds to reach higher isolation masses, and the disc forms GGs within a mass range $880-4700$\ \Mearth\ ($2.77-14.78$ \Mj). The lower upper boundary to the masses of the GGs is a consequence of the lower $\dot{M}_g$, which suppresses the growth of the planet sooner than for the DWD$_1$ case (see Eq. \ref{eqn:time_runawaygas}). Moreover, the generally less efficient pebble accretion (i.e. smaller $\dot{M}_g$, see Eq. \ref{eqn:time_pebble}) slows down the planetary growth, and the GGs are formed later than $2.5$ Myr, apart from the model $r_{\rm c} = 150$ au, $\xi = 0.015$, where a GG reached $r_{in}$ before completing the runaway gas accretion (see Tab. \ref{table:taudwd3}).

For all three $r_{\rm c}$ values the disc forms Ns, whose masses are the highest for $r_{\rm c}$ = 100 and 150 au. Globally, their mass range is $9-23$\ \Mearth\ ($0.03-0.07$ \Mj). The rest of the planets are SNs with masses $2.6-8.5$\ \Mearth, which forms within $0.15$ Myr (see Tab. \ref{table:1dwd3}). When $t_0 \geq 0.1$ Myr the disc forms only SNs,  whose masses range $1.8-8$\ \Mearth\ and whose formation times are between $0.04$ and $0.13$ Myr. The only exception is the case $t_0 = 0.1$ Myr, $r_{\rm c} = 150$ au, $\xi = 0.02$. As previously mentioned, the temperatures of the disc surrounding DWD$_3$ are lower than the DWD$_1$ case, therefore without the boost of heat provided by the post-CE phase, the pebble accretion phase is less efficient and the isolation masses are lower. This has two consequences: (i) even if the disc produces Ns in one case, their mass is not higher than $10$\ \Mearth\ ($0.03$ \Mj); (ii) all the planets, whatever $t_0$, end their evolution within $1$ au (see Tab. \ref{table:taudwd3}-\ref{table:1dwd3}).

\subsubsection{The DWD$_4$ case} 
\label{sec:dwd4system}
In the DWD$_4$ system (Tab. \ref{tab:DWDparams}), the disc formed after the last CE phase has $M_d = 0.087$ \Msun \ and $\Sigma_0 = 50, 12, 6$ g/cm$^2$ for $r_{\rm c} = 50, 100, 150$ au, respectively. The three main temperatures along the cooling track of the youngest WD, $\Teff$$_2$ = 50100, 32200, 24900 K at $t_0 = 10\tau_c$, $0.1$ and $1$ Myr, respectively. The results of the simulations are shown in Tab. \ref{table:taudwd4}, \ref{table:0.1dwd4} and \ref{table:1dwd4}.

Despite the temperature profile being similar to that of the DWD$_3$ disc (see Fig. \ref{fig:disc_T_profiles}), the DWD$_4$ disc produces GGs only for two sets of initial conditions ($t_0 = 10\tau_c$, $r_{\rm c} = 100$ au, $\xi = 0.015, 0.02$), and their masses are among the lowest produced by all four systems considered in this work. The DWD$_4$ system is surrounded by the least dense of all the discs explored before, but the region where Eq. \ref{eqn:temperature} holds can be larger than the one of the disc surrounding the DWD$_3$ system (see Fig. \ref{fig:disc_T_profiles}). Therefore the type of planets and the planetary masses it produces are in between the DWD$_3$ case (similar disc temperatures, which allow the formation of GGs) and the DWD$_2$ case (similar extension of the irradiated disc region, which allow the formation of Ns with masses higher than those of DWD$_3$). 

When $t_0 = 10\tau_c$ the disc forms only one small GG of $\sim 40$\ \Mearth ($0.12$ \Mj) when $r_{\rm c} = 100$ au. This is true also for half of the GGs produced in the case left ($\xi = 0.02$), whose masses are not higher than $90$\ \Mearth\ ($0.28$ \Mj). These planets started runaway gas accretion, but this occurred when \dMgw $\simeq 0$, therefore their growth stopped quickly. The rest of the GGs produced have masses between $140$ and $380$\ \Mearth\ ($0.44-1.19$ \Mj). No GG can form after $1$ Myr as a consequence of the low accretion rates, which stops supplying gas to the planet-forming region soon. Therefore the planet-forming region quickly loose its gas and (i) the planets have no more gas to accrete soon after reaching the critical mass, and (ii) they stops migrating because no more gas is present in the planet-forming region. Consequently, in $4$ cases over $5$ a fraction of these GGs end their evolution within $1$ and $4$ au (see Tab. \ref{table:taudwd4}). The rest of the GGs, together with the other planets produced by the disc, end their evolution within $1$ au.

When $t_0 = 10\tau_c$ the disc forms also SNs and Ns, whose masses are within the ranges $2.4-8.5$\ \Mearth\ and $9-40$\ \Mearth\ ($0.03-0.12$ \Mj), respectively. Due to the lower accretion rate with respect to the rest of the systems, i.e. the lower disc density, they form between $0.06$ and $0.4$ Myr. When $t_0 \geq 0.1$ Myr, the discs produce Ns in $3$ cases over $9$ (see Tab. \ref{table:0.1dwd4}, \ref{table:1dwd4}), and their masses are higher than those produced in the DWD$_3$ system despite the lower density of the disc of DWD$_4$. As previously mentioned, this is due to the larger extension of the irradiated region of the disc\footnote{The extension of the irradiated region of the disc is strictly related to the outcome of the CE phase, i.e. to the final masses and radii of the stars. They affect the slope of Eq. \ref{eqn:temperature_toomre} and Eq. \ref{eqn:Tb}, and thus the radial coordinate of the crossing point between the two temperature profiles.}. DWD$_4$ can host seeds initially located farther than those in the DWD$_3$ discs, which allows them to reach slightly higher isolation masses. The planets continue evolving along gas contraction until they reach $r_{in}$, and their final mass ranges between $9$ and $17$\ \Mearth\ ($0.03-0.05$ \Mj). The rest of the planets are SNs with masses within $1.6-8.5$\ \Mearth, and they forms within $0.07$ and $0.33$ Myr, therefore slower than those of the disc surrounding DWD$_3$ at $t_0 = 10\tau_c$, as the latter is characterised by an higher accretion rate.

When $t_0 = 1$ Myr, despite the lower accretion rate of the disc, the disc produces Ns when $r_{\rm c} = 150$ au, $\xi = 0.2$. The reason is the same that justify the formation of Ns when $t_0 = 0.1$ Myr. The Ns produced in this case have masses in the range $9.2-10.2$ M$_\oplus$ ($0.029-0.032$ \Mj). The rest of the planets are SNs with masses in the range $1.5-8.2$\ \Mearth, which are formed within $0.3$ and $0.65$ Myr. Apart from the set of parameters at $t_0 = 10\tau_c$ discussed above, every disc model at any $t_0$ forms planets which end their evolution within $1$ au.

\subsection{Planet formation with first generation Neptunian planets}
\label{sec:massive_seed}
In the previous sections we assumed the seeds used to start the planet formation processes formed within $0.1$ and $1$ Myr after the formation of the circumbinary discs (see Section \ref{sec:results}). When $t_0 = 10\tau_c$, we assumed the seeds to be Moon-sized first-generation planets. 
In this section we expand the previous analysis by considering first generation seeds with mass $M_0 = 10$\ \Mearth\ ($0.031$ \Mj). That is, we analyse what second-generation planets can form if the discs contain first-generation Neptunian planets that survive the evolution of the binary systems under study, with focus on whether GGs can form.

When $t_0 = 0.1$ Myr the only systems forming GGs are DWD$_1$ and DWD$_4$. In particular, the disc surrounding DWD$_1$ forms GGs when $r_{\rm c} = 150$ au and $\xi = 0.015, 0.02$, and their mass range is $1500-2300$\ \Mearth\ ($4.72-7.23$ \Mj). Given that in this case $M_0$ is larger, the seeds can start runaway gas accretion phase when \dMgw is higher than zero (see Fig. \ref{fig:accr_rates_dwd1}, bottom panel). The same considerations made when $M_0 = 0.01$\ \Mearth\ holds even in this case (see Sections \ref{sec:tracks_time}, \ref{sec:timing}, \ref{sec:formation_efficiency}), given that the discs properties did not change. Therefore, despite in this case $\dot{M}_g \simeq 2.7\dot{M}w$, when $r_{\rm c} = 150$ au the extension of the disc surrounding DWD$_1$ is enough to provide a favourable environment for the formation of GGs from seeds with $M_0 = 10$ \Mearth. Such GGs are formed within $2.6$ Myr. The Ns planets formed have masses within the range $11.2-30$\ \Mearth\ ($0.03-0.09$ \Mj), whatever the extension of the disc, and they are formed within $0.01$ and $0.07$ Myr. Given the mass of the seeds, their migration rates are high since the beginning of the simulations. Therefore all planets end their evolution within $1$ au from the centre of the disc, whatever the set of parameters adopted for the discs.

The discs surrounding the system DWD$_4$ have the lowest densities among all the discs we considered (see Section \ref{sec:dwd4system}). Nevertheless DWD$_4$ forms GGs for the disc parameters $r_{\rm c} = 100$ au, $\xi = 0.015, 0.02$. As in the case $M_0 = 0.01$\ \Mearth\, despite the disc surrounding DWD$_3$ system has similar temperatures (see Fig. \ref{fig:disc_T_profiles}) and higher densities than those surrounding DWD$_4$ system (see Section \ref{sec:dwd3system}), the latter produces discs where the irradiated region is more extended. Therefore, when $t_0 = 0.1$ Myr, DWD$_3$ does not form GGs even when $M_0 = 10$\ \Mearth. However, because of the low accretion rate, the GGs produced by DWD$_4$ have small masses: within $62$ and $75$\ \Mearth ($0.19-0.23$ \Mj). They are formed within $0.4$ Myr and their evolution ends within $1$ au. 
\\
Interestingly, when $r_{\rm c} = 150$ au, the DWD$_4$ system could form GGs too, but the planets reach the critical mass after \dMgw $= 0$, so that the planet-forming region is depleted and runaway gas accretion cannot start. This occurs whatever the value of $\xi$, and it is due to the same reason which produced the GGs for $r_{\rm c} = 100$ au. As the growth of the Ns is halted by the absence of gas in the planet-forming region, i.e. by \dMgw $= 0$, they stop migrating before reaching $r_{in}$ as well as GGs could do. Therefore, $\sim 19\%$ of the Ns end their evolution within $1$ and $4$ au. Finally, they form within $0.26$ and $0.28$ Myr and reach masses the range $27-37$\ \Mearth\ ($0.08-0.12$ \Mj). The Ns left are within $1$ au and form within $0.01$ and $0.05$ Myr with masses lower than $18.5$\ \Mearth\ ($0.06$ \Mj). 

A similar configuration of Ns, i.e. Ns reaching the critical mass without triggering runaway gas accretion, occurs also for the DWD$_2$ system. The latter is the second lowest density disc, but has a larger planet-forming area, as well as the discs surrounding the DWD$_4$ system (see Fig. \ref{fig:disc_T_profiles}). Therefore, the planets it forms when $M_0 = 10$\ \Mearth  are similar to those formed by the disc surrounding the DWD$_3$ system. However, due to the extended planet-forming area, the disc surrounding DWD$_3$ system with parameters $r_{\rm c} = 150$ au and $\xi = 0.02$ forms $3$ planets whose growth is halted by the planet-forming region being depleted during the formation process. That is, \dMgw $= 0$ and no more gas can be accreted by the planets. Such planets reach masses between $26.5$ and $28.5$\ \Mearth\ ($0.083-0.089$ \Mj), and they are formed within $0.08$ and $0.1$ Myr. The rest of Ns formed by the discs surrounding DWD$_2$ system are within $11.2-25$\ \Mearth\ ($0.03-0.08$ \Mj) and formed within $0.01$ and $0.11$ Myr. The final locations of all Ns are within $1$ au.

When $t_0 = 1$ Myr, despite the higher seed mass, no discs form GGs. This is due to \dMgw $\approx 0$ since the beginning of the simulations. The Ns formed have masses in the range $10.8-27.6$\ \Mearth\ ($0.03-0.09$ \Mj) and their final locations are within $1$ au. They are formed within $0.01$ Myr and $0.35$ Myr. The disc surrounding the DWD$_3$ system forms the smallest Ns, while the one surrounding DWD$_4$ forms the Ns with the highest masses. In particular, the circumbinary disc of DWD$_4$ with parameters $r_{\rm c} = 150$ au, $\xi = 0.015, 0.02$ forms Ns reaching the critical mass. This occurred also for the circumbinary discs of DWD$_4$ with the same parameters when $t_0 = 0.1$ Myr. However in this case such Ns are all finally located within $1$ au, as well as the Ns which does not reach M$_{\rm crit}$. That is, in this case the larger extension of the planet-forming area of the disc with respect to other systems, i.e. the irradiated area, is not sufficient to halt migration before $1$ au. The critical masses reached are within the range $26-29.9$\ \Mearth\ ($0.08-0.09$ \Mj).

\section{Discussion} 
\label{discussion}
\begin{figure*}[t]
\centering
\subfloat[][]{\includegraphics[trim=1cm 0.cm 1cm 0.1cm, clip,width=0.33\textwidth]{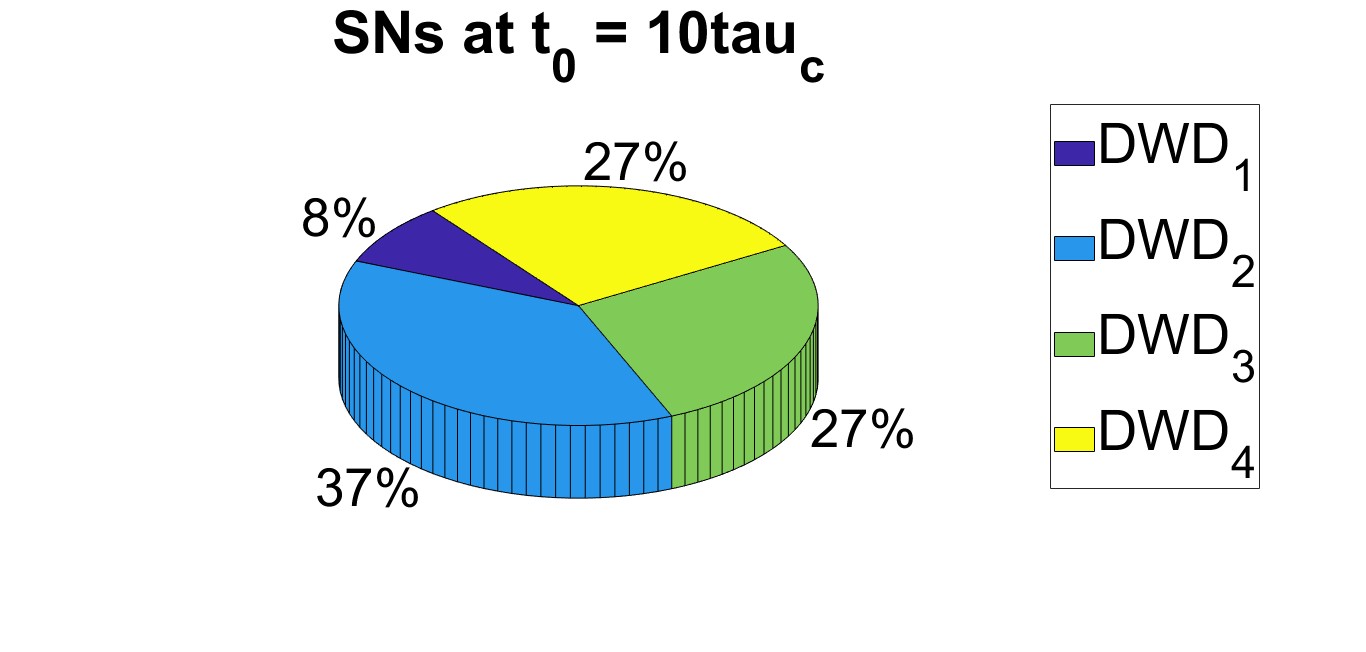}}
\subfloat[][]{\includegraphics[trim=1cm 0.cm 1cm 0.1cm, clip,width=0.33\textwidth]{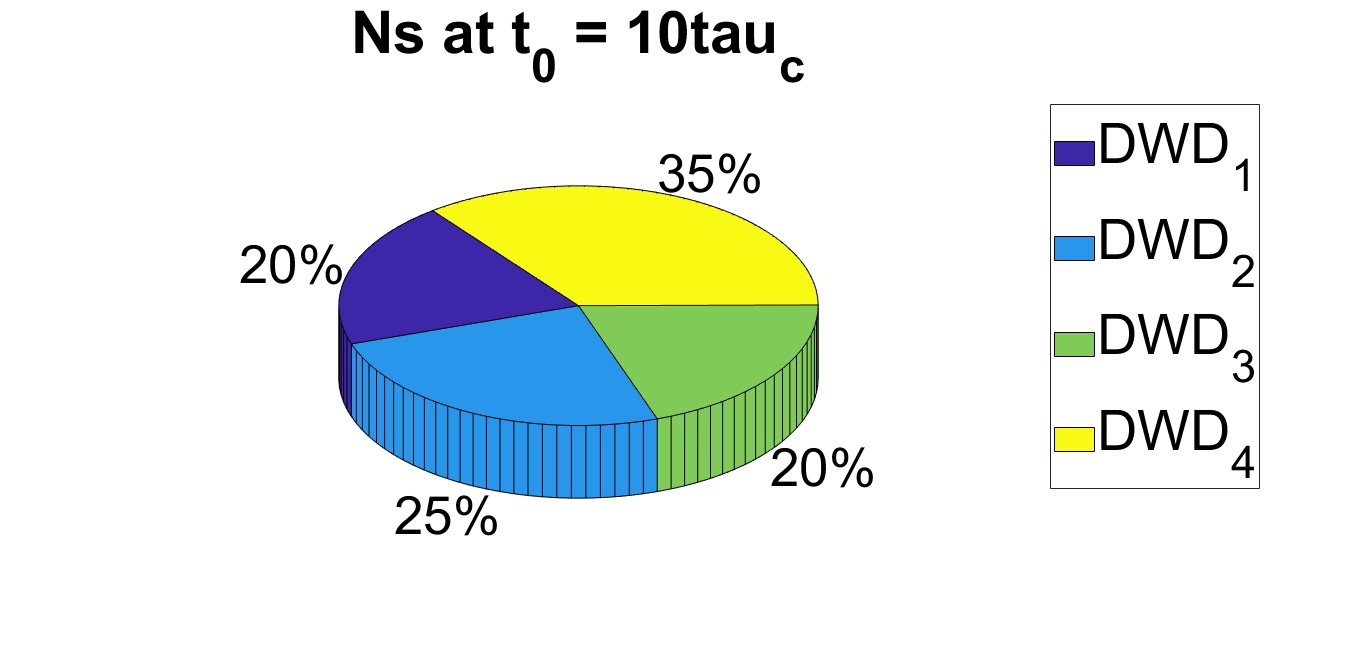}}
\subfloat[][]{\includegraphics[trim=1cm 0.cm 1cm 0.1cm, clip,width=0.33\textwidth]{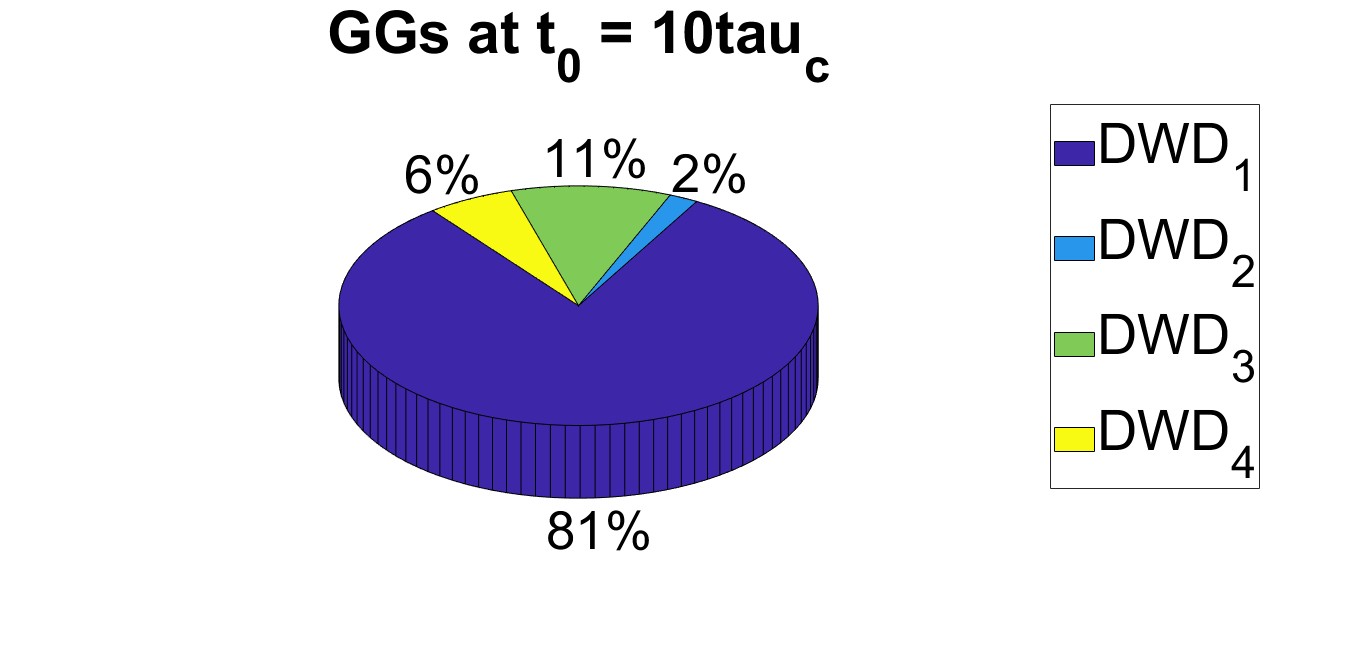}} \\
\subfloat[][]{\includegraphics[trim=1cm 0.cm 1cm 0.1cm, clip,width=0.33\textwidth]{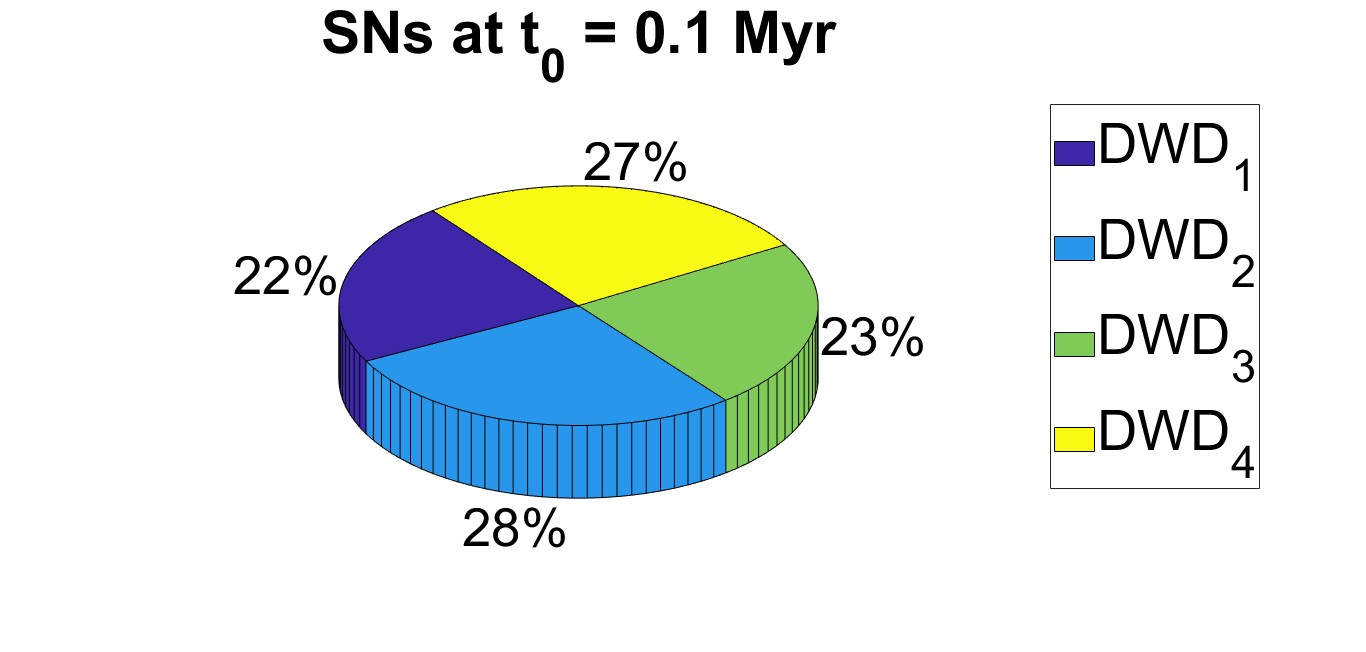}}
\subfloat[][]{\includegraphics[trim=1cm 0.cm 1cm 0.1cm, clip,width=0.33\textwidth]{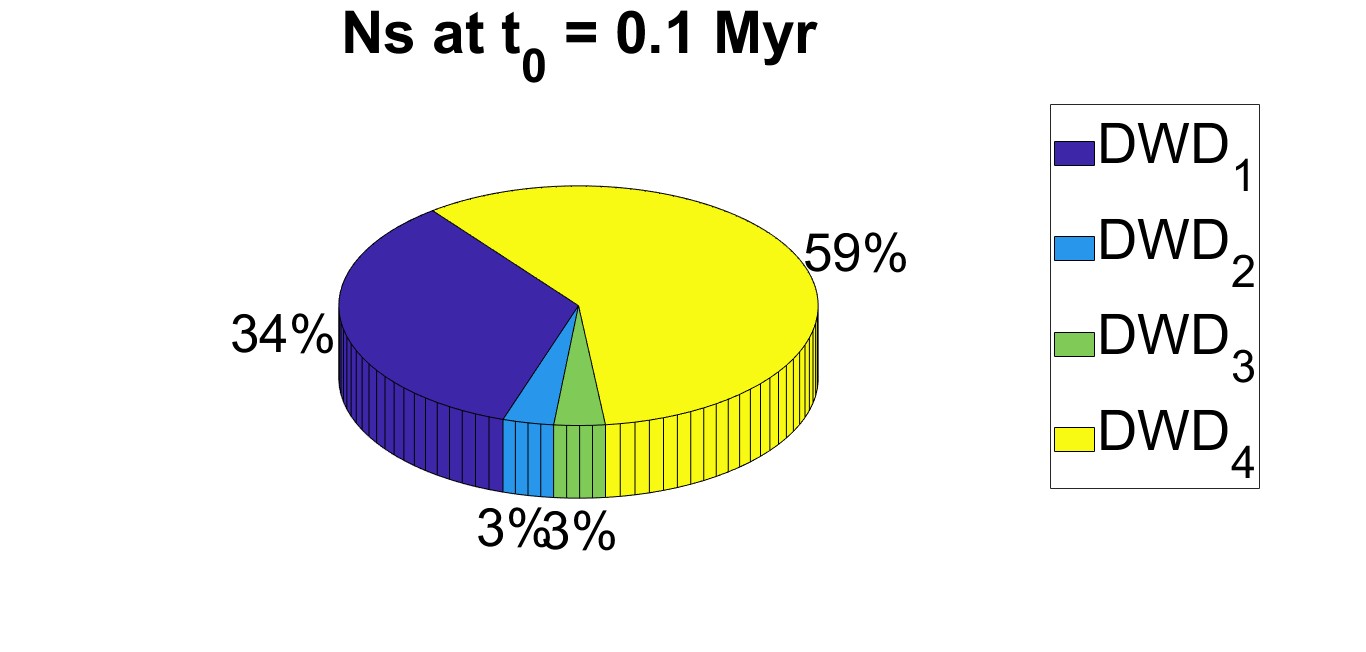}}
\subfloat[][]{\includegraphics[trim=1cm 0.cm 1cm 0.1cm, clip,width=0.33\textwidth]{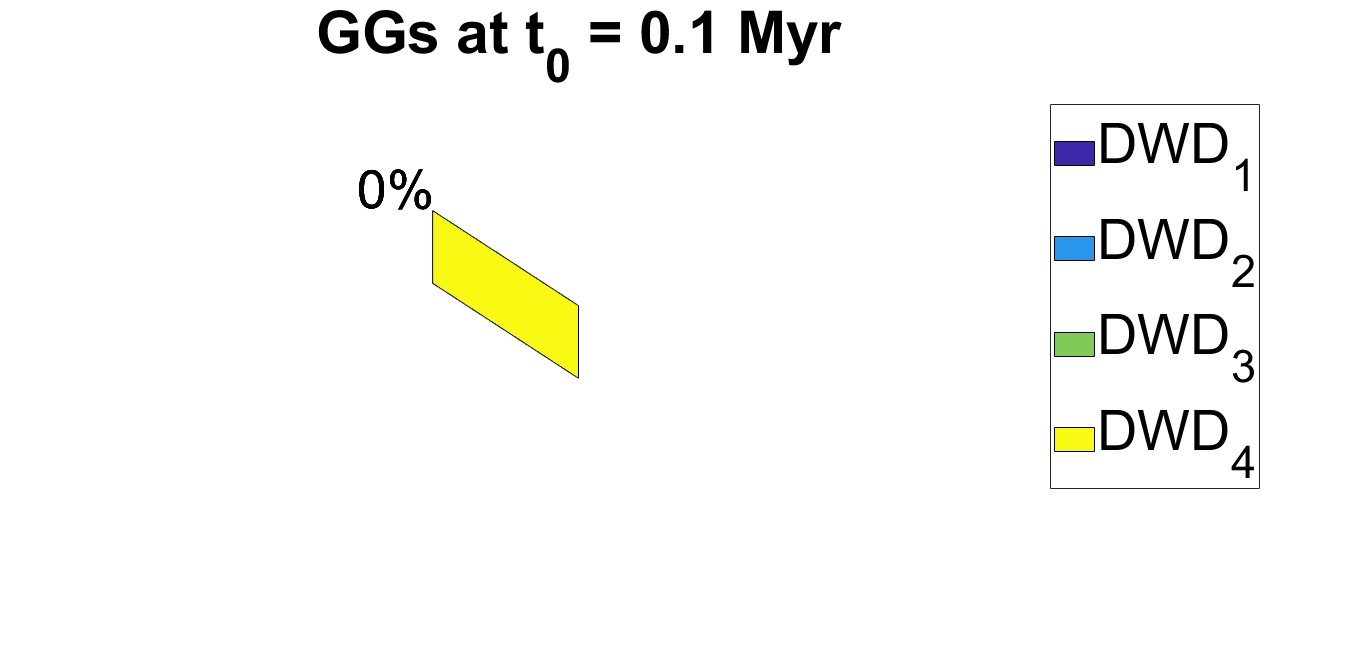}} \\
\subfloat[][]{\includegraphics[trim=1cm 0.cm 1cm 0.1cm, clip,width=0.33\textwidth]{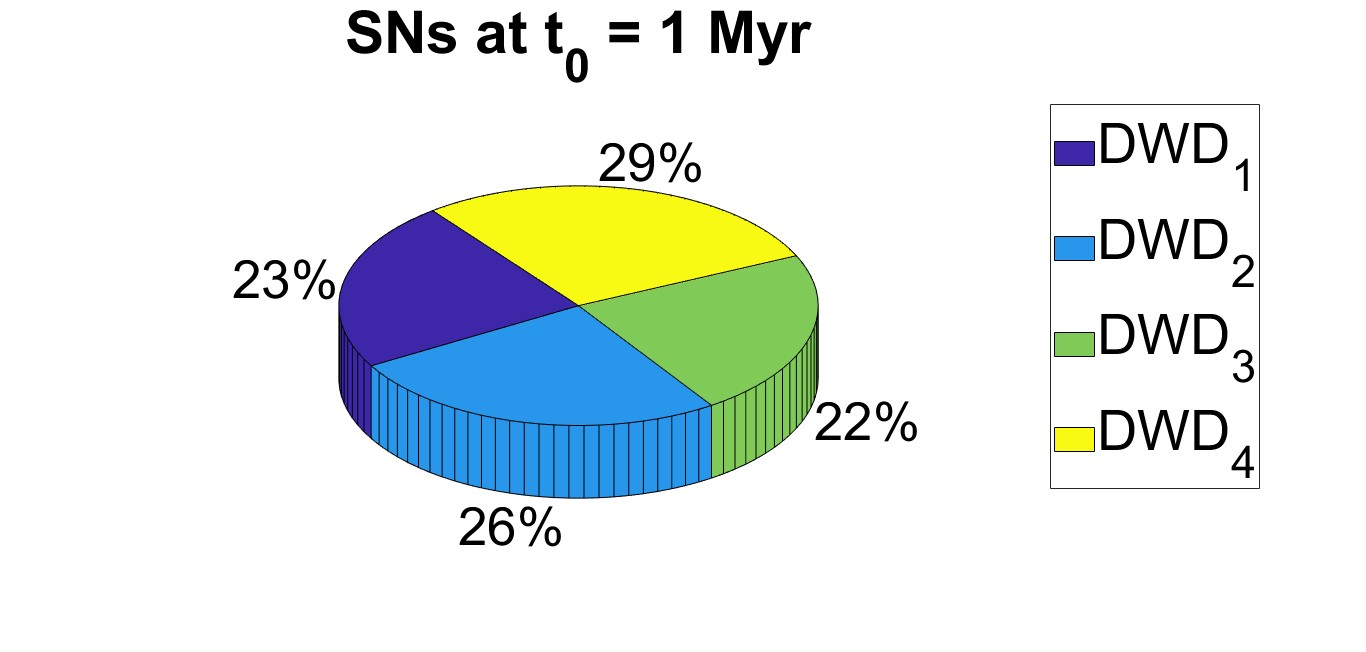}}
\subfloat[][]{\includegraphics[trim=1cm 0.cm 1cm 0.1cm, clip,width=0.33\textwidth]{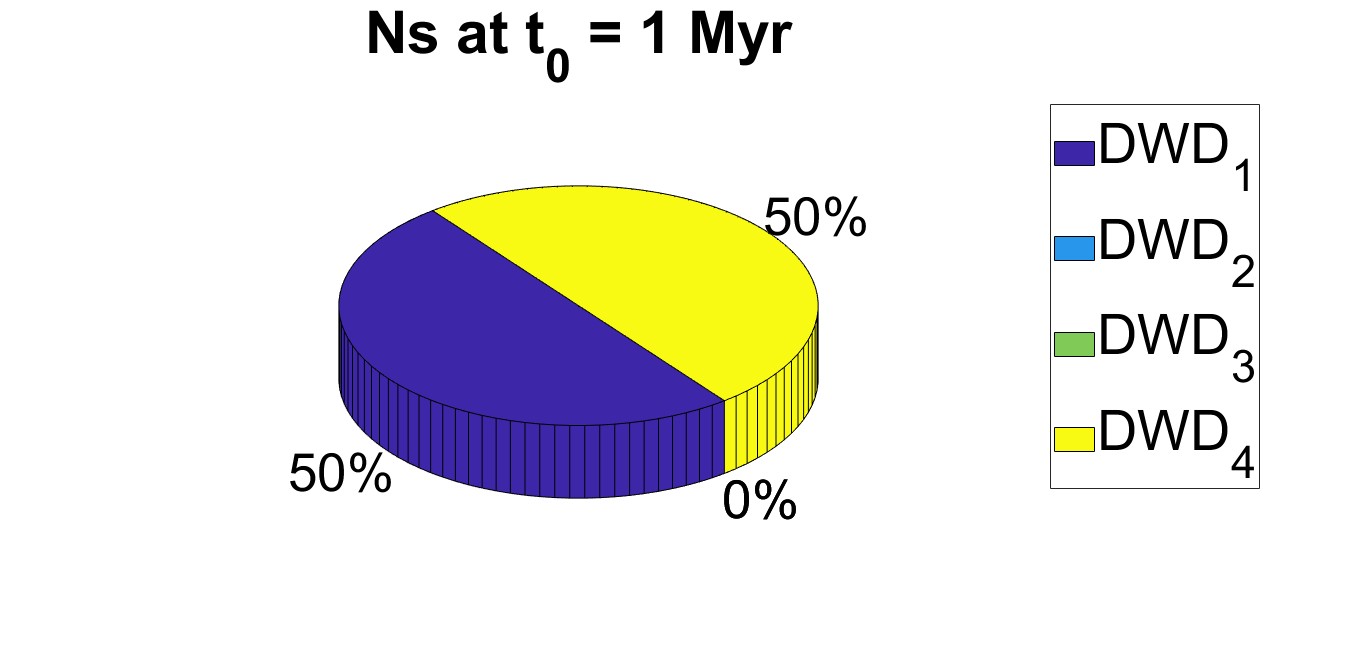}}
\subfloat[][]{\includegraphics[trim=1cm 0.cm 1cm 0.1cm, clip,width=0.33\textwidth]{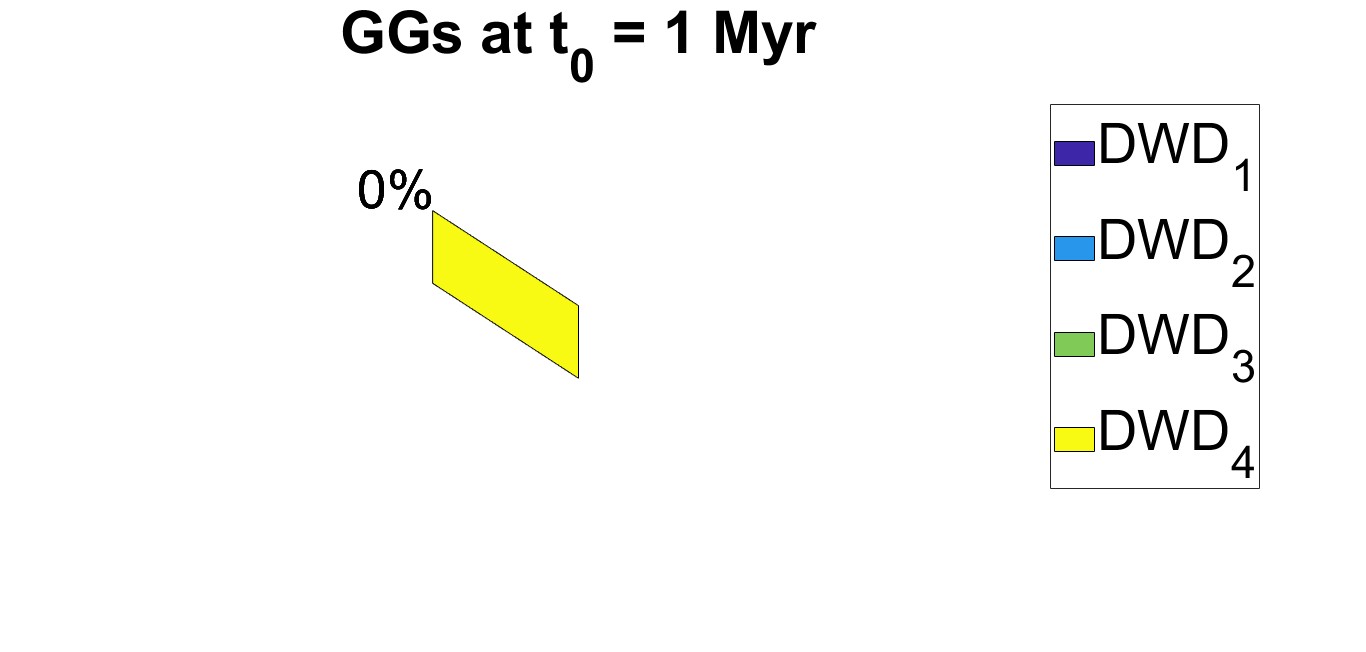}} \\
\caption{Distribution of planets among the DWD systems adopted in this work. From left to right, for each DWD system, the pie charts show the distribution of SNs, Ns and GGs, respectively. From top to bottom, the distribution of planets among the systems are shown at the different values of $t_0$ we adopted for our simulations (i.e. $10\tau_c$, $0.1$ Myr and $1$ Myr, respectively). The bottom-right and the middle-right charts show that no system forms GGs when $t_0 = 0.1, 1$ Myr (see Appendix \ref{sec:appendixA}).}
\label{fig:pie_charts_planets_distrib}
\end{figure*}

In this work we studied the planet formation process in second generation circumbinary discs around DWD systems. We assumed initial seeds of mass $0.01$\ \Mearth\ and studied their evolution for different starting time, $t_0$, of planet formation and for different disc models (Tab. \ref{table:freeparams}). Depending on the value of $t_0$, we assumed the seeds to be either of first generation origin ($t_0 = 10\tau_c$) or to form directly in the second generation disc ($t_0 = 0.1, 1$ Myr). A representation of the percentages of planets formed by our disc models is shown in Fig. \ref{fig:pie_charts_planets_distrib}. We also tested our discs with seeds of mass $10$\ \Mearth\ when $t_0 = 0.1, 1$ Myr, assuming they are first generation bodies that never underwent the runaway gas accretion process.

The gas accretion rate $\dot{M}_g$ and the photoevaporation rate $\dot{M}_w$ have a crucial role in determining the final planetary masses. If the planetary formation starts too late after the formation of the last WD of the system, the initial value $\dot{M}_g(t_0)$ might be very close to $\dot{M}_w$. This hinders the pebble accretion phase (see Eq. \ref{eqn:time_pebble}), and could inhibit the formation of GGs because of the reduced mass supplied to their planet-forming regions, i.e. \dMgw $\approx 0$ (see Eq. \ref{eqn:time_runawaygas}). In particular, an increase in $\dot{M}_w$ by an order of magnitude reduced the masses of the GGs of the case $t_0 = 10\tau_c$, $r_{\rm c} = 50$ au, $\xi = 0.02$ from $10000$\ \Mearth\ ($31.45$ \Mj) to $\sim 4200$\ \Mearth\ ($13.21$ \Mj) (see Section \ref{sec:formation_efficiency}). Such disc forms the GGs with the largest masses when $\dot{M}_w = 10^{-8}$ \Msun/yr. This particular scenario suggests that the presence of GGs of small mass or Ns around a DWD can be useful to constrain the photoevaporation rate of the disc that produced them.

The different accretion rates from system to system are strictly related to the temperature profile and to the density of the disc. Shorter and denser discs allow for a faster pebble accretion, but the initial seeds would be closer to the disc inner edge. Therefore they can only reach the lowest isolation masses possible (e.g., left panels of Fig. \ref{fig:example_dwd1}), and consequently their gas contraction phase is slow (see Eq. \ref{eqn:KHtimescale}). Moreover, during the fast pebble accretion phase the Type I migration rate quickly increases as well, leading most of the planets to the inner radius of the disc, $r_{in}$. Consequently, planetary formation ends mostly because the planets do not have more space to grow, rather than because they reach their critical masses. In this cases, most of the planets formed are SNs, and a smaller fraction are Ns. In particular, the coldest discs, i.e. when $t_0 = 1$ Myr, can only form SNs. 

The only deviation from this scenario occurs in presence of first-generation seeds of mass $10$\ \Mearth, which we used to test the planet-forming potential of the discs when the accretion rate is the lowest (i.e. $t_0 = 0.1, 1$ Myr, e.g. Fig. \ref{fig:accr_rates_dwd1}). In these cases, the initial mass is already higher than or equal to the isolation mass. Therefore the planets have more space to grow before reaching $r_{in}$, eventually reaching the critical mass, M$_{\rm crit}$. However, the accretion rate when $t_0 = 0.1, 1$ Myr is low and the planet-forming region is already depleted of gas when the planets reach M$_{\rm crit}$. Therefore the planets stop migrating toward the disc inner edge and this halts their growth. This is also one of the two mechanisms which allow having planets located farther than $1$ au at the end of their evolution. The other mechanisms is due to the planets reaching their maximum mass at the end of runaway gas accretion \citep{tanaka}. Once the runaway gas accretion starts, lower values of \dMgw lead to faster depletion of the planet-forming region, again halting the migration, and lower final masses of the GGs. The GGs with the largest masses forms when \dMgw is the largest and the disc extension are the shortest, but they reach their final location, within $1$ au, in $\sim 2$ Myr or more. If the disc extension is the largest, the GGs have smaller masses but it is more likely for them to stop migrating before $1$ au.

Another important aspect is related to the specific formation channel we adopted. In our study, we only considered the region of the disc heated by the radiation of the central system, i.e. the region of the disc where Eq. \ref{eqn:temperature} holds. This choice excluded the outer part of our disc models, but depending on the binary configuration after the last CE, such external, excluded part has a different extension. In particular, planets can have more space to grow in discs with a larger irradiated region. Therefore, seeds located farther have higher chances to trigger the gas accretion phase with respect to discs with shorter irradiated regions. Thanks to this possibility, if two discs have similar temperature profiles, the one with lower density, such as DWD$_4$, could form planets of higher masses than discs with higher density, such as DWD$_3$ (when, e.g., $r_{\rm c} = 150$ au and $t_0 = 0.1$ Myr, the planet-forming area of DWD$_3$ has an extension of $\sim 50$ au, that of DWD$_4$ has an extension of $\sim 100$ au). However, this occurs when the discs are the coldest (i.e. $t_0 = 0.1, 1$ Myr), independently on whether $M_0 = 0.01$\ \Mearth\ and $M_0 = 10$\ \Mearth.  
When temperatures are the highest possible (i.e. $t_0 = 10\tau_c$), DWD$_3$ can compensate for the shorter planet-forming area and forms planets with larger masses than the disc of DWD$_4$. As shown in Fig. \ref{fig:disc_T_profiles}, another system with large planet-forming area is DWD$_2$. However, the disc temperatures are the lowest among all the discs, independently on the starting time of planet formation. Therefore such discs form mostly SNs and low mass Ns.

It is important to note that when the temperatures of the disc are higher than the ones considered here, the isolation mass profile of the disc shift toward higher masses. This allows for faster gas contraction phases and faster GGs formation once M$_{\rm iso}$ is reached. On the other hand, given that $M_{\rm iso}/\dot{M}_g(t_0) \propto T^2$, no seed could reach the isolation mass if the temperatures of the disc are too high. The only possibility to form GGs (or even Ns) in this case would be increasing the extension of the disc, which would provide the seeds with more space to grow, or increase $\xi$, which would accelerate the pebble accretion phase. 

A last general finding is that high values of $\xi$ reduce the formation time of planetary cores. The most favourable environments for the formation of GGs are discs around stars with metallicity higher than solar metallicity that incorporate first-generation planetary material. Moreover, such GGs could be located farther from the inner radius of the disc, as an high disc metallicity would fasten the transition between the faster Type I migration regime and the slower Type II migration regime. However, the farthest GGs can be only found in discs with an extension larger than $r_{\rm c} = 100$ au, as in more compact discs the seeds cannot reach isolation masses large enough to start runaway gas accretion quickly (see Eq. \ref{eqn:time_gascontraction}).

The systems and disc models we studied confirm that it could be possible to form second-generation GGs within $\sim 2$ Myr. However, the formation of these planets requires the presence of first-generation bodies of mass of at least $0.01$\ \Mearth. In particular, if the discs are hot and dense, GGs can form in discs with extensions between $50$ and $150$ au (e.g., in our systems DWD$_1$ and DWD$_3$). Otherwise, the planet-forming area of the disc should be as large as possible in order to compensate for low density and/or low temperatures (e.g., in our systems DWD$_2$ and DWD$_4$). However, if the disc is too large, i.e., $r_{\rm c} = 150$ au, its density can be too low and can suppress the advantage for the farthest seeds to have more space to grow.

The disc models considered in this study demonstrated that it is easier to form second generation Ns and SNs than second generation GGs, i.e., our second generation circumbinary discs can form planets either after $0.1$ or $1$ Myr from  their formation. The low temperatures of the discs, especially in combination with high metallicities, allow every seed to reach the isolation mass and partially follow the gas contraction phase. However, it is unlikely these planets stop migrating before reaching the inner radius of the disc. It is more likely for this to occur for seeds with mass $M_0 = 10$\ \Mearth\. In this case, the sooner planetary formation starts, the higher is the possibility to form GGs. However, due to the low accretion rates when $t_0 = 0.1, 1$ Myr, runaway gas accretion stops soon after its onset.

\begin{sidewaysfigure*}[t]
\centering
\subfloat[][]{\includegraphics[trim=1cm 0.cm 1cm 0.1cm, clip,width=0.24\textwidth]{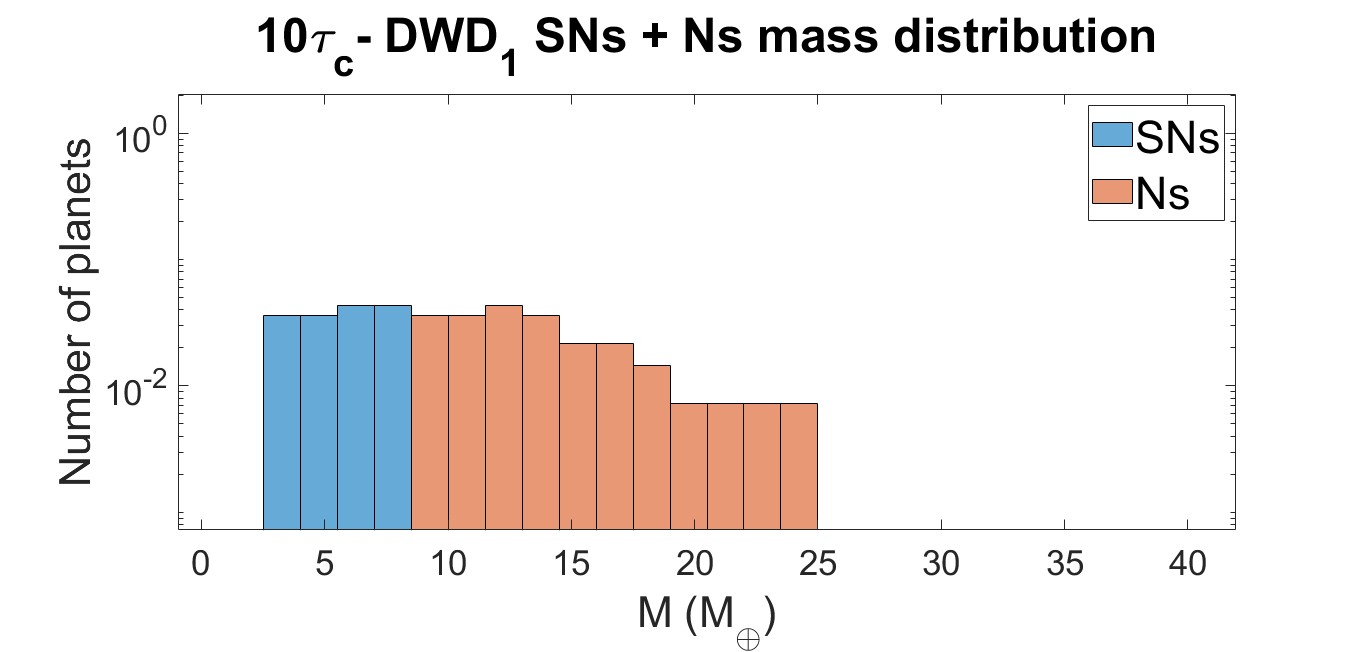}}
\subfloat[][]{\includegraphics[trim=1cm 0.cm 1cm 0.1cm, clip,width=0.24\textwidth]{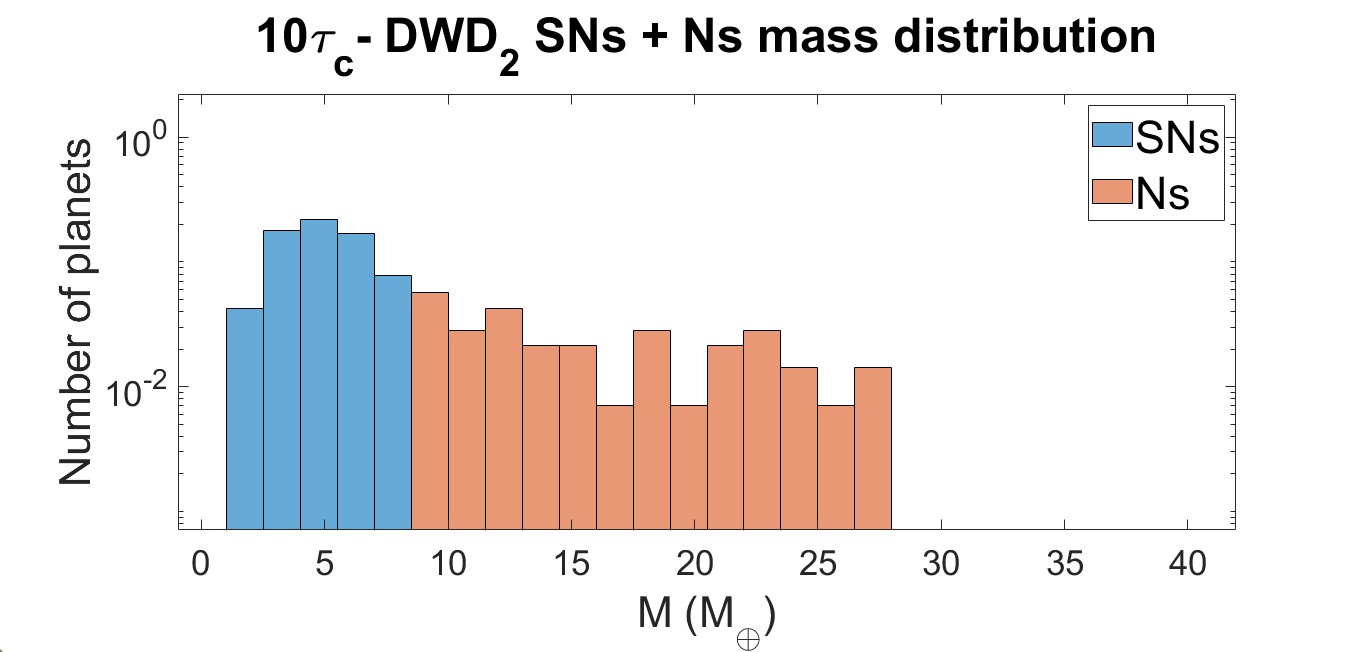}}
\subfloat[][]{\includegraphics[trim=1cm 0.cm 1cm 0.1cm, clip,width=0.24\textwidth]{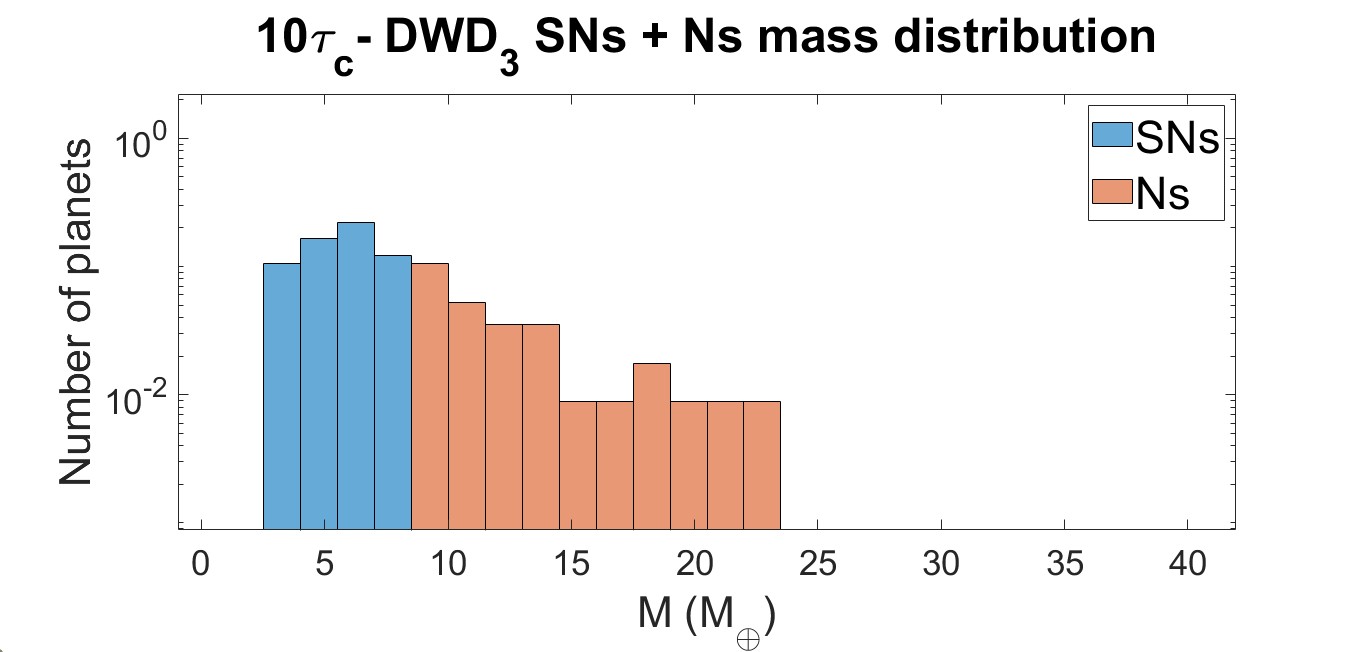}} 
\subfloat[][]{\includegraphics[trim=1cm 0.cm 1cm 0.1cm, clip,width=0.24\textwidth]{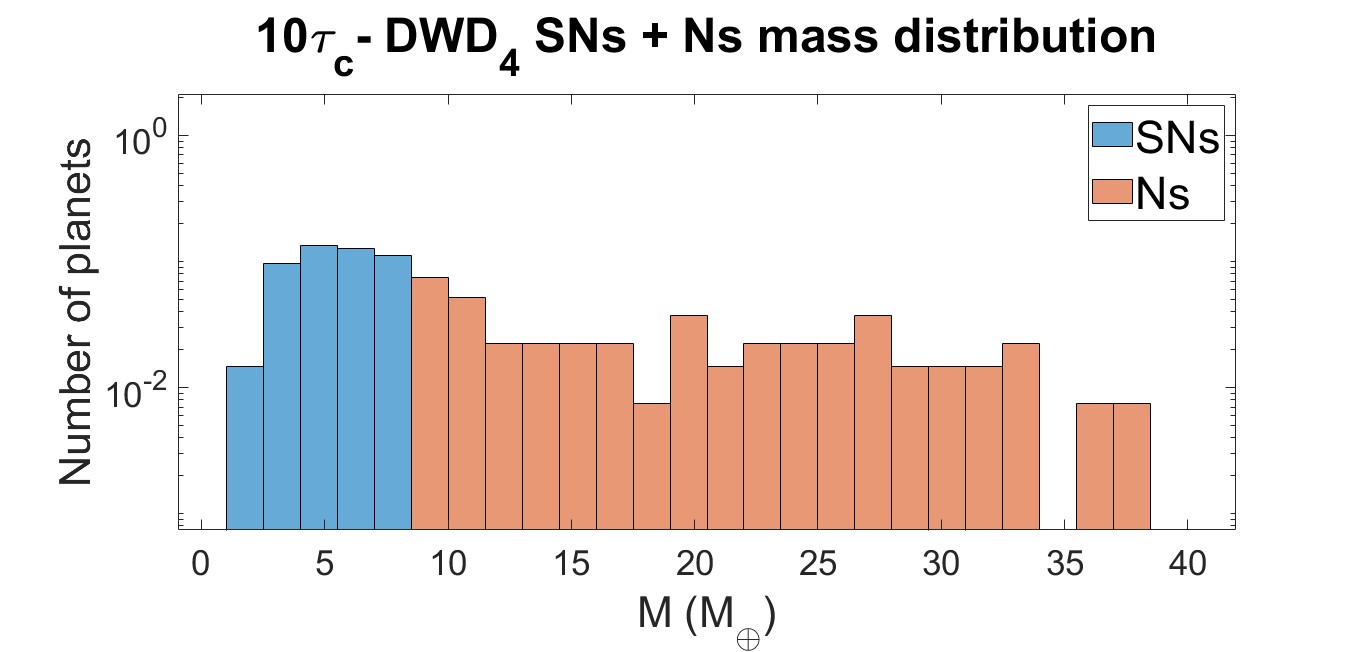}} \\
\subfloat[][]{\includegraphics[trim=1cm 0.cm 1cm 0.1cm, clip,width=0.24\textwidth]{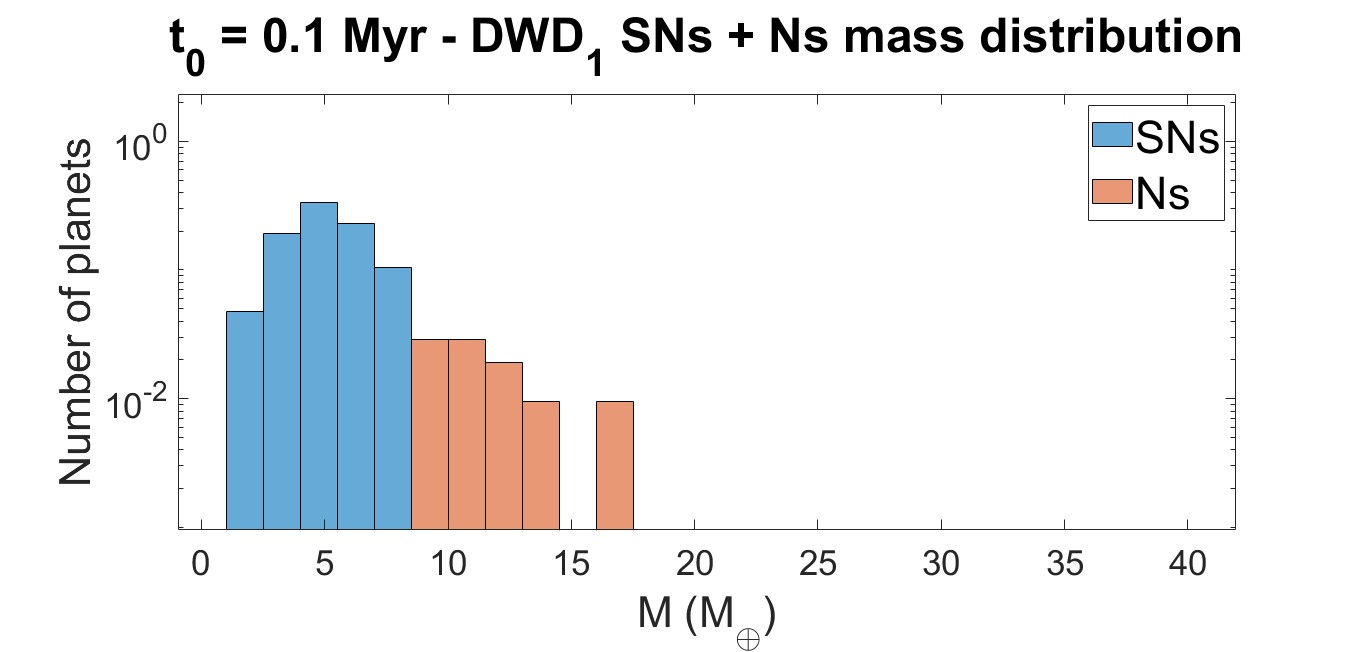}}
\subfloat[][]{\includegraphics[trim=1cm 0.cm 1cm 0.1cm, clip,width=0.24\textwidth]{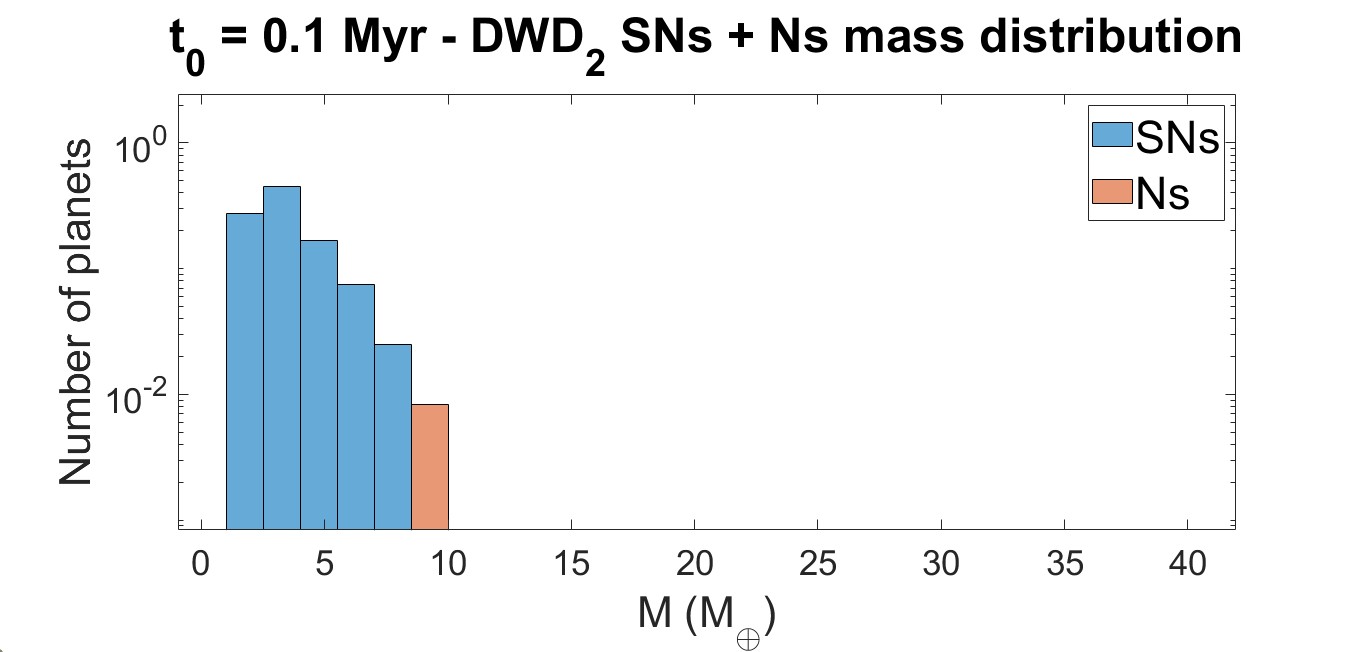}}
\subfloat[][]{\includegraphics[trim=1cm 0.cm 1cm 0.1cm, clip,width=0.24\textwidth]{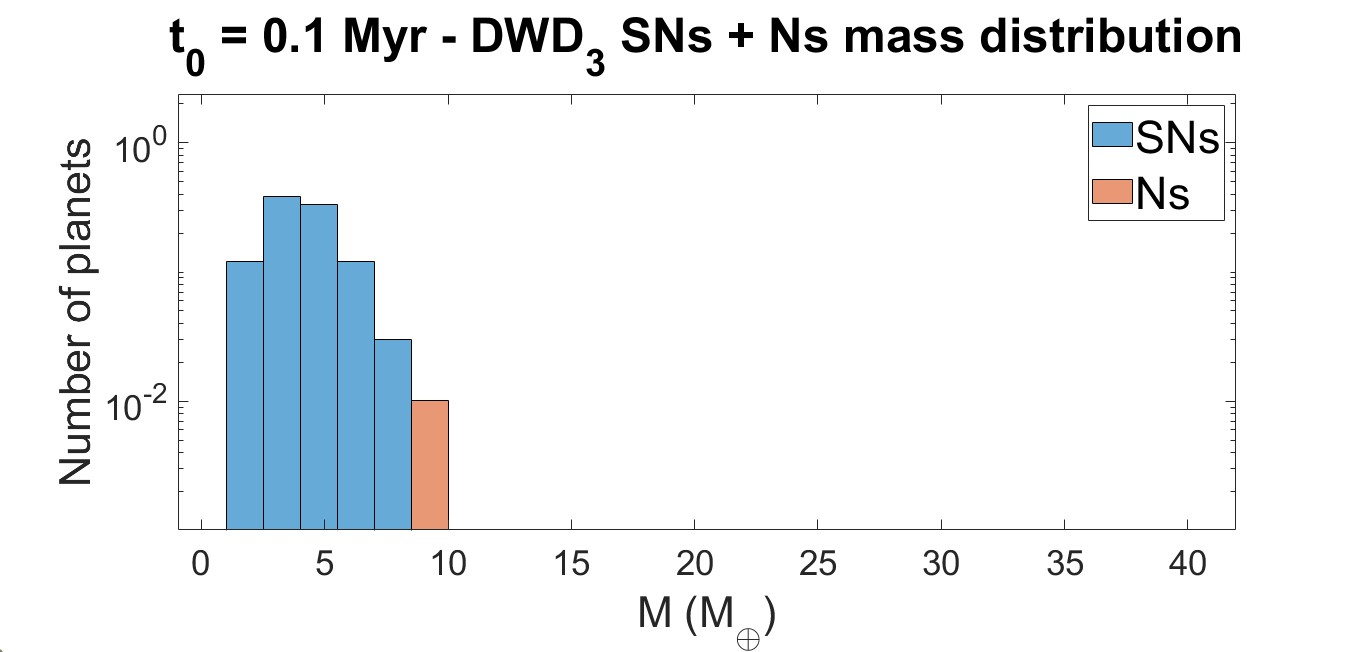}} 
\subfloat[][]{\includegraphics[trim=1cm 0.cm 1cm 0.1cm, clip,width=0.24\textwidth]{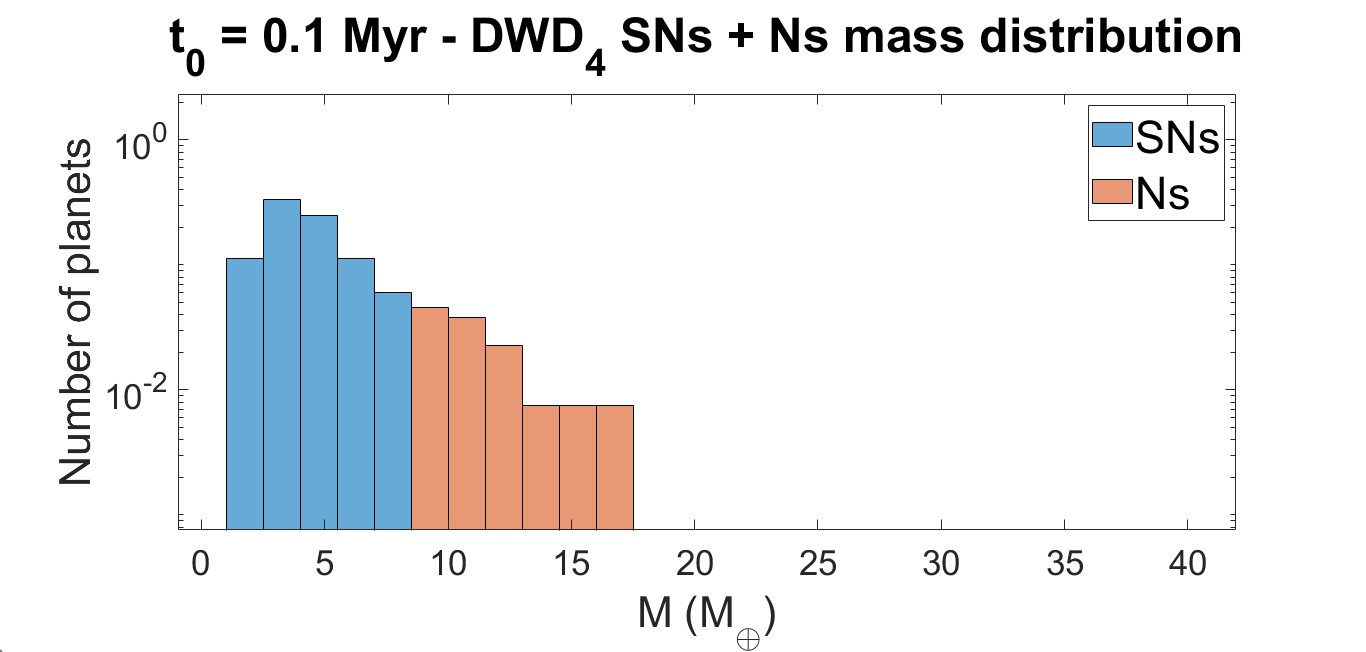}} \\
\subfloat[][]{\includegraphics[trim=1cm 0.cm 1cm 0.1cm, clip,width=0.24\textwidth]{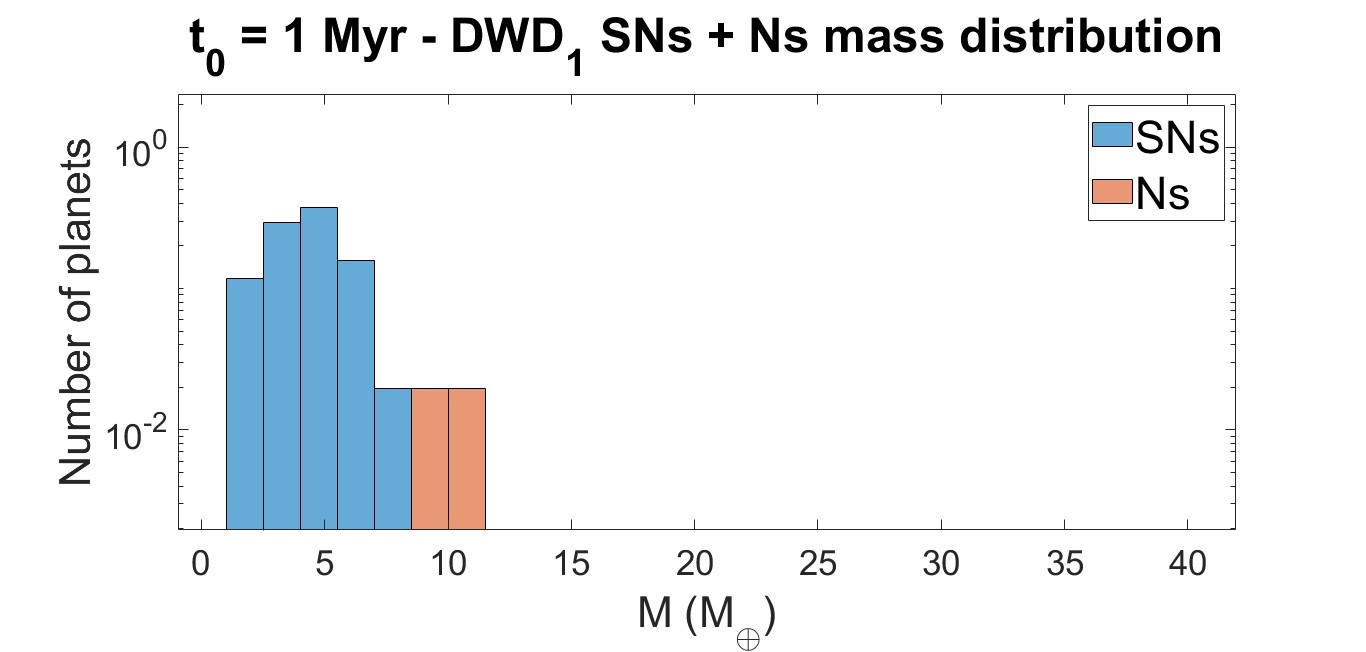}}
\subfloat[][]{\includegraphics[trim=1cm 0.cm 1cm 0.1cm, clip,width=0.24\textwidth]{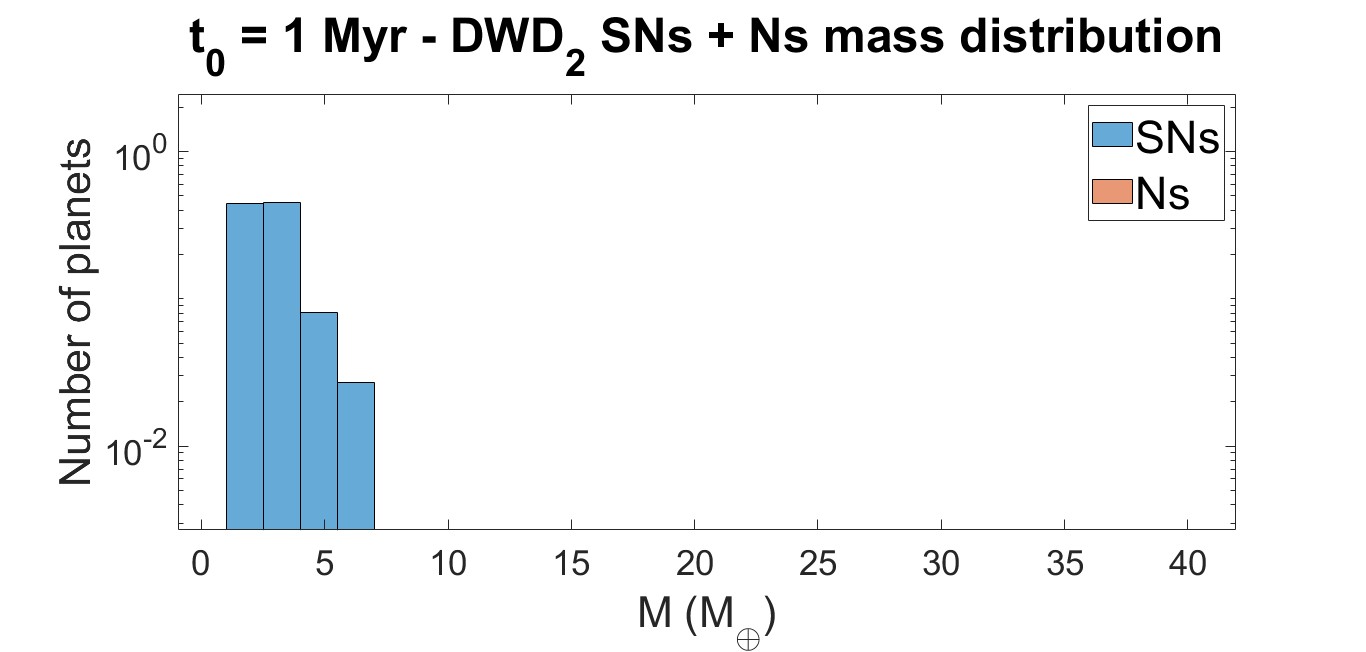}}
\subfloat[][]{\includegraphics[trim=1cm 0.cm 1cm 0.1cm, clip,width=0.24\textwidth]{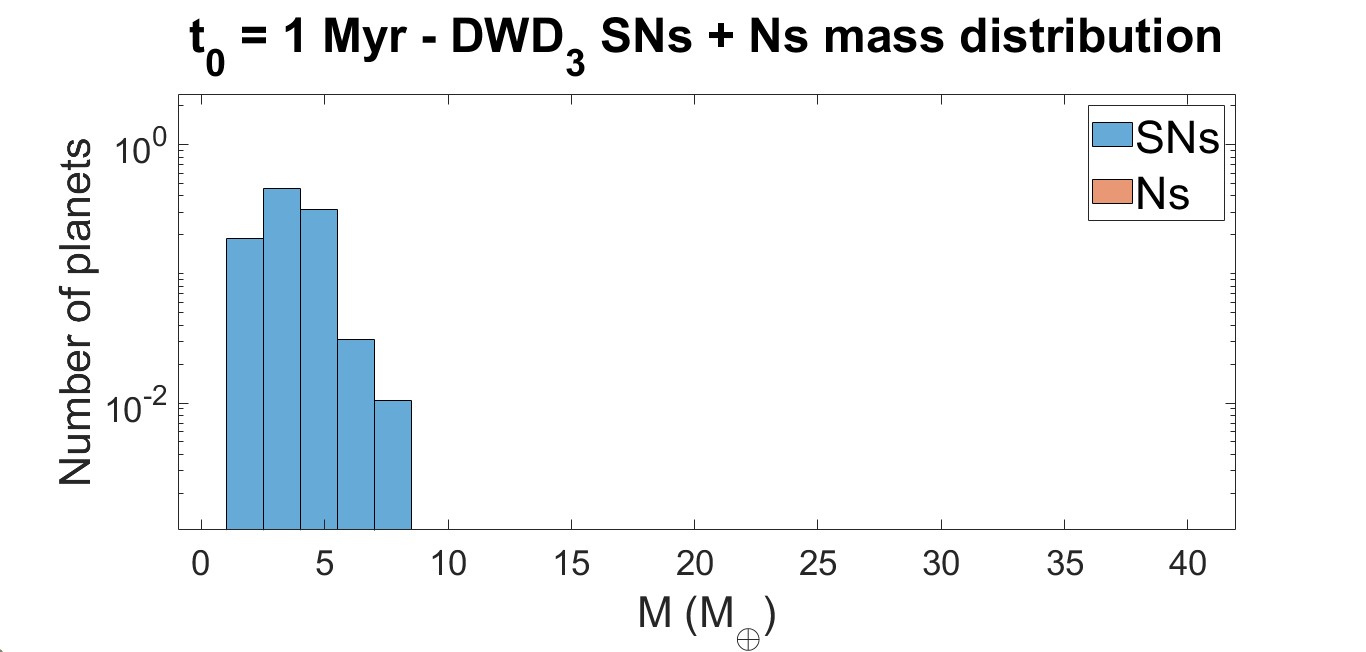}} 
\subfloat[][]{\includegraphics[trim=1cm 0.cm 1cm 0.1cm, clip,width=0.24\textwidth]{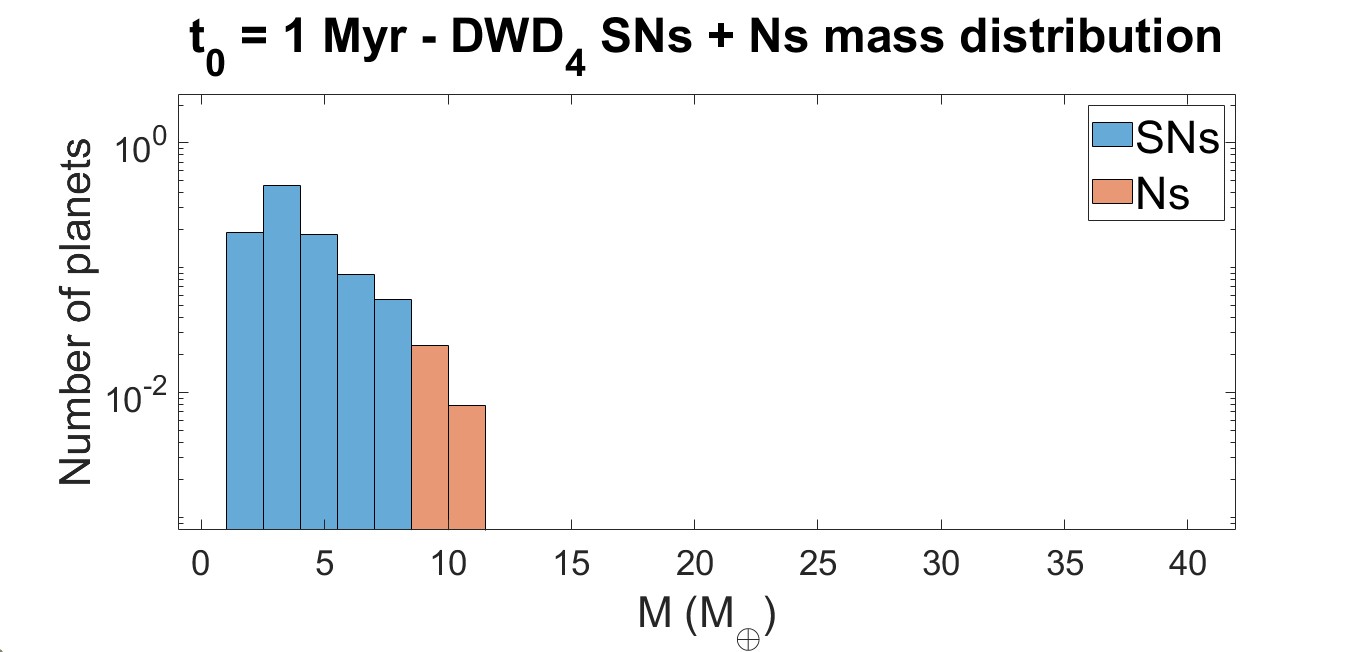}} \\
\subfloat[][]{\includegraphics[trim=1cm 0.cm 1cm 0.1cm, clip,width=0.24\textwidth]{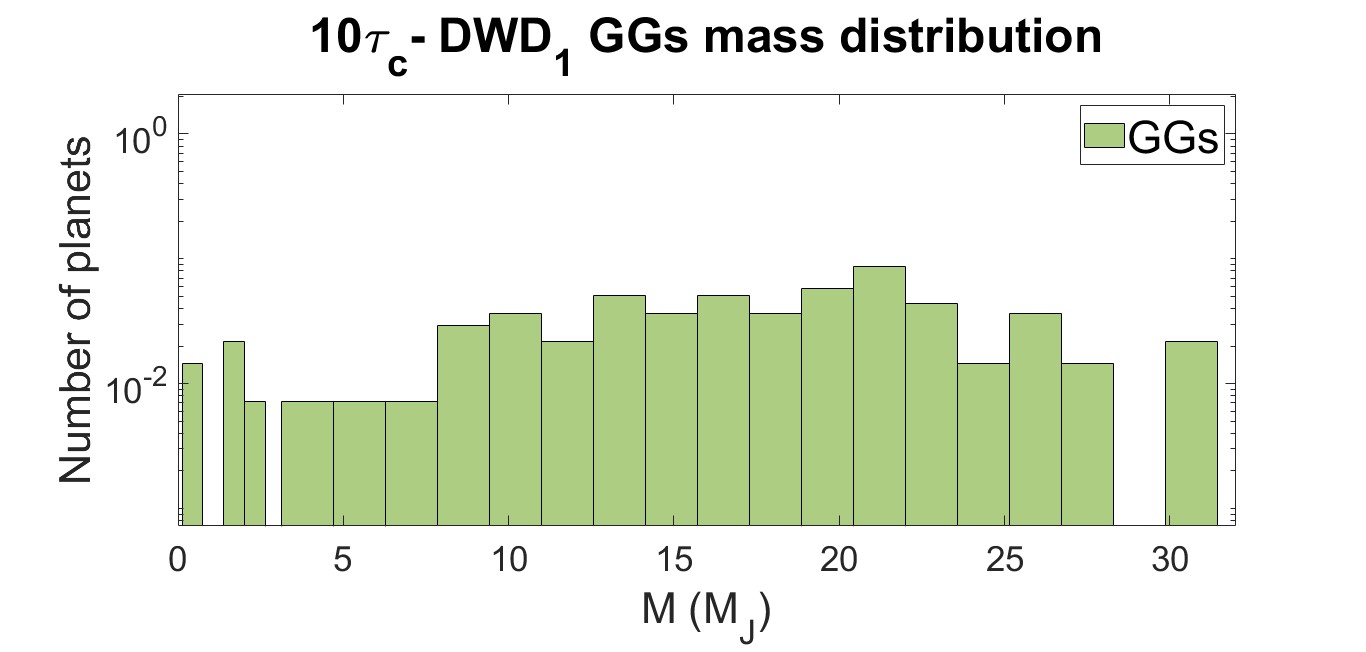}}
\subfloat[][]{\includegraphics[trim=1cm 0.cm 1cm 0.1cm, clip,width=0.24\textwidth]{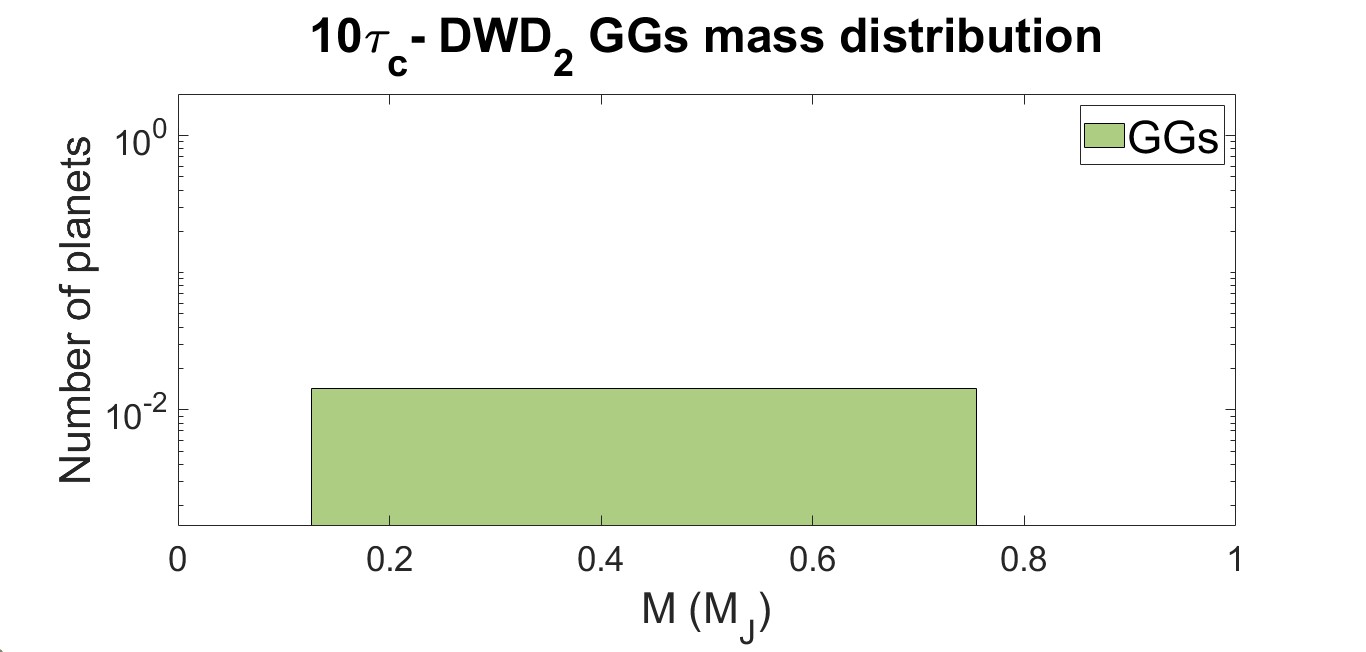}}
\subfloat[][]{\includegraphics[trim=1cm 0.cm 1cm 0.1cm, clip,width=0.24\textwidth]{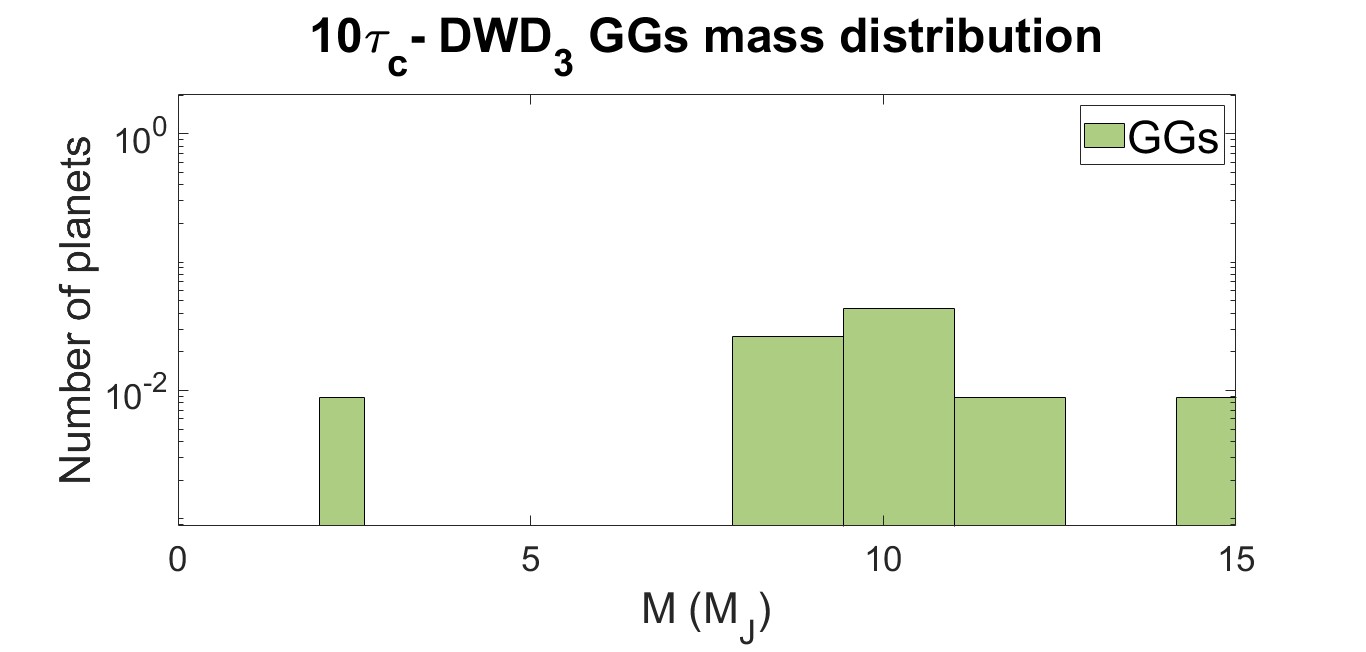}} 
\subfloat[][]{\includegraphics[trim=1cm 0.cm 1cm 0.1cm, clip,width=0.24\textwidth]{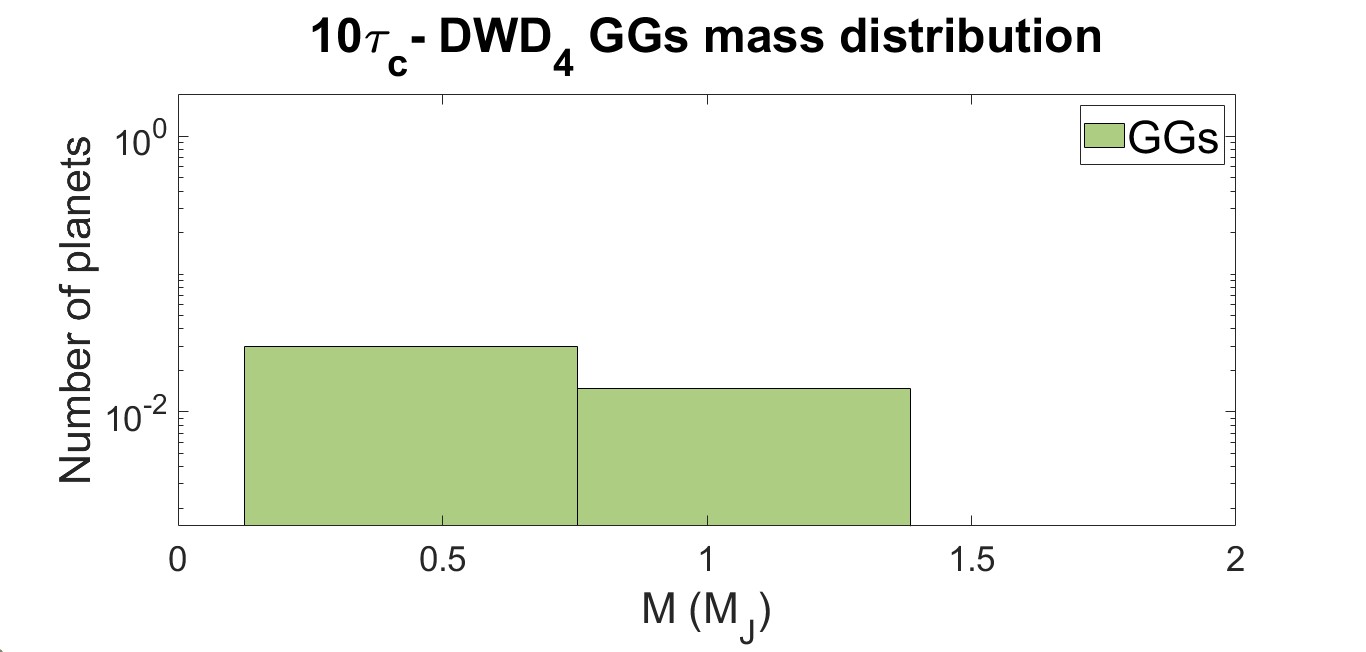}} \\
\caption{Mass distribution of the planets formed by the DWD systems analysed in this work. All histograms are normalised with respect to the total number of planets formed in each case, and the Y axis are plot in log scale. From left to right, the histograms show the mass distribution for each DWD system. From top to bottom, the first three rows of histograms show the distribution as a function of the values of $t_0$ we adopted for our simulations (i.e. $10\tau_c$, $0.1$ Myr, $1$ Myr, respectively). These histograms show only the mass distribution of SNs (blue bins) and Ns (red bins). The histograms in the bottom row show only the distributions of GGs for each system at $t_0 = 10\tau_c$, as GGs only form in this case.}
\label{fig:global_massdistrib_histograms}
\end{sidewaysfigure*}

\begin{figure*}[t]
\centering
\subfloat[][]{\includegraphics[trim=0.1cm 0.cm 1cm 0.1cm, clip,width=0.52\textwidth]{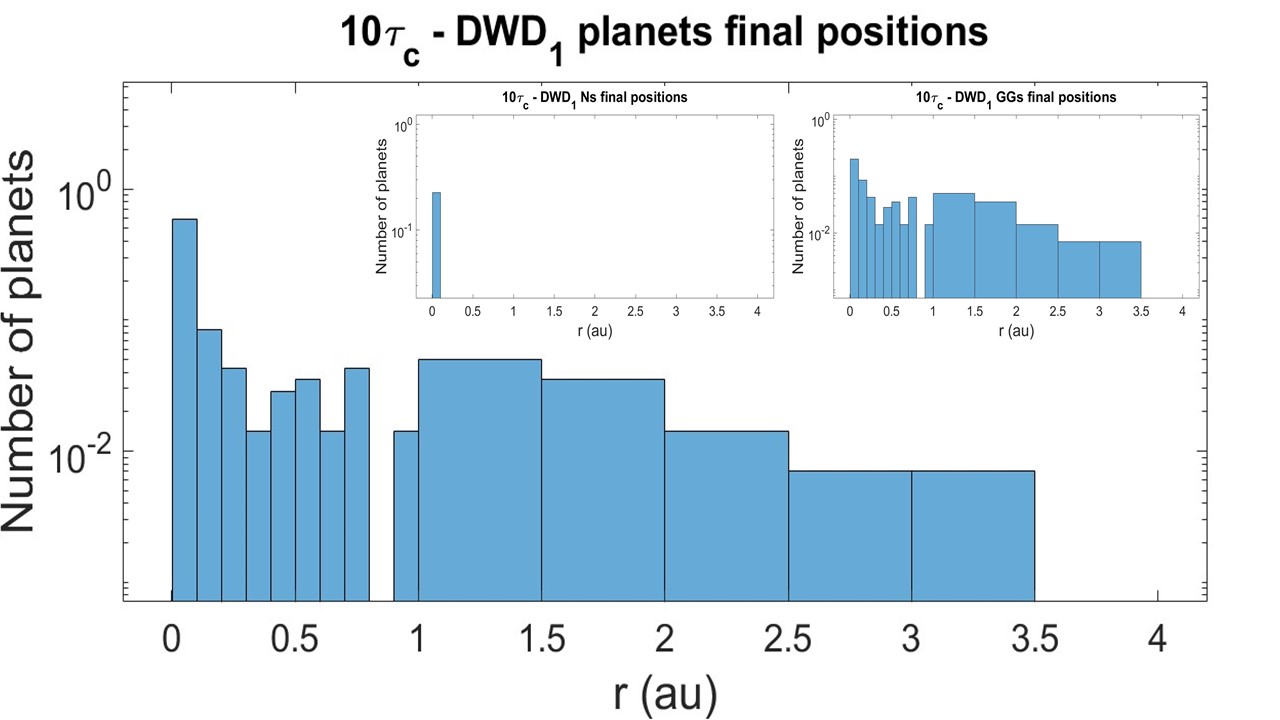}}
\subfloat[][]{\includegraphics[trim=0.1cm 0.cm 1cm 0.1cm, clip,width=0.52\textwidth]{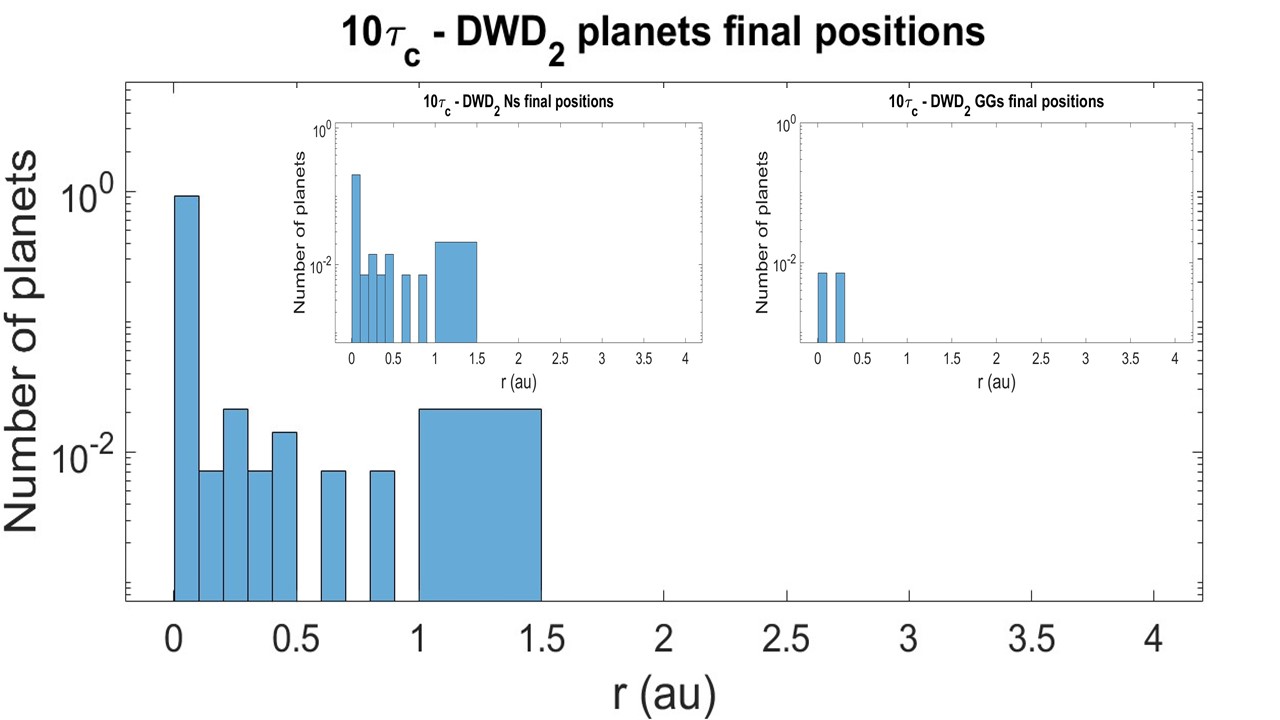}} \\
\subfloat[][]{\includegraphics[trim=0.1cm 0.cm 1cm 0.1cm, clip,width=0.52\textwidth]{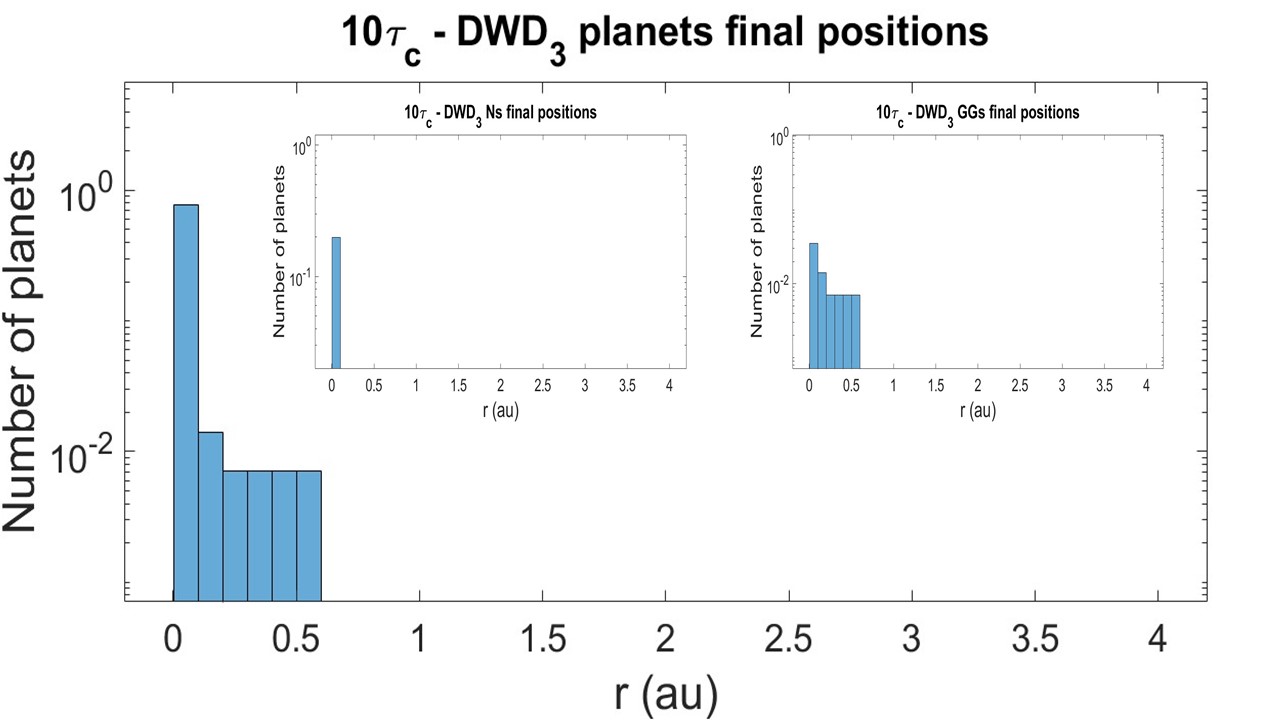}} 
\subfloat[][]{\includegraphics[trim=0.1cm 0cm 1cm 0.1cm, clip,width=0.52\textwidth]{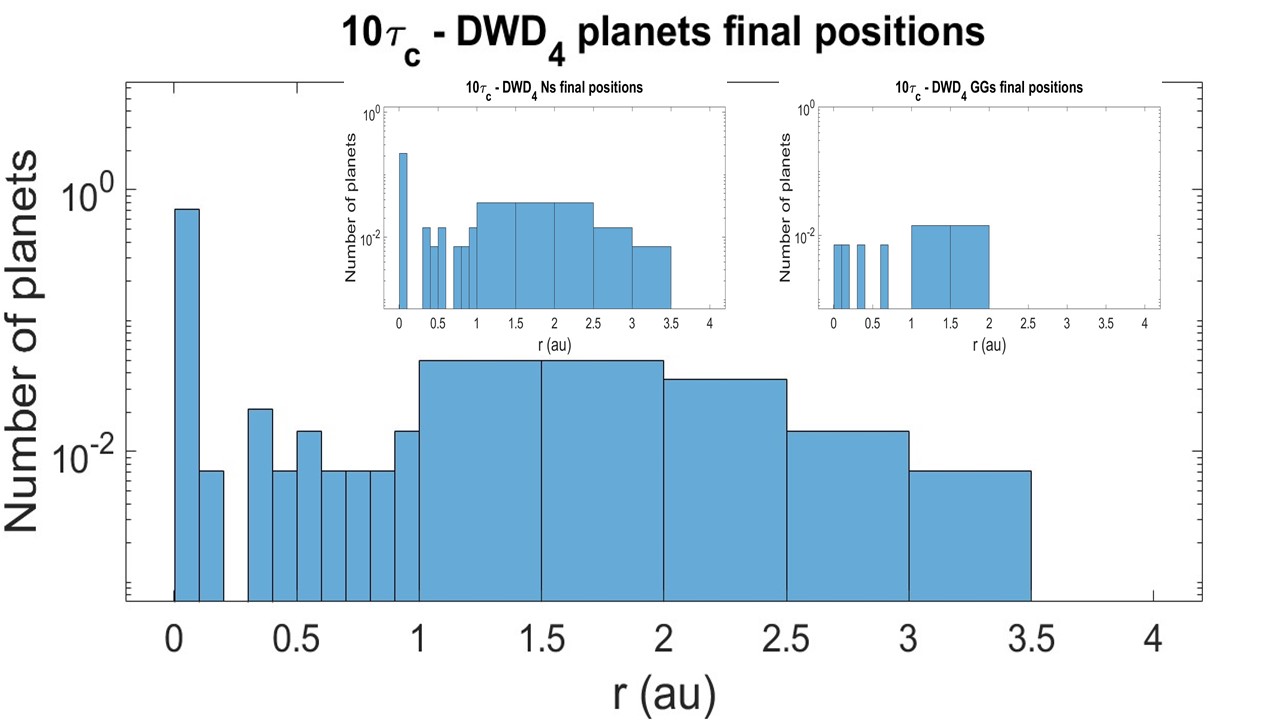}} 
\caption{Distributions of the final positions of the planets formed at $t_0 = 10\tau_c$ by the DWD systems adopted in this work. All the histograms have Y axis in log scale and each bin is normalised with respect to the total number of planets formed by each system. The largest histograms show the distribution of the final positions of all the planets for each DWD system. Each of such histograms have two smaller, overlapped histograms showing the final positions of Ns and GGs.
When $t_0 = 0.1, 1$ Myr, all planets (which consists only of SNs and Ns) migrate toward the inner radius of the disc, $r_{\rm in}$ (see Appendix \ref{sec:appendixA}). Therefore, the related histograms would only show one bin around the $r_{\rm in}$ value of the discs.}
\label{fig:positions_overlap}
\end{figure*}

\begin{figure*}[t]
\centering
\includegraphics[trim=1cm 0cm 1cm 0.2cm,clip,width=\hsize]{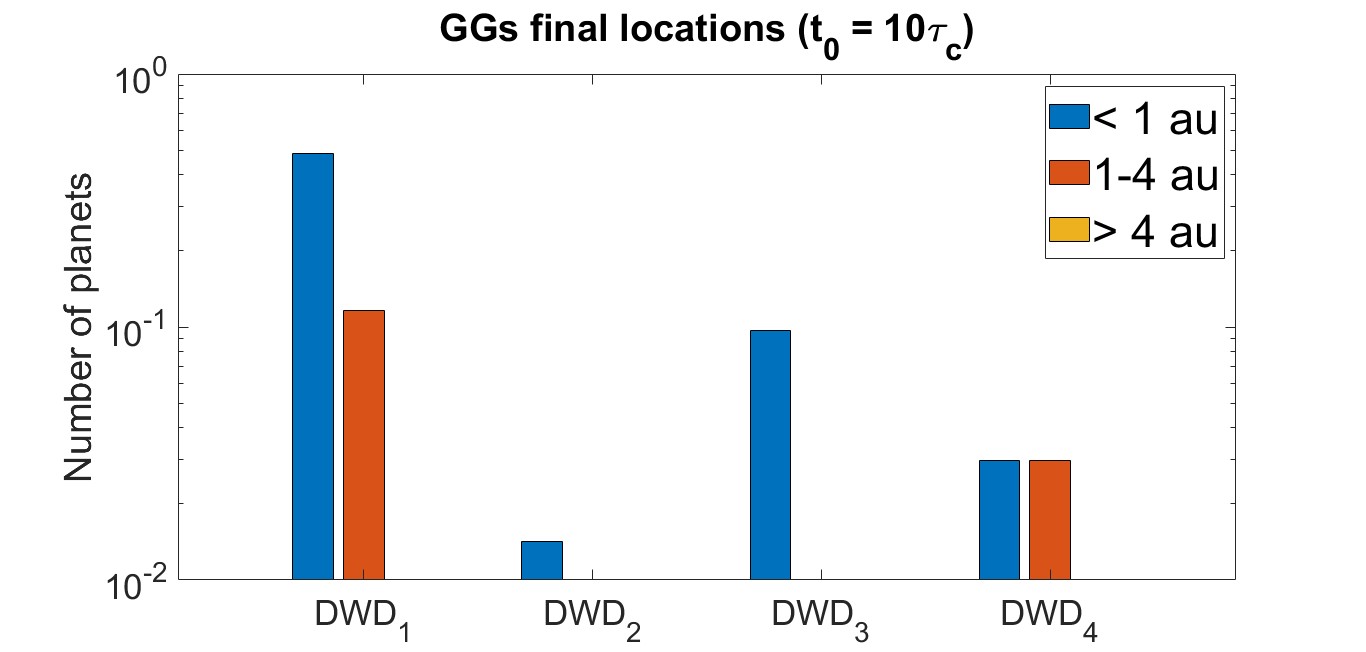}
\caption{Final positions distribution of the GGs formed at $t_0 = 10\tau_c$ by each DWD system analysed in this work. Each bar is normalized with respect to the total number of planets formed by each DWD system, and the Y axis is in log scale. The final positions are distributed among the radial intervals $r_{\rm in}-1\ \text{au}$ (blue bars), $1-4$ au (orange bars) and $4\ \text{au}-r_{\rm c}$ (yellow bars) (see Appendix \ref{sec:appendixA}).}
\label{fig:graphbar_locations_ggs}
\end{figure*}

\begin{figure*}[t]
\centering
\subfloat[][]{\includegraphics[trim=0.3cm 0.cm 1cm 0.1cm, clip,width=0.52\textwidth]{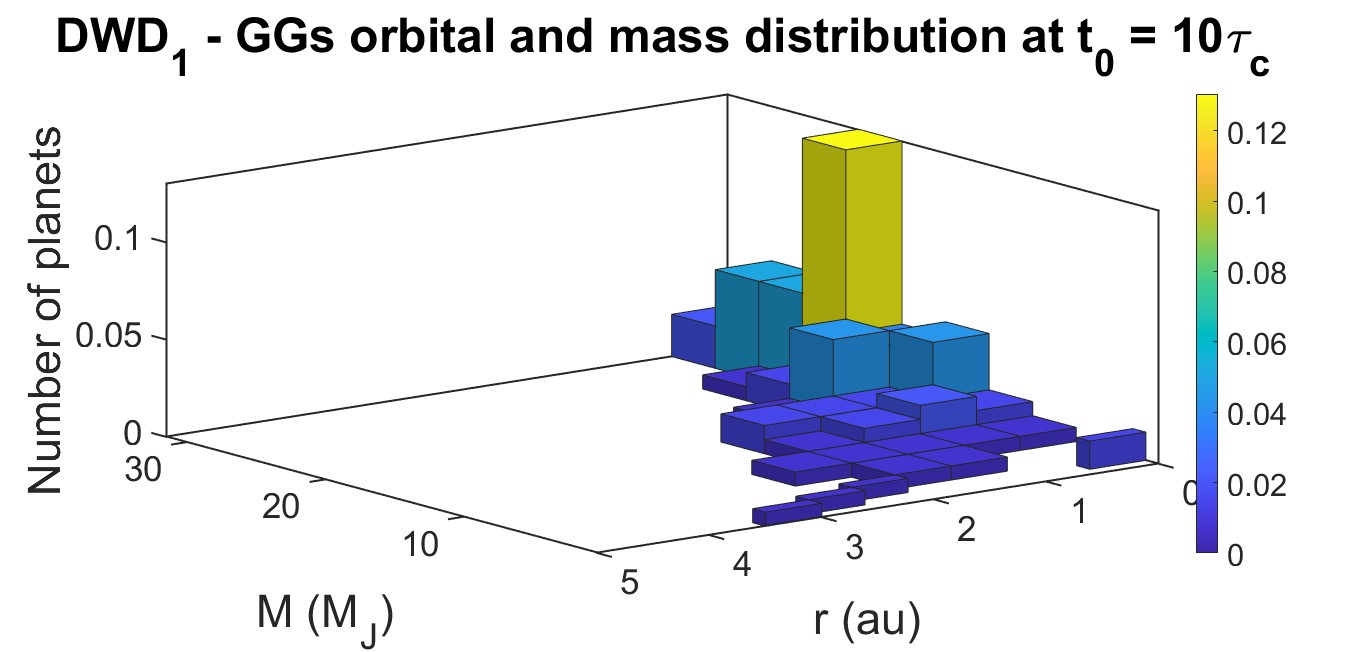}}
\subfloat[][]{\includegraphics[trim=0.3cm 0.cm 1cm 0.1cm, clip,width=0.52\textwidth]{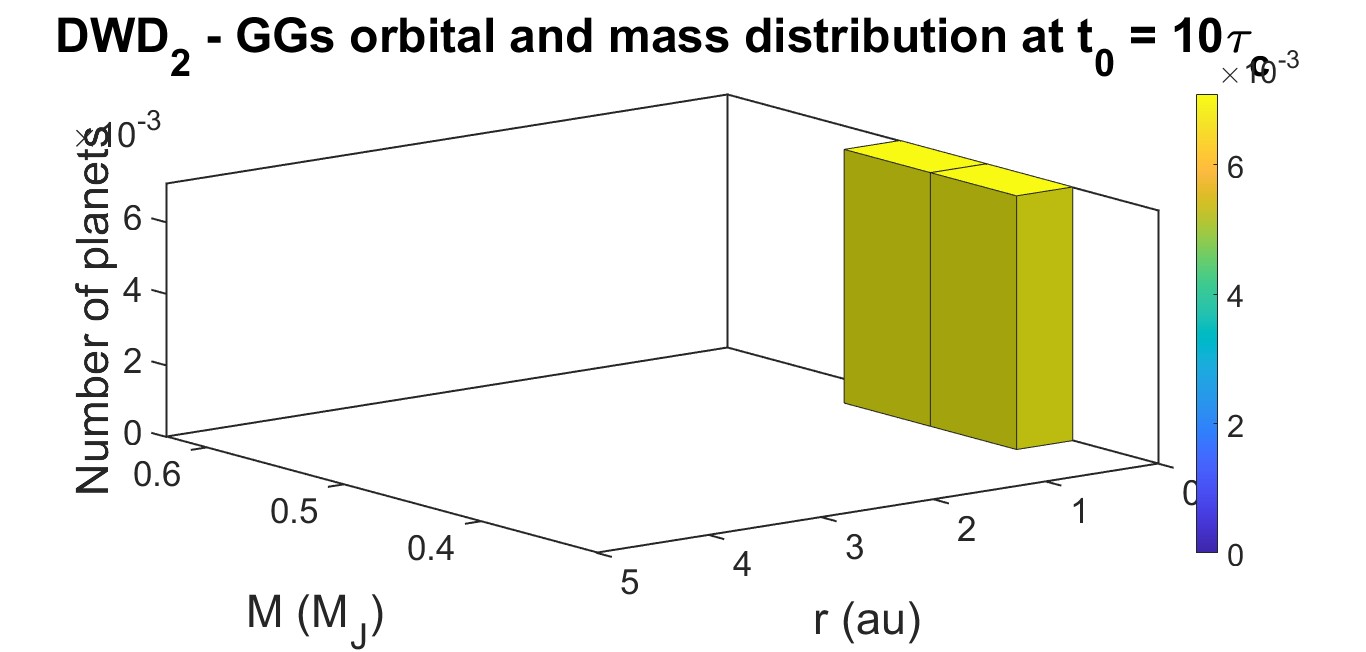}} \\
\subfloat[][]{\includegraphics[trim=0.3cm 0.cm 1cm 0.1cm, clip,width=0.52\textwidth]{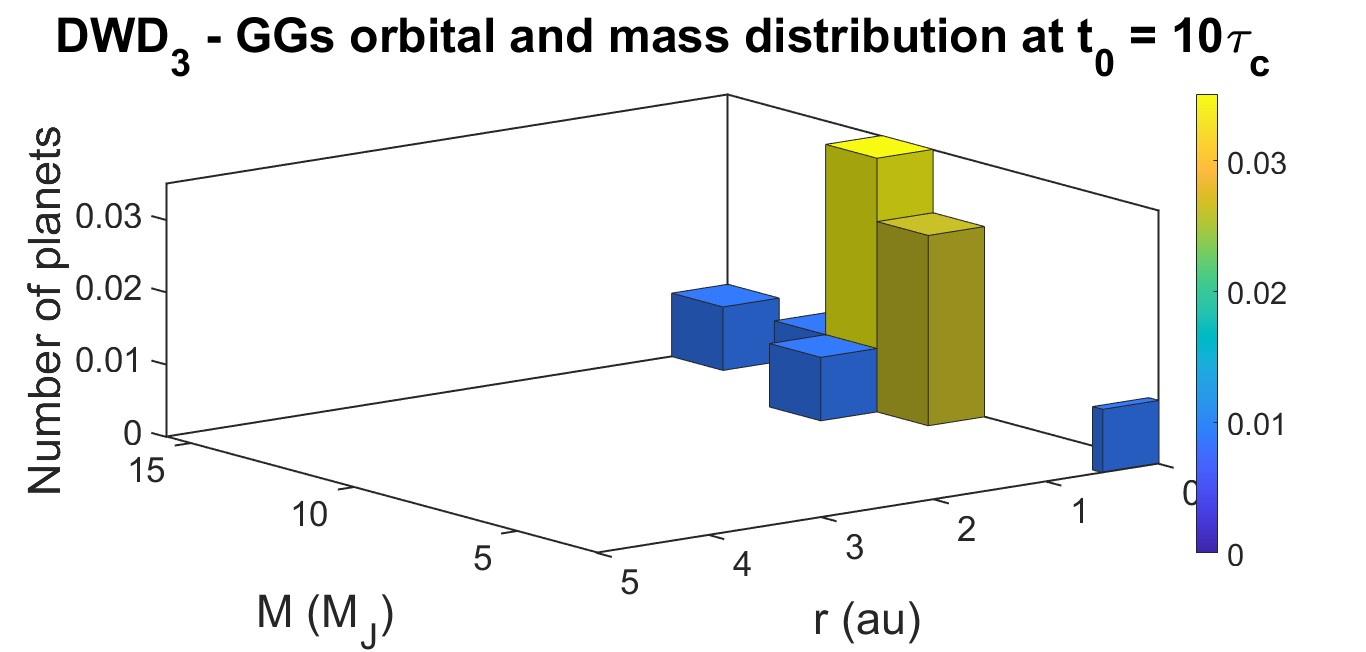}} 
\subfloat[][]{\includegraphics[trim=0.3cm 0cm 1cm 0.1cm, clip,width=0.52\textwidth]{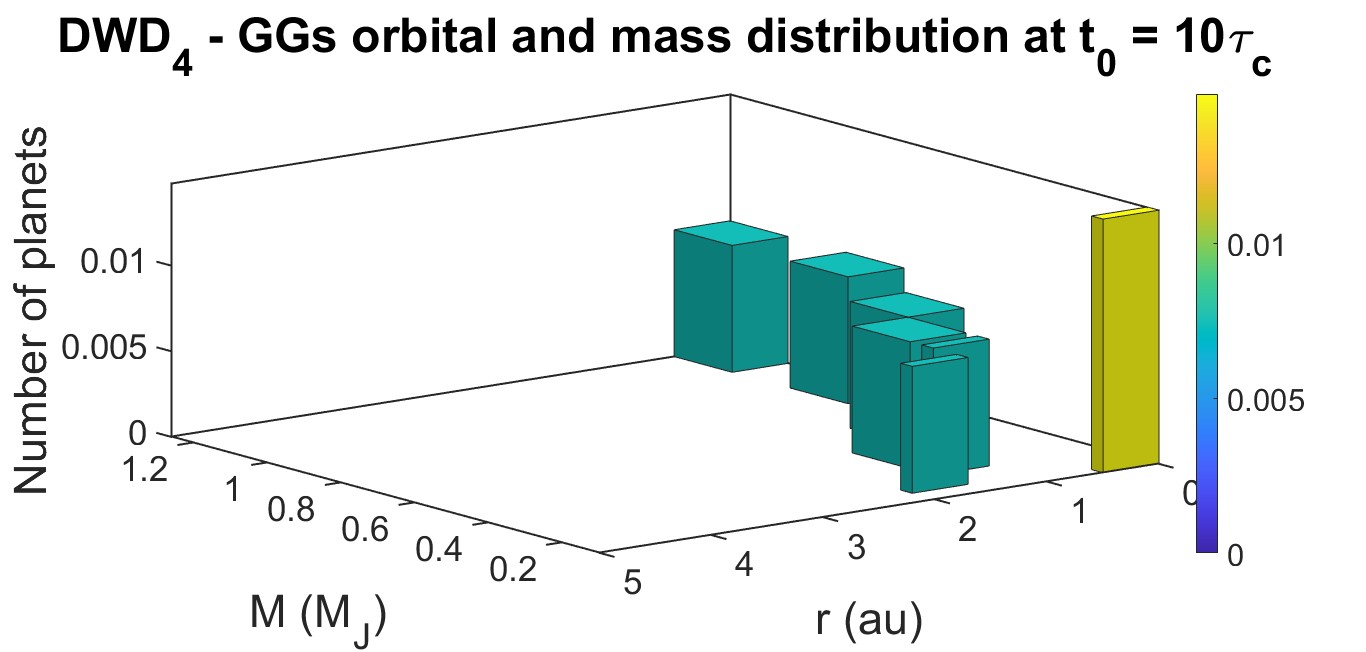}} 
\caption{Distributions of mass and final position of the GGs formed at $t_0 = 10\tau_c$ by each system analysed in this work. The X axis and the Y axis of each of the three-dimensional histogram show the radial position and the mass of the planets, respectively. The X axis values increase from right to left. The number of planets in each three-dimensional bin is normalized with respect to the total number of planets formed by each system. Following the vertical colorbar on the right of each histogram, the light-yellow bins include the largest amount of GGs, and the dark-blue bins include the smallest amount of GGs.}
\label{fig:MvsR_3dhistogram_ggs}
\end{figure*}

When considering the DWDs detectable by the LISA mission, i.e., DWD$_2$, DWD$_3$, DWD$_4$ (Tab \ref{tab:DWDparams}), the systems host giant planets whose masses range from $\sim 0.13$ \Mj\ to $\sim 15$ \Mj\ when we consider all disc temperature, metallicity, and extension combinations (see Fig. \ref{fig:global_massdistrib_histograms}). Aside from the planetary mass, the other parameter relevant for the detection is the orbital separation from the central binary. Following the results by \cite{dani2019} giant planets with masses larger than 0.27 \Mj, and separations less than $\sim$ 4 au, could be detected. In our analysis, immediately after the onset of the disc at t$_0 = 10\tau_c$, all three systems are capable of producing GG planets whose semi-major axes fall within 4 au (Tab. \ref{table:taudwd2}, \ref{table:taudwd3}, \ref{table:taudwd4} and Fig. \ref{fig:positions_overlap}, \ref{fig:graphbar_locations_ggs}). In particular,  DWD$_3$ is the LISA system whose environment can produce GGs with the largest masses (see Fig. \ref{fig:MvsR_3dhistogram_ggs}). These planets are located within $1$ au, and their masses fall between the range $M = 2.77-14.78$ \Mj\ (see Fig. \ref{fig:global_massdistrib_histograms}, third column). These characteristics make DWD$_3$ the best system among the ones analysed for producing second generation planets that could be detected by LISA. 
However, here we do not account for the distance of the systems from Earth, from which the amplitude of the monochromatic GW depend \citep{Korol2020}. Furthermore, while DWD$_2$ has a binary separations which produces GW signal within the LISA band, DWD$_3$ and DWD$_4$ will need to further shrink their orbit for their GW to be detectable by LISA. 
For such the above planetary detectability considerations are meant to be indicative, based on previous analysis \citep{dani2019,Katz2022}. A focused Bayesian analysis on the detection significance of these systems will be performed in future studies.

\section{Conclusions}
\label{sec:conclusions} 
The results of this study argue that second generation circumbinary discs around DWDs are capable of forming second generation \emph{Magrathea} planets. However, the possibility that the formation of the planets with the heaviest masses, i.e., GGs, occurs exclusively in second generation discs appears unfavoured. Rather, these planets are most likely originated by first generation bodies whose growth process has been restarted by the disc of stellar material produced by the binary evolution. Even if our models do not form GGs at $t_0 \geq 0.1$ Myr, this does not exclude that such they could form in second generation discs, as we restricted our analysis only to the region of the disc mainly heated by stellar radiation without exploring the scenario where these planets form by disc instability.

Our results point toward a key role of the accretion rate of the disc, which in the adopted treatment depends mainly on the disc temperature and density profiles as we focus on a fixed viscosity value following \cite{johansen}. However, viscosity plays a similar role to density and temperature in affecting the disc accretion rate and the planet formation tracks, especially in the early stages of the disc lifetime. 
In particular, when planet formation starts, the higher is the accretion rate with respect to the photoevaporation rate of the disc, the easier it is for the planets to reach the critical mass. Moreover, the higher is the disc metallicity, the faster is pebble accretion, meaning that the seeds can reach the isolation mass more quickly. A combination of high accretion rate and metallicity defines the best environment for planet formation, and in particular for giants formation.

In general, our disc models are characterised by steep isolation mass profiles. The latter facilitates the formation of SN and N planets, which are the most common planets. On the other hand, the steepness of the isolation mass profile hinders the outer seeds to reach the isolation mass. When they reach it and they complete their gas contraction phase, the accretion rate of the disc is already too low with respect to the photoevaporation rate, and the planet-forming region quickly run out of gas.

Most of the second generation planets we found reach the inner regions of the disc (within $1$ au from the inner edge of the disc) while they are still undergoing the gas accretion phase. Therefore, they should be observed close to the binary system. In particular, planets formed from first generation seeds can reach the critical mass more easily if the planet formation starts soon after the onset of the disc and the disc metallicity and density are high. In this case they could be found within $1$ and $4$ au. 

The GGs formed by our systems reached very high masses for some of the parameters we used for our discs. However, considering the global population of GGs for each system, it is less likely to find planets with masses within $\sim 1000-2500$\ \Mearth\ ($\sim 3.14-7.86$ \Mj) than with masses within $2500-7500$\ \Mearth\ ($\sim 7.86-23.59$ \Mj) and $40-640$\ \Mearth\ ($0.12-2.01$ \Mj). In particular, the best scenario for LISA mission observations, i.e., the DWD$_3$ system, shows that $\sim 78.57\%$ of GGs have masses larger than $0.27$ \Mj, and all of them reach a final location within $1$ au. 
Among the systems which can be observed by LISA, DWD$_4$ is the only one forming  GGs whose final locations are within $1$ and $4$ au. In particular $50\%$ of such GGs are finally located within $1$ au and $50\%$ are located within $1$ and $4$ au. Moreover, $\sim 44.44\%$ of them have masses larger than $0.27$ \Mj.

The analysis presented here serves as the first step towards addressing planetary formation at the end of binary evolution, when both stars begin their cooling down process. This work is set within a more comprehensive analysis to both study the orbital stability of the newly formed planetary systems \citep{Nigioni}, and account for the potential presence of first-generation planets (or planetesimals) within the second-generation disc, throughout specific long-term evolution modelling \citep{Columba2022}. These theoretical developments will allow us to further deepen our knowledge on the events occurring in the final stages of a circumbinary system, and to have an improved vision on the occurrence rates of \emph{Magrathea} exoplanets detectable with LISA.

\begin{acknowledgements}
The authors thanks Jacques Kluska and Valeriya Korol for the helpful discussions.
D.T. acknowledges the support of the Italian National Institute of Astrophysics (INAF) through the INAF Main Stream project ``Ariel and the astrochemical link between circumstellar discs and planets'' (CUP: C54I19000700005) and of the Italian Space Agency (ASI) through the ASI-INAF agreement  no. 2021-5-HH.0.
C.D. acknowledges financial support from the grant CEX2021-001131-S funded by MCIN/AEI/ 10.13039/501100011033 and the Group project Ref. PID2019-110689RB-I00/AEI/10.13039/501100011033. 
\end{acknowledgements}

\bibliographystyle{aa} 
\bibliography{Biblio}

\begin{appendix}
\section{Tables of results}
\label{sec:appendixA}

\begin{table}[p]
\caption{Formation times, fraction and positions of the planets produced by the circumbinary disc of DWD$_1$ when $t_0 = 0.1$ Myr.}
\label{table:0.1dwd1}
\centering
\begin{tabular}{|c|c|c|r|c|} 
\hline
\multicolumn{5}{|c|}{\rule{0pt}{3.2ex}\large{$t_0= 0.1$ Myr,\ $\Teff$$_2 = 45700$ K}} \\ [1.5ex]
\cline{1-5}
\hline
\hline
$\xi$ & $\Delta t_1$ & $\Delta t_g$ & \makecell[c]{$n_F$} & $n_p$\\ 
& [Myr] & [Myr] & \makecell[c]{[\%]} & [\%] \\
\hline\hline
\multicolumn{5}{|c|}{$\boldsymbol{r_{\rm c}} = 50$ au} \\ 
\hline

\cline{1-3} 
&   &   & \makecell[l]{\textbf{SN}} 100 & \makecell[l]{$\bf{\Delta_1}$} 100 \\

0.01 & 0.05 & $-$ & \makecell[l]{\textbf{N}} $-$ & \makecell[l]{$\bf{\Delta_{14}}$} $-$ \\

 &  &  & \makecell[l]{\textbf{GG}} $-$ & \makecell[l]{$\bf{\Delta_4}$} $-$ \\
\hline

\cline{1-3} 
&  &  & \makecell[l]{\textbf{SN}} 100 & \makecell[l]{$\bf{\Delta_1}$} 100 \\

0.015 & 0.04 & $-$ & \makecell[l]{\textbf{N}} $-$ & \makecell[l]{$\bf{\Delta_{14}}$} $-$ \\

 &  &  & \makecell[l]{\textbf{GG}} $-$ & \makecell[l]{$\bf{\Delta_4}$} $-$ \\
\hline

\cline{1-3} 
&  &  & \makecell[l]{\textbf{SN}} 100 & \makecell[l]{$\bf{\Delta_1}$}  100 \\

0.02 & 0.03 & $-$ & \makecell[l]{\textbf{N}} $-$ & \makecell[l]{$\bf{\Delta_{14}}$} $-$ \\

&  &  & \makecell[l]{\textbf{GG}} $-$ & \makecell[l]{$\bf{\Delta_4}$}  $-$ \\
\hline\hline
\multicolumn{5}{|c|}{$\boldsymbol{r_{\rm c}} = 100$ au} \\ 
\hline\hline

\cline{1-3} 
&  &  & \makecell[l]{\textbf{SN}} 100 & \makecell[l]{$\bf{\Delta_1}$}  100 \\

0.01 & 0.08 & $-$ & \makecell[l]{\textbf{N}} $-$ & \makecell[l]{$\bf{\Delta_{14}}$}  $-$ \\

&  &  & \makecell[l]{\textbf{GG}} $-$ & \makecell[l]{$\bf{\Delta_4}$}  $-$ \\
\hline

\cline{1-3} 
&  &  & \makecell[l]{\textbf{SN}} 83 & \makecell[l]{$\bf{\Delta_1}$}  100 \\

0.015 & 0.07 & $-$ & \makecell[l]{\textbf{N}} 17 & \makecell[l]{$\bf{\Delta_{14}}$}  $-$ \\

&  &  & \makecell[l]{\textbf{GG}} $-$ & \makecell[l]{$\bf{\Delta_4}$}  $-$ \\
\hline

\cline{1-3}
&  &  & \makecell[l]{\textbf{SN}} 83 & \makecell[l]{$\bf{\Delta_1}$}  100 \\

0.02 & 0.05 & $-$ & \makecell[l]{\textbf{N}} 17 & \makecell[l]{$\bf{\Delta_{14}}$}  $-$ \\

&  &  & \makecell[l]{\textbf{GG}} $-$ & \makecell[l]{$\bf{\Delta_4}$}  $-$ \\ 
\hline\hline
\multicolumn{5}{|c|}{$\boldsymbol{r_{\rm c}} = 150$ au} \\ 
\hline\hline

\cline{1-3} 
 &  &  & \makecell[l]{\textbf{SN}} 92 & \makecell[l]{$\bf{\Delta_1}$}  100 \\

0.01 & 0.13 & $-$ & \makecell[l]{\textbf{N}} 8 & \makecell[l]{$\bf{\Delta_{14}}$}  $-$ \\

 &  &  & \makecell[l]{\textbf{GG}} $-$ & \makecell[l]{$\bf{\Delta_4}$}  $-$ \\
\hline

\cline{1-3}
&  &  & \makecell[l]{\textbf{SN}} 77 & \makecell[l]{$\bf{\Delta_1}$}  100 \\

0.015 & 0.1 & $-$ & \makecell[l]{\textbf{N}} 23 & \makecell[l]{$\bf{\Delta_{14}}$} $-$ \\

&  &  & \makecell[l]{\textbf{GG}} $-$ & \makecell[l]{$\bf{\Delta_4}$}  $-$ \\
\hline

\cline{1-3} 
&  &  & \makecell[l]{\textbf{SN}} 77 &\makecell[l]{ $\bf{\Delta_1}$}  100 \\

0.02 & 0.08 & $-$ & \makecell[l]{\textbf{N}} 23 & \makecell[l]{$\bf{\Delta_{14}}$}  $-$ \\

 &  &  & \makecell[l]{\textbf{GG}} $-$ & \makecell[l]{$\bf{\Delta_4}$}  $-$ \\ 
\hline
\end{tabular}
\tablefoot{\\The values related to the vertically arranged items of columns $n_F$ and $n_p$ are presented on their right, respectively. See Section \ref{sec:classesdefinition} for the definition of the three classes of planets. Following, a brief description of the parameters reported in the table (see Section \ref{sec:discmodel}, \ref{sec:accretionmodel} and \ref{sec:timing} for details).\\
\tablefoottext{a}{$\Delta t_1$: formation time of the first planet reaching the end of its formation process.} \\
\tablefoottext{b}{$\Delta t_g$: formation time of the first gas giant planet.}\\
\tablefoottext{c}{$n_F$: percentage of (i) Sub-Neptunes (SN), (ii) Neptunes (N), (iii) Gas Giants (GG) formed with respect to the total number of initial seeds.} \\
\tablefoottext{d}{$n_p$: percentage of planets located within the ranges $\Delta_1 = r_{in}-1$ au, $\Delta_{14} = 1-4$ au and $\Delta_4 = 4\ \text{AU}-r_c$.}
} 
\end{table}

\begin{table}[p]
\caption{Formation times, fraction and positions of the planets produced by the circumbinary disc of DWD$_1$ when $t_0 = 1$ Myr.}
\label{table:1dwd1}
\centering
\begin{tabular}{|c|c|c|r|c|} 
\hline
\multicolumn{5}{|c|}{\rule{0pt}{3.2ex}\large{$t_0= 1$ Myr,\ $\Teff$$_2 = 33950$ K}} \\ [1.5ex]
\cline{1-5}
\hline
\hline
$\xi$ & $\Delta t_1$ & $\Delta t_g$ & \makecell[c]{$n_F$} & $n_p$\\ 
& [Myr] & [Myr] & \makecell[c]{[\%]} & [\%] \\
\hline\hline
\multicolumn{5}{|c|}{$\boldsymbol{r_{\rm c}} = 50$ au} \\ 
\hline

\cline{1-3}
&  &   & \makecell[l]{\textbf{SN}} 100 & \makecell[l]{$\bf{\Delta_1}$}  100 \\

0.01 & 0.25 & $-$ & \makecell[l]{\textbf{N}} $-$ & \makecell[l]{$\bf{\Delta_{14}}$}  $-$ \\

 &  &  & \makecell[l]{\textbf{GG}} $-$ & \makecell[l]{$\bf{\Delta_4}$}  $-$ \\
\hline

\cline{1-3}
&  &  & \makecell[l]{\textbf{SN}} 100 & \makecell[l]{$\bf{\Delta_1}$}  100 \\

0.015 & 0.18 & $-$ & \makecell[l]{\textbf{N}} $-$ & \makecell[l]{$\bf{\Delta_{14}}$}  $-$ \\

&  &  & \makecell[l]{\textbf{GG}} $-$ & \makecell[l]{$\bf{\Delta_4}$}  $-$ \\
\hline

\cline{1-3} 
&  &  & \makecell[l]{\textbf{SN}} 100 & \makecell[l]{$\bf{\Delta_1}$}  100 \\

0.02 & 0.15 & $-$ & \makecell[l]{\textbf{N}} $-$ & \makecell[l]{$\bf{\Delta_{14}}$}  $-$ \\

 &  &  & \makecell[l]{\textbf{GG}} $-$ & \makecell[l]{$\bf{\Delta_4}$}  $-$ \\
\hline\hline
\multicolumn{5}{|c|}{$\boldsymbol{r_{\rm c}} = 100$ au} \\ 
\hline\hline

\cline{1-3} 
&  &  & \makecell[l]{\textbf{SN}} 100 & \makecell[l]{$\bf{\Delta_1}$}  100 \\

0.01 & 0.3 & $-$ & \makecell[l]{\textbf{N}} $-$ & \makecell[l]{$\bf{\Delta_{14}}$}  $-$ \\

&  &  & \makecell[l]{\textbf{GG}} $-$ & \makecell[l]{$\bf{\Delta_4}$}  $-$ \\
\hline

\cline{1-3} 
&  &  & \makecell[l]{\textbf{SN}} 100 & \makecell[l]{$\bf{\Delta_1}$}  100 \\

0.015 & 0.2 & $-$ & \makecell[l]{\textbf{N}} $-$ & \makecell[l]{$\bf{\Delta_{14}}$}  $-$ \\

 &  &  & \makecell[l]{\textbf{GG}} $-$ & \makecell[l]{$\bf{\Delta_4}$}  $-$ \\
\hline

\cline{1-3} 
&  &  & \makecell[l]{\textbf{SN}} 100 & \makecell[l]{$\bf{\Delta_1}$}  100 \\

0.02 & 0.15 & $-$ & \makecell[l]{\textbf{N}} $-$ & \makecell[l]{$\bf{\Delta_{14}}$}  $-$ \\

&  &  & \makecell[l]{\textbf{GG}} $-$ & \makecell[l]{$\bf{\Delta_4}$}  $-$ \\ 
\hline\hline
\multicolumn{5}{|c|}{$\boldsymbol{r_{\rm c}} = 150$ au} \\ 
\hline\hline

\cline{1-3} 
&  &  & \makecell[l]{\textbf{SN}} 100 & \makecell[l]{$\bf{\Delta_1}$}  100 \\

0.01 & 0.35 & $-$ & \makecell[l]{\textbf{N}} $-$ & \makecell[l]{$\bf{\Delta_{14}}$}  $-$ \\

&  &  & \makecell[l]{\textbf{GG}} $-$ & \makecell[l]{$\bf{\Delta_4}$}  $-$ \\
\hline

\cline{1-3} 
&  &  & \makecell[l]{\textbf{SN}} 100 & \makecell[l]{$\bf{\Delta_1}$}  100 \\

0.015 & 0.25 & $-$ & \makecell[l]{\textbf{N}} $-$ & \makecell[l]{$\bf{\Delta_{14}}$}  $-$ \\

&  &  & \makecell[l]{\textbf{GG}} $-$ & \makecell[l]{$\bf{\Delta_4}$}  $-$ \\
\hline

\cline{1-3}
&  &  & \makecell[l]{\textbf{SN}} 100 & \makecell[l]{$\bf{\Delta_1}$}  100 \\

0.02 & 0.15 & $-$ & \makecell[l]{\textbf{N}} $-$ & \makecell[l]{$\bf{\Delta_{14}}$}  $-$ \\

 &  &  & \makecell[l]{\textbf{GG}} $-$ & \makecell[l]{$\bf{\Delta_4}$}  $-$ \\ 
\hline
\end{tabular}
\tablefoot{\\The values related to the vertically arranged items of columns $n_F$ and $n_p$ are presented on their right, respectively. See Section \ref{sec:classesdefinition} for the definition of the three classes of planets. Following, a brief description of the parameters reported in the table (see Section \ref{sec:discmodel}, \ref{sec:accretionmodel} and \ref{sec:timing} for details).\\
\tablefoottext{a}{$\Delta t_1$: formation time of the first planet reaching the end of its formation process.} \\
\tablefoottext{b}{$\Delta t_g$: formation time of the first gas giant planet.}\\
\tablefoottext{c}{$n_F$: percentage of (i) Sub-Neptunes (SN), (ii) Neptunes (N), (iii) Gas Giants (GG) formed with respect to the total number of initial seeds.} \\
\tablefoottext{d}{$n_p$: percentage of planets located within the ranges $\Delta_1 = r_{in}-1$ au, $\Delta_{14} = 1-4$ au and $\Delta_4 = 4\ \text{AU}-r_c$.}
} 
\end{table}

\begin{table}[p]
\caption{Formation times, fraction and positions of the planets produced by the circumbinary disc of DWD$_2$ when $t_0 = 10\tau_c$.}
\label{table:taudwd2}
\centering
\begin{tabular}{|c|c|c|r|c|} 
\hline
\multicolumn{5}{|c|}{\rule{0pt}{3.2ex}\large{$t_0= 10\tau_c$,\ $\Teff$$_2 = 58400$ K}} \\ [1.5ex]
\cline{1-5}
\hline
\hline
$\xi$ & $\Delta t_1$ & $\Delta t_g$ & \makecell[c]{$n_F$} & $n_p$\\ 
& [Myr] & [Myr] & \makecell[c]{[\%]} & [\%] \\
\hline\hline
\multicolumn{5}{|c|}{$\boldsymbol{r_{\rm c}} = 50$ au} \\ 
\hline

\cline{1-3} 
&   &   & \makecell[l]{\textbf{SN}} 100 & \makecell[l]{$\bf{\Delta_1}$}  100 \\

0.01 & 0.09 & $-$ & \makecell[l]{\textbf{N}} $-$ & \makecell[l]{$\bf{\Delta_{14}}$}  $-$ \\

 &  &  & \makecell[l]{\textbf{GG}} $-$ & \makecell[l]{$\bf{\Delta_4}$}  $-$ \\
\hline

\cline{1-3}
&  &  & \makecell[l]{\textbf{SN}} 92 & \makecell[l]{$\bf{\Delta_1}$} 100 \\

0.015 & 0.06 & $-$ & \makecell[l]{\textbf{N}} 8 & \makecell[l]{$\bf{\Delta_{14}}$}  $-$ \\

 &  &  & \makecell[l]{\textbf{GG}} $-$ & \makecell[l]{$\bf{\Delta_4}$}  $-$ \\
\hline

\cline{1-3}
 &  &  & \makecell[l]{\textbf{SN}} 83 & \makecell[l]{$\bf{\Delta_1}$}  100 \\

0.02 & 0.05 & $-$ & \makecell[l]{\textbf{N}} 17 & \makecell[l]{$\bf{\Delta_{14}}$}  $-$ \\

 &  &  & \makecell[l]{\textbf{GG}} $-$ & \makecell[l]{$\bf{\Delta_4}$}  $-$ \\
\hline\hline
\multicolumn{5}{|c|}{$\boldsymbol{r_{\rm c}} = 100$ au} \\ 
\hline\hline

\cline{1-3}
 &  &  & \makecell[l]{\textbf{SN}} 81 & \makecell[l]{$\bf{\Delta_1}$}  100 \\

0.01 & 0.2 & $-$ & \makecell[l]{\textbf{N}} 19 & \makecell[l]{$\bf{\Delta_{14}}$}  $-$ \\

&  &  & \makecell[l]{\textbf{GG}} $-$ & \makecell[l]{$\bf{\Delta_4}$} $-$ \\
\hline

\cline{1-3} 
&  &  & \makecell[l]{\textbf{SN}} 63 & \makecell[l]{$\bf{\Delta_1}$} 100 \\

0.015 & 0.16 & $-$ & \makecell[l]{\textbf{N}} 37 & \makecell[l]{$\bf{\Delta_{14}}$}  $-$ \\

&  &  & \makecell[l]{\textbf{GG}} $-$ & \makecell[l]{$\bf{\Delta_4}$}  $-$ \\
\hline

\cline{1-3}
&  &  & \makecell[l]{\textbf{SN}} 63 & \makecell[l]{$\bf{\Delta_1}$} 100 \\

0.02 & 0.13 & 1 & \makecell[l]{\textbf{N}} 25 & \makecell[l]{$\bf{\Delta_{14}}$}  $-$ \\

&  &  & \makecell[l]{\textbf{GG}} 12 & \makecell[l]{$\bf{\Delta_4}$}  $-$ \\ 
\hline\hline
\multicolumn{5}{|c|}{$\boldsymbol{r_{\rm c}} = 150$ au} \\ 
\hline\hline

\cline{1-3} 
&  &  & \makecell[l]{\textbf{SN}} 63 & \makecell[l]{$\bf{\Delta_1}$}  100 \\

0.01 & 0.34 & $-$ & \makecell[l]{\textbf{N}} 37 & \makecell[l]{$\bf{\Delta_{14}}$}  $-$ \\

&  &  & \makecell[l]{\textbf{GG}} $-$ & \makecell[l]{$\bf{\Delta_4}$}  $-$ \\
\hline

\cline{1-3}
&  &  & \makecell[l]{\textbf{SN}} 53 & \makecell[l]{$\bf{\Delta_1}$}  100 \\

0.015 & 0.25 & $-$ & \makecell[l]{\textbf{N}} 47 & \makecell[l]{$\bf{\Delta_{14}}$}  $-$ \\

&  &  & \makecell[l]{\textbf{GG}} $-$ & \makecell[l]{$\bf{\Delta_4}$}  $-$ \\
\hline

\cline{1-3} 
&  &  & \makecell[l]{\textbf{SN}} 47 & \makecell[l]{$\bf{\Delta_1}$}  84 \\

0.02 & 0.2 & $-$ & \makecell[l]{\textbf{N}} 53 & \makecell[l]{$\bf{\Delta_{14}}$}  16 \\

&  &  & \makecell[l]{\textbf{GG}} $-$ & \makecell[l]{$\bf{\Delta_4}$}  $-$ \\ 
\hline
\end{tabular}
\tablefoot{\\The values related to the vertically arranged items of columns $n_F$ and $n_p$ are presented on their right, respectively. See Section \ref{sec:classesdefinition} for the definition of the three classes of planets. Following, a brief description of the parameters reported in the table (see Section \ref{sec:discmodel}, \ref{sec:accretionmodel} and \ref{sec:timing} for details).\\
\tablefoottext{a}{$\Delta t_1$: formation time of the first planet reaching the end of its formation process.} \\
\tablefoottext{b}{$\Delta t_g$: formation time of the first gas giant planet.}\\
\tablefoottext{c}{$n_F$: percentage of (i) Sub-Neptunes (SN), (ii) Neptunes (N), (iii) Gas Giants (GG) formed with respect to the total number of initial seeds.} \\
\tablefoottext{d}{$n_p$: percentage of planets located within the ranges $\Delta_1 = r_{in}-1$ au, $\Delta_{14} = 1-4$ au and $\Delta_4 = 4\ \text{AU}-r_c$.}
} 
\end{table}

\begin{table}[p]
\caption{Formation times, fraction and positions of the planets produced by the circumbinary disc of DWD$_2$ when $t_0 = 0.1$ Myr.}
\label{table:0.1dwd2}
\centering
\begin{tabular}{|c|c|c|r|c|} 
\hline
\multicolumn{5}{|c|}{\rule{0pt}{3.2ex}\large{$t_0= 0.1$ Myr,\ $\Teff$$_2 = 32000$ K}} \\ [1.5ex]
\cline{1-5}
\hline
\hline
$\xi$ & $\Delta t_1$ & $\Delta t_g$ & \makecell[c]{$n_F$} & $n_p$\\ 
& [Myr] & [Myr] & \makecell[c]{[\%]} & [\%] \\
\hline\hline
\multicolumn{5}{|c|}{$\boldsymbol{r_{\rm c}} = 50$ au} \\ 
\hline

\cline{1-3} 
&   &   & \makecell[l]{\textbf{SN}} 100 & \makecell[l]{$\bf{\Delta_1}$}  100 \\

0.01 & 0.08 & $-$ & \makecell[l]{\textbf{N}} $-$ & \makecell[l]{$\bf{\Delta_{14}}$}  $-$ \\

&  &  & \makecell[l]{\textbf{GG}} $-$ & \makecell[l]{$\bf{\Delta_4}$}  $-$ \\
\hline

\cline{1-3} 
&  &  & \makecell[l]{\textbf{SN}} 100 & \makecell[l]{$\bf{\Delta_1}$}  100 \\

0.015 & 0.07 & $-$ & \makecell[l]{\textbf{N}} $-$ & \makecell[l]{$\bf{\Delta_{14}}$}  $-$ \\

&  &  & \makecell[l]{\textbf{GG}} $-$ & \makecell[l]{$\bf{\Delta_4}$} $-$ \\
\hline

\cline{1-3}
&  &  & \makecell[l]{\textbf{SN}} 100 & \makecell[l]{$\bf{\Delta_1}$}  100 \\

0.02 & 0.05 & $-$ & \makecell[l]{\textbf{N}} $-$ & \makecell[l]{$\bf{\Delta_{14}}$}  $-$ \\

&  &  & \makecell[l]{\textbf{GG}} $-$ & \makecell[l]{$\bf{\Delta_4}$}  $-$ \\
\hline\hline
\multicolumn{5}{|c|}{$\boldsymbol{r_{\rm c}} = 100$ au} \\ 
\hline\hline

\cline{1-3} 
&  &  & \makecell[l]{\textbf{SN}} 100 & \makecell[l]{$\bf{\Delta_1}$}  100 \\

0.01 & 0.18 & $-$ & \makecell[l]{\textbf{N}} $-$ & \makecell[l]{$\bf{\Delta_{14}}$}  $-$ \\

 &  &  & \makecell[l]{\textbf{GG}} $-$ & \makecell[l]{$\bf{\Delta_4}$}  $-$ \\
\hline

\cline{1-3} 
&  &  & \makecell[l]{\textbf{SN}} 100 & \makecell[l]{$\bf{\Delta_1}$}  100 \\

0.015 & 0.14 & $-$ & \makecell[l]{\textbf{N}} $-$ & \makecell[l]{$\bf{\Delta_{14}}$}  $-$ \\

&  &  & \makecell[l]{\textbf{GG}} $-$ & \makecell[l]{$\bf{\Delta_4}$} $-$ \\
\hline

\cline{1-3}
&  &  & \makecell[l]{\textbf{SN}} 100 & \makecell[l]{$\bf{\Delta_1}$}  100 \\

0.02 & 0.1 & $-$ & \makecell[l]{\textbf{N}} $-$ & \makecell[l]{$\bf{\Delta_{14}}$}  $-$ \\

&  &  & \makecell[l]{\textbf{GG}} $-$ & \makecell[l]{$\bf{\Delta_4}$}  $-$ \\ 
\hline\hline
\multicolumn{5}{|c|}{$\boldsymbol{r_{\rm c}} = 150$ au} \\ 
\hline\hline

\cline{1-3} 
&  &  & \makecell[l]{\textbf{SN}} 100 & \makecell[l]{$\bf{\Delta_1}$}  100 \\

0.01 & 0.28 & $-$ & \makecell[l]{\textbf{N}} $-$ & \makecell[l]{$\bf{\Delta_{14}}$}  $-$ \\

&  &  & \makecell[l]{\textbf{GG}} $-$ & \makecell[l]{$\bf{\Delta_4}$}  $-$ \\
\hline

\cline{1-3}
&  &  & \makecell[l]{\textbf{SN}} 100 & \makecell[l]{$\bf{\Delta_1}$}  100 \\

0.015 & 0.2 & $-$ & \makecell[l]{\textbf{N}} $-$ & \makecell[l]{$\bf{\Delta_{14}}$}  $-$ \\

&  &  & \makecell[l]{\textbf{GG}} $-$ & \makecell[l]{$\bf{\Delta_4}$}  $-$ \\
\hline

\cline{1-3} 
&  &  & \makecell[l]{\textbf{SN}} 100 & \makecell[l]{$\bf{\Delta_1}$}  100 \\

0.02 & 0.17 & $-$ & \makecell[l]{\textbf{N}} $-$ & \makecell[l]{$\bf{\Delta_{14}}$}  $-$ \\

&  &  & \makecell[l]{\textbf{GG}} $-$ & \makecell[l]{$\bf{\Delta_4}$}  $-$ \\ 
\hline
\end{tabular}
\tablefoot{\\The values related to the vertically arranged items of columns $n_F$ and $n_p$ are presented on their right, respectively. See Section \ref{sec:classesdefinition} for the definition of the three classes of planets. Following, a brief description of the parameters reported in the table (see Section \ref{sec:discmodel}, \ref{sec:accretionmodel} and \ref{sec:timing} for details).\\
\tablefoottext{a}{$\Delta t_1$: formation time of the first planet reaching the end of its formation process.} \\
\tablefoottext{b}{$\Delta t_g$: formation time of the first gas giant planet.}\\
\tablefoottext{c}{$n_F$: percentage of (i) Sub-Neptunes (SN), (ii) Neptunes (N), (iii) Gas Giants (GG) formed with respect to the total number of initial seeds.} \\
\tablefoottext{d}{$n_p$: percentage of planets located within the ranges $\Delta_1 = r_{in}-1$ au, $\Delta_{14} = 1-4$ au and $\Delta_4 = 4\ \text{AU}-r_c$.}
} 
\end{table}

\begin{table}[p]
\caption{Formation times, fraction and positions of the planets produced by the circumbinary disc of DWD$_2$ when $t_0 = 1$ Myr.}
\label{table:1dwd2}
\centering
\begin{tabular}{|c|c|c|r|c|} 
\hline
\multicolumn{5}{|c|}{\rule{0pt}{3.2ex}\large{$t_0= 1$ Myr,\ $\Teff$$_2 = 25000$ K}} \\ [1.5ex]
\cline{1-5}
\hline
\hline
$\xi$ & $\Delta t_1$ & $\Delta t_g$ & \makecell[c]{$n_F$} & $n_p$\\ 
& [Myr] & [Myr] & \makecell[c]{[\%]} & [\%] \\
\hline\hline
\multicolumn{5}{|c|}{$\boldsymbol{r_{\rm c}} = 50$ au} \\ 
\hline

\cline{1-3} 
&   &   & \makecell[l]{\textbf{SN}} 100 & \makecell[l]{$\bf{\Delta_1}$}  100 \\

0.01 & 0.4 & $-$ & \makecell[l]{\textbf{N}} $-$ & \makecell[l]{$\bf{\Delta_{14}}$}  $-$ \\

&  &  & \makecell[l]{\textbf{GG}} $-$ & \makecell[l]{$\bf{\Delta_4}$}  $-$ \\
\hline

\cline{1-3}
&  &  & \makecell[l]{\textbf{SN}} 100 & \makecell[l]{$\bf{\Delta_1}$}  100 \\

0.015 & 0.3 & $-$ & \makecell[l]{\textbf{N}} $-$ & \makecell[l]{$\bf{\Delta_{14}}$}  $-$ \\

&  &  & \makecell[l]{\textbf{GG}} $-$ & \makecell[l]{$\bf{\Delta_4}$} $-$ \\
\hline

\cline{1-3} 
&  &  & \makecell[l]{\textbf{SN}} 100 & \makecell[l]{$\bf{\Delta_1}$} 100 \\

0.02 & 0.3 & $-$ & \makecell[l]{\textbf{N}} $-$ & \makecell[l]{$\bf{\Delta_{14}}$} $-$ \\

&  &  & \makecell[l]{\textbf{GG}} $-$ & \makecell[l]{$\bf{\Delta_4}$} $-$ \\
\hline\hline
\multicolumn{5}{|c|}{$\boldsymbol{r_{\rm c}} = 100$ au} \\ 
\hline\hline

\cline{1-3} 
&  &  & \makecell[l]{\textbf{SN}} 100 & \makecell[l]{$\bf{\Delta_1}$} 100 \\

0.01 & 0.4 & $-$ & \makecell[l]{\textbf{N}} $-$ & \makecell[l]{$\bf{\Delta_{14}}$} $-$ \\

&  &  & \makecell[l]{\textbf{GG}} $-$ & \makecell[l]{$\bf{\Delta_4}$} $-$ \\
\hline

\cline{1-3} 
&  &  & \makecell[l]{\textbf{SN}} 100 & \makecell[l]{$\bf{\Delta_1}$} 100 \\

0.015 & 0.3 & $-$ & \makecell[l]{\textbf{N}} $-$ & \makecell[l]{$\bf{\Delta_{14}}$} $-$ \\

 &  &  & \makecell[l]{\textbf{GG}} $-$ & \makecell[l]{$\bf{\Delta_4}$} $-$ \\
\hline

\cline{1-3}
&  &  & \makecell[l]{\textbf{SN}} 100 & \makecell[l]{$\bf{\Delta_1}$} 100 \\

0.02 & 0.3 & $-$ & \makecell[l]{\textbf{N}} $-$ & \makecell[l]{$\bf{\Delta_{14}}$} $-$ \\

&  &  & \makecell[l]{\textbf{GG}} $-$ & \makecell[l]{$\bf{\Delta_4}$} $-$ \\ 
\hline\hline
\multicolumn{5}{|c|}{$\boldsymbol{r_{\rm c}} = 150$ au} \\ 
\hline\hline

\cline{1-3} 
&  &  & \makecell[l]{\textbf{SN}} 100 & \makecell[l]{$\bf{\Delta_1}$} 100 \\

0.01 & 0.45 & $-$ & \makecell[l]{\textbf{N}} $-$ & \makecell[l]{$\bf{\Delta_{14}}$} $-$ \\

&  &  & \makecell[l]{\textbf{GG}} $-$ & \makecell[l]{$\bf{\Delta_4}$} $-$ \\
\hline

\cline{1-3} 
&  &  & \makecell[l]{\textbf{SN}} 100 & \makecell[l]{$\bf{\Delta_1}$} 100 \\

0.015 & 0.35 & $-$ & \makecell[l]{\textbf{N}} $-$ & \makecell[l]{$\bf{\Delta_{14}}$} $-$ \\

&  &  & \makecell[l]{\textbf{GG}} $-$ & \makecell[l]{$\bf{\Delta_4}$} $-$ \\
\hline

\cline{1-3} 
&  &  & \makecell[l]{\textbf{SN}} 100 & \makecell[l]{$\bf{\Delta_1}$} 100 \\

0.02 & 0.3 & $-$ & \makecell[l]{\textbf{N}} $-$ & \makecell[l]{$\bf{\Delta_{14}}$} $-$ \\

&  &  & \makecell[l]{\textbf{GG}} $-$ & \makecell[l]{$\bf{\Delta_4}$} $-$ \\ 
\hline
\end{tabular}
\tablefoot{\\The values related to the vertically arranged items of columns $n_F$ and $n_p$ are presented on their right, respectively. See Section \ref{sec:classesdefinition} for the definition of the three classes of planets. Following, a brief description of the parameters reported in the table (see Section \ref{sec:discmodel}, \ref{sec:accretionmodel} and \ref{sec:timing} for details).\\
\tablefoottext{a}{$\Delta t_1$: formation time of the first planet reaching the end of its formation process.} \\
\tablefoottext{b}{$\Delta t_g$: formation time of the first gas giant planet.}\\
\tablefoottext{c}{$n_F$: percentage of (i) Sub-Neptunes (SN), (ii) Neptunes (N), (iii) Gas Giants (GG) formed with respect to the total number of initial seeds.} \\
\tablefoottext{d}{$n_p$: percentage of planets located within the ranges $\Delta_1 = r_{in}-1$ au, $\Delta_{14} = 1-4$ au and $\Delta_4 = 4\ \text{AU}-r_c$.}
} 
\end{table}

\begin{table}[p]
\caption{Formation times, fraction and positions of the planets produced by the circumbinary disc of DWD$_3$ when $t_0 = 10\tau_c$.}
\label{table:taudwd3}
\centering
\begin{tabular}{|c|c|c|r|c|} 
\hline
\multicolumn{5}{|c|}{\rule{0pt}{3.2ex}\large{$t_0= 10\tau_c$,\ $\Teff$$_2 = 52000$ K}} \\ [1.5ex]
\cline{1-5}
\hline
\hline
$\xi$ & $\Delta t_1$ & $\Delta t_g$ & \makecell[c]{$n_F$} & $n_p$\\ 
& [Myr] & [Myr] & \makecell[c]{[\%]} & [\%] \\
\hline\hline
\multicolumn{5}{|c|}{$\boldsymbol{r_{\rm c}} = 50$ au} \\ 
\hline

\cline{1-3}
&   &   & \makecell[l]{\textbf{SN}} 100 & \makecell[l]{$\bf{\Delta_1}$} 100 \\

0.01 & 0.04 & $-$ & \makecell[l]{\textbf{N}} $-$ & \makecell[l]{$\bf{\Delta_{14}}$} $-$ \\

&  &  & \makecell[l]{\textbf{GG}} $-$ & \makecell[l]{$\bf{\Delta_4}$} $-$ \\
\hline

\cline{1-3} 
&  &  & \makecell[l]{\textbf{SN}} 100 & \makecell[l]{$\bf{\Delta_1}$} 100 \\

0.015 & 0.03 & $-$ & \makecell[l]{\textbf{N}} $-$ & \makecell[l]{$\bf{\Delta_{14}}$} $-$ \\

&  &  & \makecell[l]{\textbf{GG}} $-$ & \makecell[l]{$\bf{\Delta_4}$} $-$ \\
\hline

\cline{1-3}
&  &  & \makecell[l]{\textbf{SN}} 73 & \makecell[l]{$\bf{\Delta_1}$} 100 \\

0.02 & 0.02 & $-$ & \makecell[l]{\textbf{N}} 27 & \makecell[l]{$\bf{\Delta_{14}}$} $-$ \\

&  &  & \makecell[l]{\textbf{GG}} $-$ & \makecell[l]{$\bf{\Delta_4}$} $-$ \\
\hline\hline
\multicolumn{5}{|c|}{$\boldsymbol{r_{\rm c}} = 100$ au} \\ 
\hline\hline

\cline{1-3}
&  &  & \makecell[l]{\textbf{SN}} 77 & \makecell[l]{$\bf{\Delta_1}$} 100 \\

0.01 & 0.09 & $-$ & \makecell[l]{\textbf{N}} 23 & \makecell[l]{$\bf{\Delta_{14}}$} $-$ \\

&  &  & \makecell[l]{\textbf{GG}} $-$ & \makecell[l]{$\bf{\Delta_4}$} $-$ \\
\hline

\cline{1-3} 
&  &  & \makecell[l]{\textbf{SN}} 62 & \makecell[l]{$\bf{\Delta_1}$} 100 \\

0.015 & 0.06 & $-$ & \makecell[l]{\textbf{N}} 38 & \makecell[l]{$\bf{\Delta_{14}}$} $-$ \\

&  &  & \makecell[l]{\textbf{GG}} $-$ & \makecell[l]{$\bf{\Delta_4}$} $-$ \\
\hline

\cline{1-3}
&  &  & \makecell[l]{\textbf{SN}} 39 & \makecell[l]{$\bf{\Delta_1}$} 100\\

0.02 & 0.05 & 2.7 & \makecell[l]{\textbf{N}} 46 & \makecell[l]{$\bf{\Delta_{14}}$} $-$ \\

 &  &  & \makecell[l]{\textbf{GG}} 15 &\makecell[l]{ $\bf{\Delta_4}$} $-$ \\ 
\hline\hline
\multicolumn{5}{|c|}{$\boldsymbol{r_{\rm c}} = 150$ au} \\ 
\hline\hline

\cline{1-3} 
&  &  & \makecell[l]{\textbf{SN}} 67 & \makecell[l]{$\bf{\Delta_1}$} 100 \\

0.01 & 0.15 & $-$ & \makecell[l]{\textbf{N}} 33 & \makecell[l]{$\bf{\Delta_{14}}$} $-$ \\

 &  &  & \makecell[l]{\textbf{GG}} $-$ & \makecell[l]{$\bf{\Delta_4}$} $-$ \\
\hline

\cline{1-3}
&  &  & \makecell[l]{\textbf{SN}} 46 & \makecell[l]{$\bf{\Delta_1}$} 100 \\

0.015 & 0.11 & 0.65 & \makecell[l]{\textbf{N}} 27 & \makecell[l]{$\bf{\Delta_{14}}$} $-$ \\

&  &  & \makecell[l]{\textbf{GG}} 27 & \makecell[l]{$\bf{\Delta_4}$} $-$ \\
\hline

\cline{1-3}
&  &  & \makecell[l]{\textbf{SN}} 27 & \makecell[l]{$\bf{\Delta_1}$} 100 \\

0.02 & 0.09 &3  & \makecell[l]{\textbf{N}} 40 & \makecell[l]{$\bf{\Delta_{14}}$} $-$ \\

&  &  & \makecell[l]{\textbf{GG}} 33 & \makecell[l]{$\bf{\Delta_4}$} $-$ \\ 
\hline
\end{tabular}
\tablefoot{\\The values related to the vertically arranged items of columns $n_F$ and $n_p$ are presented on their right, respectively. See Section \ref{sec:classesdefinition} for the definition of the three classes of planets. Following, a brief description of the parameters reported in the table (see Section \ref{sec:discmodel}, \ref{sec:accretionmodel} and \ref{sec:timing} for details).\\
\tablefoottext{a}{$\Delta t_1$: formation time of the first planet reaching the end of its formation process.} \\
\tablefoottext{b}{$\Delta t_g$: formation time of the first gas giant planet.}\\
\tablefoottext{c}{$n_F$: percentage of (i) Sub-Neptunes (SN), (ii) Neptunes (N), (iii) Gas Giants (GG) formed with respect to the total number of initial seeds.} \\
\tablefoottext{d}{$n_p$: percentage of planets located within the ranges $\Delta_1 = r_{in}-1$ au, $\Delta_{14} = 1-4$ au and $\Delta_4 = 4\ \text{AU}-r_c$.}
} 
\end{table}

\begin{table}[p]
\caption{Formation times, fraction and positions of the planets produced by the circumbinary disc of DWD$_3$ when $t_0 = 0.1$ Myr.}
\label{table:0.1dwd3}
\centering
\begin{tabular}{|c|c|c|r|c|} 
\hline
\multicolumn{5}{|c|}{\rule{0pt}{3.2ex}\large{$t_0= 0.1$ Myr,\ $\Teff$$_2 = 32200$ K}} \\ [1.5ex]
\cline{1-5}
\hline
\hline
$\xi$ & $\Delta t_1$ & $\Delta t_g$ & \makecell[c]{$n_F$} & $n_p$\\ 
& [Myr] & [Myr] & \makecell[c]{[\%]} & [\%] \\
\hline\hline
\multicolumn{5}{|c|}{$\boldsymbol{r_{\rm c}} = 50$ au} \\ 
\hline

\cline{1-3} 
&   &   & \makecell[l]{\textbf{SN}} 100 & \makecell[l]{$\bf{\Delta_1}$} 100 \\

0.01 & 0.05 & $-$ & \makecell[l]{\textbf{N}} $-$ & \makecell[l]{$\bf{\Delta_{14}}$} $-$ \\

&  &  & \makecell[l]{\textbf{GG}} $-$ & \makecell[l]{$\bf{\Delta_4}$} $-$ \\
\hline

\cline{1-3} 
&  &  & \makecell[l]{\textbf{SN}} 100 & \makecell[l]{$\bf{\Delta_1}$} 100 \\

0.015 & 0.04 & $-$ & \makecell[l]{\textbf{N}} $-$ & \makecell[l]{$\bf{\Delta_{14}}$} $-$ \\

 &  &  & \makecell[l]{\textbf{GG}} $-$ & \makecell[l]{$\bf{\Delta_4}$} $-$ \\
\hline

\cline{1-3} 
&  &  & \makecell[l]{\textbf{SN}} 100 & \makecell[l]{$\bf{\Delta_1}$} 100 \\

0.02 & 0.04 & $-$ & \makecell[l]{\textbf{N}} $-$ & \makecell[l]{$\bf{\Delta_{14}}$} $-$ \\

&  &  & \makecell[l]{\textbf{GG}} $-$ & \makecell[l]{$\bf{\Delta_4}$} $-$ \\
\hline\hline
\multicolumn{5}{|c|}{$\boldsymbol{r_{\rm c}} = 100$ au} \\ 
\hline\hline

\cline{1-3} 
&  &  & \makecell[l]{\textbf{SN}} 100 & \makecell[l]{$\bf{\Delta_1}$} 100 \\

0.01 & 0.08 & $-$ & \makecell[l]{\textbf{N}} $-$ & \makecell[l]{$\bf{\Delta_{14}}$}  $-$ \\

&  &  & \makecell[l]{\textbf{GG}} $-$ & \makecell[l]{$\bf{\Delta_4}$} $-$ \\
\hline

\cline{1-3} 
&  &  & \makecell[l]{\textbf{SN}} 100 & \makecell[l]{$\bf{\Delta_1}$} 100 \\

0.015 & 0.07 & $-$ & \makecell[l]{\textbf{N}} $-$ & \makecell[l]{$\bf{\Delta_{14}}$} $-$ \\

&  &  & \makecell[l]{\textbf{GG}} $-$ & \makecell[l]{$\bf{\Delta_4}$} $-$ \\
\hline

\cline{1-3} 
&  &  & \makecell[l]{\textbf{SN}} 100 & \makecell[l]{$\bf{\Delta_1}$} 100 \\

0.02 & 0.06 & $-$ & \makecell[l]{\textbf{N}} $-$ & \makecell[l]{$\bf{\Delta_{14}}$} $-$ \\

 &  &  & \makecell[l]{\textbf{GG}} $-$ & \makecell[l]{$\bf{\Delta_4}$} $-$ \\ 
\hline\hline
\multicolumn{5}{|c|}{$\boldsymbol{r_{\rm c}} = 150$ au} \\ 
\hline\hline

\cline{1-3} 
 &  &  & \makecell[l]{\textbf{SN}} 100 & \makecell[l]{$\bf{\Delta_1}$} 100 \\

0.01 & 0.13 & $-$ & \makecell[l]{\textbf{N}} $-$ & \makecell[l]{$\bf{\Delta_{14}}$} $-$ \\

 &  &  & \makecell[l]{\textbf{GG}} $-$ & \makecell[l]{$\bf{\Delta_4}$} $-$ \\
\hline

\cline{1-3} 
&  &  & \makecell[l]{\textbf{SN}} 100 & \makecell[l]{$\bf{\Delta_1}$} 100 \\

0.015 & 0.09 & $-$ & \makecell[l]{\textbf{N}} $-$ & \makecell[l]{$\bf{\Delta_{14}}$} $-$ \\

 &  &  & \makecell[l]{\textbf{GG}} $-$ & \makecell[l]{$\bf{\Delta_4}$} $-$ \\
\hline

\cline{1-3}
&  &  & \makecell[l]{\textbf{SN}} 92 & \makecell[l]{$\bf{\Delta_1}$} 100 \\

0.02 & 0.07 & $-$ & \makecell[l]{\textbf{N}} 8 & \makecell[l]{$\bf{\Delta_{14}}$} $-$ \\

 &  &  & \makecell[l]{\textbf{GG}} $-$ & \makecell[l]{$\bf{\Delta_4}$} $-$ \\ 
\hline
\end{tabular}
\tablefoot{\\The values related to the vertically arranged items of columns $n_F$ and $n_p$ are presented on their right, respectively. See Section \ref{sec:classesdefinition} for the definition of the three classes of planets. Following, a brief description of the parameters reported in the table (see Section \ref{sec:discmodel}, \ref{sec:accretionmodel} and \ref{sec:timing} for details).\\
\tablefoottext{a}{$\Delta t_1$: formation time of the first planet reaching the end of its formation process.} \\
\tablefoottext{b}{$\Delta t_g$: formation time of the first gas giant planet.}\\
\tablefoottext{c}{$n_F$: percentage of (i) Sub-Neptunes (SN), (ii) Neptunes (N), (iii) Gas Giants (GG) formed with respect to the total number of initial seeds.} \\
\tablefoottext{d}{$n_p$: percentage of planets located within the ranges $\Delta_1 = r_{in}-1$ au, $\Delta_{14} = 1-4$ au and $\Delta_4 = 4\ \text{AU}-r_c$.}
} 
\end{table}

\begin{table}[p]
\caption{Formation times, fraction and positions of the planets produced by the circumbinary disc of DWD$_3$ when $t_0 = 1$ Myr.}
\label{table:1dwd3}
\centering
\begin{tabular}{|c|c|c|r|c|} 
\hline
\multicolumn{5}{|c|}{\rule{0pt}{3.2ex}\large{$t_0= 1$ Myr,\ $\Teff$$_2 = 24900$ K}} \\ [1.5ex]
\cline{1-5}
\hline
\hline
$\xi$ & $\Delta t_1$ & $\Delta t_g$ & \makecell[c]{$n_F$} & $n_p$\\ 
& [Myr] & [Myr] & \makecell[c]{[\%]} & [\%] \\
\hline\hline
\multicolumn{5}{|c|}{$\boldsymbol{r_{\rm c}} = 50$ au} \\ 
\hline

\cline{1-3} 
&   &   & \makecell[l]{\textbf{SN}} 100 & \makecell[l]{$\bf{\Delta_1}$} 100 \\

0.01 & 0.13 & $-$ & \makecell[l]{\textbf{N}} $-$ & \makecell[l]{$\bf{\Delta_{14}}$} $-$ \\

&  &  & \makecell[l]{\textbf{GG}} $-$ & \makecell[l]{$\bf{\Delta_4}$} $-$ \\
\hline

\cline{1-3} 
&  &  & \makecell[l]{\textbf{SN}} 100 & \makecell[l]{$\bf{\Delta_1}$} 100 \\

0.015 & 0.13 & $-$ & \makecell[l]{\textbf{N}} $-$ & \makecell[l]{$\bf{\Delta_{14}}$} $-$ \\

&  &  & \makecell[l]{\textbf{GG}} $-$ & \makecell[l]{$\bf{\Delta_4}$} $-$ \\
\hline

\cline{1-3} 
&  &  & \makecell[l]{\textbf{SN}} 100 &\makecell[l]{$\bf{\Delta_1}$} 100 \\

0.02 & 0.12 & $-$ & \makecell[l]{\textbf{N}} $-$ & \makecell[l]{$\bf{\Delta_{14}}$} $-$ \\

&  &  & \makecell[l]{\textbf{GG}} $-$ & \makecell[l]{$\bf{\Delta_4}$} $-$ \\
\hline\hline
\multicolumn{5}{|c|}{$\boldsymbol{r_{\rm c}} = 100$ au} \\
\hline\hline

\cline{1-3} 
&  &  & \makecell[l]{\textbf{SN}} 100 & \makecell[l]{$\bf{\Delta_1}$} 100 \\

0.01 & 0.2 & $-$ & \makecell[l]{\textbf{N}} $-$ & \makecell[l]{$\bf{\Delta_{14}}$} $-$ \\

&  &  & \makecell[l]{\textbf{GG}} $-$ & \makecell[l]{$\bf{\Delta_4}$} $-$ \\
\hline

\cline{1-3} 
&  &  & \makecell[l]{\textbf{SN}} 100 & \makecell[l]{$\bf{\Delta_1}$} 100 \\

0.015 & 0.15 & $-$ & \makecell[l]{\textbf{N}} $-$ & \makecell[l]{$\bf{\Delta_{14}}$} $-$ \\

 &  &  & \makecell[l]{\textbf{GG}} $-$ & \makecell[l]{$\bf{\Delta_4}$} $-$ \\
\hline

\cline{1-3} 
 &  &  & \makecell[l]{\textbf{SN}} 100 & \makecell[l]{$\bf{\Delta_1}$} 100 \\

0.02 & 0.13 & $-$ & \makecell[l]{\textbf{N}} $-$ & \makecell[l]{$\bf{\Delta_{14}}$} $-$ \\

 &  &  & \makecell[l]{\textbf{GG}} $-$ & \makecell[l]{$\bf{\Delta_4}$} $-$ \\ 
\hline\hline
\multicolumn{5}{|c|}{$\boldsymbol{r_{\rm c}} = 150$ au} \\ 
\hline\hline

\cline{1-3} 
&  &  & \makecell[l]{\textbf{SN}} 100 & \makecell[l]{$\bf{\Delta_1}$} 100 \\

0.01 & 0.3 & $-$ & \makecell[l]{\textbf{N}} $-$ & \makecell[l]{$\bf{\Delta_{14}}$} $-$ \\

&  &  & \makecell[l]{\textbf{GG}} $-$ & \makecell[l]{$\bf{\Delta_4}$} $-$ \\
\hline

\cline{1-3} 
&  &  & \makecell[l]{\textbf{SN}} 100 & \makecell[c]{$\bf{\Delta_1}$} 100 \\

0.015 & 0.25 & $-$ & \makecell[l]{\textbf{N}} $-$ & \makecell[l]{$\bf{\Delta_{14}}$} $-$ \\

 &  &  & \makecell[l]{\textbf{GG}} $-$ & \makecell[l]{$\bf{\Delta_4}$} $-$ \\
\hline

\cline{1-3} 
 &  &  & \makecell[l]{\textbf{SN}} 100 & \makecell[l]{$\bf{\Delta_1}$} 100 \\

0.02 & 0.16 & $-$ & \makecell[l]{\textbf{N}} $-$ & \makecell[l]{$\bf{\Delta_{14}}$} $-$ \\

 &  &  & \makecell[l]{\textbf{GG}} $-$ & \makecell[l]{$\bf{\Delta_4}$} $-$ \\ 
\hline
\end{tabular}
\tablefoot{\\The values related to the vertically arranged items of columns $n_F$ and $n_p$ are presented on their right, respectively. See Section \ref{sec:classesdefinition} for the definition of the three classes of planets. Following, a brief description of the parameters reported in the table (see Section \ref{sec:discmodel}, \ref{sec:accretionmodel} and \ref{sec:timing} for details).\\
\tablefoottext{a}{$\Delta t_1$: formation time of the first planet reaching the end of its formation process.} \\
\tablefoottext{b}{$\Delta t_g$: formation time of the first gas giant planet.}\\
\tablefoottext{c}{$n_F$: percentage of (i) Sub-Neptunes (SN), (ii) Neptunes (N), (iii) Gas Giants (GG) formed with respect to the total number of initial seeds.} \\
\tablefoottext{d}{$n_p$: percentage of planets located within the ranges $\Delta_1 = r_{in}-1$ au, $\Delta_{14} = 1-4$ au and $\Delta_4 = 4\ \text{AU}-r_c$.}
} 
\end{table}

\begin{table}[p]
\caption{Formation times, fraction and positions of the planets produced by the circumbinary disc of DWD$_4$ when $t_0 = 10\tau_c$.}
\label{table:taudwd4}
\centering
\begin{tabular}{|c|c|c|r|c|} 
\hline
\multicolumn{5}{|c|}{\rule{0pt}{3.2ex}\large{$t_0= 10\tau_c$,\ $\Teff$$_2 = 50100$ K}} \\ [1.5ex]
\cline{1-5}
\hline
\hline
$\xi$ & $\Delta t_1$ & $\Delta t_g$ & \makecell[c]{$n_F$} & $n_p$\\ 
& [Myr] & [Myr] & \makecell[c]{[\%]} & [\%] \\
\hline\hline
\multicolumn{5}{|c|}{$\boldsymbol{r_{\rm c}} = 50$ au} \\ 
\hline

\cline{1-3} 
&   &   & \makecell[l]{\textbf{SN}} 85 & \makecell[l]{$\bf{\Delta_1}$} 100 \\

0.01 & 0.1 & $-$ & \makecell[l]{\textbf{N}} 15 & \makecell[l]{$\bf{\Delta_{14}}$} $-$ \\

 &  &  & \makecell[l]{\textbf{GG}} $-$ & \makecell[l]{$\bf{\Delta_4}$} $-$ \\
\hline

\cline{1-3}
&  &  & \makecell[l]{\textbf{SN}} 77 & \makecell[l]{$\bf{\Delta_1}$} 100 \\

0.015 & 0.07 & $-$ & \makecell[l]{\textbf{N}} 23 & \makecell[l]{$\bf{\Delta_{14}}$} $-$ \\

 &  &  & \makecell[l]{\textbf{GG}} $-$ & \makecell[l]{$\bf{\Delta_4}$} $-$ \\
\hline

\cline{1-3} 
&  &  & \makecell[l]{\textbf{SN}} 62 & \makecell[l]{$\bf{\Delta_1}$} 100 \\

0.02 & 0.06 & $-$ & \makecell[l]{\textbf{N}} 38 & \makecell[l]{$\bf{\Delta_{14}}$} $-$ \\

&  &  & \makecell[l]{\textbf{GG}} $-$ & \makecell[l]{$\bf{\Delta_4}$} $-$ \\
\hline\hline
\multicolumn{5}{|c|}{$\boldsymbol{r_{\rm c}} = 100$ au} \\ 
\hline\hline

\cline{1-3} 
&  &  & \makecell[l]{\textbf{SN}} 56 & \makecell[l]{$\bf{\Delta_1}$} 100 \\

0.01 & 0.25 & $-$ & \makecell[l]{\textbf{N}} 44 & \makecell[l]{$\bf{\Delta_{14}}$} $-$ \\

&  &  & \makecell[l]{\textbf{GG}} $-$ & \makecell[l]{$\bf{\Delta_4}$} $-$ \\
\hline

\cline{1-3} 
&  &  & \makecell[l]{\textbf{SN}} 45 & \makecell[l]{$\bf{\Delta_1}$} 95 \\

0.015 & 0.18 & 1.2 & \makecell[l]{\textbf{N}} 50 & \makecell[l]{$\bf{\Delta_{14}}$} 5 \\

&  &  & \makecell[l]{\textbf{GG}} 5 & \makecell[l]{$\bf{\Delta_4}$} $-$ \\
\hline

\cline{1-3} 
&  &  & \makecell[l]{\textbf{SN}} 33 & \makecell[l]{$\bf{\Delta_1}$} 72 \\

0.02 & 0.14 & 1 & \makecell[l]{\textbf{N}} 28 & \makecell[l]{$\bf{\Delta_{14}}$} 28 \\

 &  &  & \makecell[l]{\textbf{GG}} 39 &\makecell[l]{ $\bf{\Delta_4}$} $-$ \\ 
\hline\hline
\multicolumn{5}{|c|}{$\boldsymbol{r_{\rm c}} = 150$ au} \\
\hline\hline

\cline{1-3} 
&  &  & \makecell[l]{\textbf{SN}} 48 & \makecell[l]{$\bf{\Delta_1}$} 100 \\

0.01 & 0.4 & $-$ & \makecell[l]{\textbf{N}} 33 & \makecell[l]{$\bf{\Delta_{14}}$} $-$ \\

&  &  & \makecell[l]{\textbf{GG}} $-$ & \makecell[l]{$\bf{\Delta_4}$} $-$ \\
\hline

\cline{1-3} 
&  &  & \makecell[l]{\textbf{SN}} 33 & \makecell[l]{$\bf{\Delta_1}$} 67 \\

0.015 & 0.3 & $-$ & \makecell[l]{\textbf{N}} 67 & \makecell[l]{$\bf{\Delta_{14}}$} 33 \\

 &  &  & \makecell[l]{\textbf{GG}} $-$ & \makecell[l]{$\bf{\Delta_4}$} $-$ \\
\hline

\cline{1-3} 
&  &  & \makecell[l]{\textbf{SN}} 24 & \makecell[l]{$\bf{\Delta_1}$} 57 \\

0.02 & 0.22 & $-$ & \makecell[l]{\textbf{N}} 76 & \makecell[l]{$\bf{\Delta_{14}}$} 43 \\

&  &  & \makecell[l]{\textbf{GG}} $-$ & \makecell[l]{$\bf{\Delta_4}$} $-$ \\ 
\hline
\end{tabular}
\tablefoot{\\The values related to the vertically arranged items of columns $n_F$ and $n_p$ are presented on their right, respectively. See Section \ref{sec:classesdefinition} for the definition of the three classes of planets. Following, a brief description of the parameters reported in the table (see Section \ref{sec:discmodel}, \ref{sec:accretionmodel} and \ref{sec:timing} for details).\\
\tablefoottext{a}{$\Delta t_1$: formation time of the first planet reaching the end of its formation process.} \\
\tablefoottext{b}{$\Delta t_g$: formation time of the first gas giant planet.}\\
\tablefoottext{c}{$n_F$: percentage of (i) Sub-Neptunes (SN), (ii) Neptunes (N), (iii) Gas Giants (GG) formed with respect to the total number of initial seeds.} \\
\tablefoottext{d}{$n_p$: percentage of planets located within the ranges $\Delta_1 = r_{in}-1$ au, $\Delta_{14} = 1-4$ au and $\Delta_4 = 4\ \text{AU}-r_c$.}
} 
\end{table}

\begin{table}[p]
\caption{Formation times, fraction and positions of the planets produced by the circumbinary disc of DWD$_4$ when $t_0 = 0.1$ Myr.}
\label{table:0.1dwd4}
\centering
\begin{tabular}{|c|c|c|r|c|} 
\hline
\multicolumn{5}{|c|}{\rule{0pt}{3.2ex}\large{$t_0= 0.1$ Myr,\ $\Teff$$_2 = 32200$ K}} \\ [1.5ex]
\cline{1-5}
\hline
\hline
$\xi$ & $\Delta t_1$ & $\Delta t_g$ & \makecell[c]{$n_F$} & $n_p$\\ 
& [Myr] & [Myr] & \makecell[c]{[\%]} & [\%] \\
\hline\hline
\multicolumn{5}{|c|}{$\boldsymbol{r_{\rm c}} = 50$ au} \\ 
\hline

\cline{1-3} 
&   &   & \makecell[l]{\textbf{SN}} 100 & \makecell[l]{$\bf{\Delta_1}$} 100 \\

0.01 & 0.11 & $-$ & \makecell[l]{\textbf{N}} $-$ & \makecell[l]{$\bf{\Delta_{14}}$} $-$ \\

 &  &  & \makecell[l]{\textbf{GG}} $-$ & \makecell[l]{$\bf{\Delta_4}$} $-$ \\
\hline

\cline{1-3}
&  &  & \makecell[l]{\textbf{SN}} 100 & \makecell[l]{$\bf{\Delta_1}$} 100 \\

0.015 & 0.08 & $-$ & \makecell[l]{\textbf{N}} $-$ & \makecell[l]{$\bf{\Delta_{14}}$} $-$ \\

 &  &  & \makecell[l]{\textbf{GG}} $-$ & \makecell[l]{$\bf{\Delta_4}$} $-$ \\
\hline

\cline{1-3} 
&  &  & \makecell[l]{\textbf{SN}} 100 & \makecell[l]{$\bf{\Delta_1}$} 100 \\

0.02 & 0.07 & $-$ & \makecell[l]{\textbf{N}} $-$ & \makecell[l]{$\bf{\Delta_{14}}$} $-$ \\

&  &  & \makecell[l]{\textbf{GG}} $-$ & \makecell[l]{$\bf{\Delta_4}$} $-$ \\
\hline\hline
\multicolumn{5}{|c|}{$\boldsymbol{r_{\rm c}} = 100$ au} \\ 
\hline\hline

\cline{1-3}
&  &  & \makecell[l]{\textbf{SN}} 100 & \makecell[l]{$\bf{\Delta_1}$} 100 \\

0.01 & 0.21 & $-$ & \makecell[l]{\textbf{N}} $-$ & \makecell[l]{$\bf{\Delta_{14}}$} $-$ \\

 &  &  & \makecell[l]{\textbf{GG}} $-$ & \makecell[l]{$\bf{\Delta_4}$} $-$ \\
\hline

\cline{1-3} 
&  &  & \makecell[l]{\textbf{SN}} 100 & \makecell[l]{$\bf{\Delta_1}$} 100 \\

0.015 & 0.15 & $-$ & \makecell[l]{\textbf{N}} $-$ & \makecell[l]{$\bf{\Delta_{14}}$} $-$ \\

 &  &  & \makecell[l]{\textbf{GG}} $-$ & \makecell[l]{$\bf{\Delta_4}$} $-$ \\
\hline

\cline{1-3} 
&  &  & \makecell[l]{\textbf{SN}} 73 & \makecell[l]{$\bf{\Delta_1}$} 100 \\

0.02 & 0.13 & $-$ & \makecell[l]{\textbf{N}} 27 & \makecell[l]{$\bf{\Delta_{14}}$} $-$ \\

 &  &  & \makecell[l]{\textbf{GG}} $-$ & \makecell[l]{$\bf{\Delta_4}$} $-$ \\ 
\hline\hline
\multicolumn{5}{|c|}{$\boldsymbol{r_{\rm c}} = 150$ au} \\ 
\hline\hline

\cline{1-3} 
&  &  & \makecell[l]{\textbf{SN}} 100 & \makecell[l]{$\bf{\Delta_1}$} 100 \\

0.01 & 0.33 & $-$ & \makecell[l]{\textbf{N}} $-$ & \makecell[l]{$\bf{\Delta_{14}}$} $-$ \\

 &  &  & \makecell[l]{\textbf{GG}} $-$ & \makecell[l]{$\bf{\Delta_4}$} $-$ \\
\hline

\cline{1-3} 
&  &  & \makecell[l]{\textbf{SN}} 72 & \makecell[l]{$\bf{\Delta_1}$} 100 \\

0.015 & 0.25 & $-$ & \makecell[l]{\textbf{N}} 28 & \makecell[l]{$\bf{\Delta_{14}}$} $-$ \\

&  &  & \makecell[l]{\textbf{GG}} $-$ & \makecell[l]{$\bf{\Delta_4}$} $-$ \\
\hline

\cline{1-3}
&  &  & \makecell[l]{\textbf{SN}} 61 & \makecell[l]{$\bf{\Delta_1}$} 100 \\

0.02 & 0.19 & $-$ & \makecell[l]{\textbf{N}} 39 & \makecell[l]{$\bf{\Delta_{14}}$} $-$ \\

 &  &  & \makecell[l]{\textbf{GG}} $-$ & \makecell[l]{$\bf{\Delta_4}$} $-$ \\ 
\hline
\end{tabular}
\tablefoot{\\The values related to the vertically arranged items of columns $n_F$ and $n_p$ are presented on their right, respectively. See Section \ref{sec:classesdefinition} for the definition of the three classes of planets. Following, a brief description of the parameters reported in the table (see Section \ref{sec:discmodel}, \ref{sec:accretionmodel} and \ref{sec:timing} for details).\\
\tablefoottext{a}{$\Delta t_1$: formation time of the first planet reaching the end of its formation process.} \\
\tablefoottext{b}{$\Delta t_g$: formation time of the first gas giant planet.}\\
\tablefoottext{c}{$n_F$: percentage of (i) Sub-Neptunes (SN), (ii) Neptunes (N), (iii) Gas Giants (GG) formed with respect to the total number of initial seeds.} \\
\tablefoottext{d}{$n_p$: percentage of planets located within the ranges $\Delta_1 = r_{in}-1$ au, $\Delta_{14} = 1-4$ au and $\Delta_4 = 4\ \text{AU}-r_c$.}
} 
\end{table}

\begin{table}[p]
\caption{Formation times, fraction and positions of the planets produced by the circumbinary disc of DWD$_4$ when $t_0 = 1$ Myr.}
\label{table:1dwd4}
\centering
\begin{tabular}{|c|c|c|r|c|} 
\hline
\multicolumn{5}{|c|}{\rule{0pt}{3.2ex}\large{$t_0= 1$ Myr,\ $\Teff$$_2 = 24900$ K}} \\ [1.5ex]
\cline{1-5}
\hline
\hline
$\xi$ & $\Delta t_1$ & $\Delta t_g$ & \makecell[c]{$n_F$} & $n_p$\\ 
& [Myr] & [Myr] & \makecell[c]{[\%]} & [\%] \\
\hline\hline
\multicolumn{5}{|c|}{$\boldsymbol{r_{\rm c}} = 50$ au} \\ 
\hline

\cline{1-3} 
&   &   & \makecell[l]{\textbf{SN}} 100 & \makecell[l]{$\bf{\Delta_1}$} 100 \\

0.01 & 0.55 & $-$ & \makecell[l]{\textbf{N}} $-$ & \makecell[l]{$\bf{\Delta_{14}}$} $-$ \\

&  &  & \makecell[l]{\textbf{GG}} $-$ & \makecell[l]{$\bf{\Delta_4}$} $-$ \\
\hline

\cline{1-3} 
&  &  & \makecell[l]{\textbf{SN}} 100 & \makecell[l]{$\bf{\Delta_1}$} 100 \\

0.015 & 0.4 & $-$ & \makecell[l]{\textbf{N}} $-$ & \makecell[l]{$\bf{\Delta_{14}}$} $-$ \\

&  &  & \makecell[l]{\textbf{GG}} $-$ & \makecell[l]{$\bf{\Delta_4}$} $-$ \\
\hline

\cline{1-3}
&  &  & \makecell[l]{\textbf{SN}} 100 & \makecell[l]{$\bf{\Delta_1}$} 100 \\

0.02 & 0.3 & $-$ & \makecell[l]{\textbf{N}} $-$ & \makecell[l]{$\bf{\Delta_{14}}$} $-$ \\

&  &  & \makecell[l]{\textbf{GG}} $-$ & \makecell[l]{$\bf{\Delta_4}$} $-$ \\
\hline\hline
\multicolumn{5}{|c|}{$\boldsymbol{r_{\rm c}} = 100$ au} \\ 
\hline\hline

\cline{1-3} 
&  &  & \makecell[l]{\textbf{SN}} 100 & \makecell[l]{$\bf{\Delta_1}$} 100 \\

0.01 & 0.6 & $-$ & \makecell[l]{\textbf{N}} $-$ & \makecell[l]{$\bf{\Delta_{14}}$} $-$ \\

&  &  & \makecell[l]{\textbf{GG}} $-$ & \makecell[l]{$\bf{\Delta_4}$} $-$ \\
\hline

\cline{1-3} 
&  &  & \makecell[l]{\textbf{SN}} 100 & \makecell[l]{$\bf{\Delta_1}$} 100 \\

0.015 & 0.4 & $-$ & \makecell[l]{\textbf{N}} $-$ & \makecell[l]{$\bf{\Delta_{14}}$} $-$ \\

&  &  & \makecell[l]{\textbf{GG}} $-$ & \makecell[l]{$\bf{\Delta_4}$} $-$ \\
\hline

\cline{1-3} 
&  &  & \makecell[l]{\textbf{SN}} 100 & \makecell[l]{$\bf{\Delta_1}$} 100 \\

0.02 & 0.35 & $-$ & \makecell[l]{\textbf{N}} $-$ & \makecell[l]{$\bf{\Delta_{14}}$} $-$ \\

&  &  & \makecell[l]{\textbf{GG}} $-$ & \makecell[l]{$\bf{\Delta_4}$} $-$ \\ 
\hline\hline
\multicolumn{5}{|c|}{$\boldsymbol{r_{\rm c}} = 150$ au} \\
\hline\hline

\cline{1-3} 
&  &  & \makecell[l]{\textbf{SN}} 100 & \makecell[l]{$\bf{\Delta_1}$} 100 \\

0.01 & 0.65 & $-$ & \makecell[l]{\textbf{N}} $-$ & \makecell[l]{$\bf{\Delta_{14}}$} $-$ \\

&  &  & \makecell[l]{\textbf{GG}} $-$ & \makecell[l]{$\bf{\Delta_4}$} $-$ \\
\hline

\cline{1-3}
&  &  & \makecell[l]{\textbf{SN}} 100 & \makecell[l]{$\bf{\Delta_1}$} 100 \\

0.015 & 0.45 & $-$ & \makecell[l]{\textbf{N}} $-$ & \makecell[l]{$\bf{\Delta_{14}}$} $-$ \\

&  &  & \makecell[l]{\textbf{GG}} $-$ & \makecell[l]{$\bf{\Delta_4}$} $-$ \\
\hline

\cline{1-3} 
&  &  & \makecell[l]{\textbf{SN}} 82 & \makecell[l]{$\bf{\Delta_1}$} 100 \\

0.02 & 0.4 & $-$ & \makecell[l]{\textbf{N}} 18 & \makecell[l]{$\bf{\Delta_{14}}$} $-$ \\

&  &  & \makecell[l]{\textbf{GG}} $-$ & \makecell[l]{$\bf{\Delta_4}$} $-$ \\ 
\hline
\end{tabular}
\tablefoot{\\The values related to the vertically arranged items of columns $n_F$ and $n_p$ are presented on their right, respectively. See Section \ref{sec:classesdefinition} for the definition of the three classes of planets. Following, a brief description of the parameters reported in the table (see Section \ref{sec:discmodel}, \ref{sec:accretionmodel} and \ref{sec:timing} for details).\\
\tablefoottext{a}{$\Delta t_1$: formation time of the first planet reaching the end of its formation process.} \\
\tablefoottext{b}{$\Delta t_g$: formation time of the first gas giant planet.}\\
\tablefoottext{c}{$n_F$: percentage of (i) Sub-Neptunes (SN), (ii) Neptunes (N), (iii) Gas Giants (GG) formed with respect to the total number of initial seeds.} \\
\tablefoottext{d}{$n_p$: percentage of planets located within the ranges $\Delta_1 = r_{in}-1$ au, $\Delta_{14} = 1-4$ au and $\Delta_4 = 4\ \text{AU}-r_c$.}
} 
\end{table}

\end{appendix}

\end{document}